\documentclass[twocolumn, nofootinbib, prd]{revtex4}
\usepackage{graphicx}
\usepackage{amsmath}
\usepackage{amssymb}
\usepackage{verbatim}
\usepackage{color}
\usepackage{hyperref}

    \renewcommand{\dbltopfraction}{0.9}	
    \renewcommand{\dblfloatpagefraction}{0.6}	

\begin{document}
\title{Rational Orbits around Charged Black Holes}
\author{Vedant Misra${}^{*}$ and Janna Levin${}^{**,!}$}
\affiliation{${}^{*}$Physics Department, Columbia University,
New York, NY 10027}
\affiliation{${}^{**}$Department of Physics and Astronomy, Barnard
College of Columbia University, 3009 Broadway, New York, NY 10027}
\affiliation{${}^{!}$Institute for Strings, Cosmology and Astroparticle
  Physics, Columbia University, New York, NY 10027}
\affiliation{vedant@phys.columbia.edu}
\affiliation{janna@astro.columbia.edu}

\widetext

\begin{abstract}
We show that all eccentric
timelike orbits in Reissner-Nordstr\"{o}m spacetime 
can be classified using a taxonomy that draws upon an isomorphism between
periodic orbits and the set of rational numbers.  By virtue of the fact
that the rationals are dense, the taxonomy can be used to
approximate aperiodic orbits with periodic orbits.  This may help
reduce computational overhead for calculations in gravitational wave
astronomy.  Our dynamical systems approach 
enables us to study orbits for both charged and uncharged particles
in spite of the fact that charged particle orbits around a charged black hole do not
admit a simple one-dimensional effective potential description.  Finally,
we show that comparing periodic orbits in the RN and Schwarzschild geometries
enables us to distinguish charged and uncharged spacetimes by looking
only at the orbital dynamics.
\end{abstract}

\maketitle
\vfill
\eject

\section{Introduction}

Black holes are useful
large-scale laboratories for testing general relativity in strong 
gravitational fields.  While directly observing black holes
proves difficult because they emit no electromagnetic radiation, 
black hole pairs can be detected via the gravitational radiation they may emit.
The terrestrial network of interferometric
gravitational wave detectors and the proposed space-based detector
LISA (Laser Interferometer Space Antenna), are expected to
detect gravitational radiation and launch an
era of gravitational wave astronomy that would make use of 
direct observations of black holes. 

Because gravitational waves are shaped by the motion of
massive celestial objects, extracting astrophysically meaningful
information from them requires a comprehensive theoretical
understanding of the sources' underlying dynamics \cite{flanagan, glampedakis2,
drasco1, drasco2, drasco3, lang}.  
For example, one popular processing method, 
matched filtering, makes use of
a template signal, generated using theoretical predictions, to
find signals in noisy detector output
\cite{{glampedakis},{grishchuk}}.
Template signal generation is an example of the type of
computationally expensive process encountered in gravitational 
wave astronomy when studying aperiodic orbits in the strong-field regime.  

A method for approximating aperiodic orbits with periodic orbits, which could
cut down significantly on computational expenses, was introduced
in an earlier paper \cite{levin}.  
The approximation method takes the form of a taxonomy that assigns
to each periodic orbit a rational number.  By virtue of the fact
that the rationals are dense, the taxonomy can be used to approximate
aperiodic orbits with periodic orbits to arbitrary precision.  Because
periodic orbits might have Fourier series that converge more rapidly
than those of aperiodic orbits, and because for periodic orbits the
evolution of a geometry's conserved quantities may be
interdependent, calculations pertaining to periodic orbits might be less 
computationally intensive than those for aperiodic orbits \cite{levin}.

The taxonomy was applied to the Kerr
geometry in \cite{levin} and \cite{levin_el} and to black hole pairs
in \cite{levin_d1} and \cite{levin_d2}.  Here, we will extend the approach to
the Reissner-Nordstr\"{o}m (RN) solution to the Einstein field
equations, which describes the gravitational field of a static,
non-rotating, electrically charged, spherically symmetric body
\cite{carroll}.  Studying a geometry of this type is less astrophysically
motivated than studying its electrically neutral counterpart.
Were it to form in spite of
the fact that the electromagnetic repulsion in compressing an
electrically charged mass is greater by about 40 orders of magnitude
than the gravitational attraction, it would 
neutralize its own electric charge if enough opposite charge were available.
Nonetheless, in the spirit of being prepared for the unexpected, and in
support of the ambitious gravitational wave experiments coming online, we will not
presumptively exclude any possible sources.  
Were gravitational waves from a charged black hole
candidate detected experimentally, a thorough understanding of the RN
orbits could help identify their source.

At the other extreme, microscopic black holes that might form in
accelerator experiments may be charged. A pair of black
holes that scatter and then evaporate might be described by a
scattering amplitude that is a sum over these classical paths. The
solutions we describe might find application in particle physics as
well as astrophysics.

Both Kerr and RN orbits bear many
qualitative similarities to orbits in Schwarzschild spacetime,
including ``zoom-whirl" behavior \cite{barack, glampedakis}, 
in which the test particle zooms
away from the central mass quasielliptically to successive apastra
separated by nearly circular whirls.
It is this behavior which the taxonomy exploits.  
As was the case for Kerr orbits before this taxonomy was introduced,
no unifying framework for making general claims about orbits has 
been applied to the RN spacetime before.

\section{Reissner-Nordstr\"{o}m Effective Potential}
We use
the effective potential formulation of the RN spacetime  
to calculate various interesting dynamical properties of
the geometry. In an appendix we detail the Hamiltonian formulation
that is used to generate the orbits pictured throughout the paper.
We begin with the RN metric,
\begin{equation}
	\label{eqn:metric}
	g_{\mu\nu}=\left(
	\begin{array}{cccc}
		-\Delta & & & \\
		& \Delta^{-1} & & \\
		& & r^2 & \\
		& & & r^2 \sin^2\theta
	\end{array}
	\right),
\end{equation}
where the horizon function $\Delta$ is
\begin{equation}
	\Delta= 1 - \frac{2}{r} + \frac{Q^2}{r^2}.
\end{equation}
We have assumed that the central magnetic charge is zero, which would otherwise change
the horizon function's third term to $(Q^2+P^2)/r^2$ .
We use geometrized units and measure $r$ and the central charge $Q$
in units of ${M}$.

A first constant of motion is always
\begin{equation}
	g_{\mu \nu} \frac{dx^\mu}{d \tau} \frac{dx^\nu}{d \tau} = \kappa, \
\end{equation}
\vspace{-.5cm}
\begin{equation}
	\kappa= \bigg\{
		\begin{array}{l l}
			-1 & \quad \mbox{for timelike geodesics}\\
			0 & \quad \mbox{for null geodesics}\\ 
		\end{array} \ . \nonumber
\end{equation}
Explicitly, for timelike orbits, this gives
\begin{equation}
	\label{eqn:simplified}
	-\Delta\, \dot{t}^2 + 
	\frac{1}{\Delta}\,\dot{r}^2 +
	r^2\, \dot{\theta}^2 +
	r^2 \sin^2 \theta\, \dot{\varphi}^2 
	= -1,
\end{equation}
where an overdot indicates differentiation with respect to an affine parameter $\tau$.

To derive expressions for the conserved quantities we write the Lagrangian for a charged particle, in which the final term is derived using the fact that the only nonvanishing component of the vector potential $A_\mu$ is $A_0$ \cite{chandra}:
\begin{equation}
	\mathcal{L}= \frac{1}{2} 
	\big(-
	\Delta \,\dot{t}^2 +
	\frac{1}{\Delta}\,\dot{r}^2 +
	r^2 \,\dot{\theta}^2 +
	r^2 \sin^2 \theta \,\dot{\varphi}^2 
	\big)
	- \frac{Q Q_*}{r} \dot{t}.
\end{equation}
$Q_*$ is the charge per unit mass of the test particle.  
Geodesics in the RN geometry are constrained to a plane.  Since the
geometry is spherically symmetric, we can set $\theta=\pi/2$
and $\dot{\theta}=0$ so that every geodesic is equatorial.

Using
\begin{equation}
	\frac{d}{d\tau}\left(\frac{\partial\mathcal{L}}{\partial\dot{q}}\right) - 
	\frac{d\mathcal{L}}{dq}=0,
\end{equation}
yields the constants of motion:
\begin{equation}
	\label{eqn:motion}
	p_t = -\Delta \dot t - \frac{Q Q_*}{r} = -E \hspace{0.5in}
	p_\varphi = r^2 \dot \varphi = L. 
\end{equation}

Applying (\ref{eqn:motion}) to (\ref{eqn:simplified}) and simplifying gives us 
\begin{equation}
	\label{eqn:rdotpot}
	\frac{1}{2}\dot{r}^2 + V_{\textrm{eff}} =\mathcal{E}_{\textrm{eff}},
\end{equation}
where 
\begin{widetext}
\begin{eqnarray}
	\label{eqn:potential}
	V_{\textrm{eff}} &=& 
	\left(- 1 + EQQ_* \right)\frac{1}{r} + 
	\left(\frac{ L^2 + Q^2-Q^2Q_*^2 }{2}\right)\frac{1}{r^2} + 
	 \left(-L^2 \right)\frac{1}{r^3} + 
	\left(\frac{Q^2L^2}{2} \right)\frac{1}{r^4} \\
	\mathcal{E}_{\textrm{eff}} &= & \frac{E^2-1}{2}.
\end{eqnarray}
\end{widetext}

Due to the geometry's spherical symmetry, 
the effective potential is one-dimensional -- just as in the
Schwarzschild case -- but with one caveat: the test particle's charge
$Q_*$ is coupled to the energy E.  This prohibits us from writing $V_{\textrm{eff}}$
in a form that is independent of energy without also writing $\mathcal{E}_{\textrm{eff}}$ in a form that is
not constant. First we will consider the case
when $Q_*=0$ so that we can define a true one-dimensional effective
potential.   Applying this condition to (\ref{eqn:potential}) gives us the effective
RN potential for an uncharged particle:
\begin{align}
	\label{eqn:potential2}
	V_{\textrm{eff}} &=-\frac{1}{r}+\frac{L^2 + Q^2}{2
          r^2}-\frac{L^2}{r^3} + \frac{Q^2L^2}{2 r^4}\nonumber \\
        &=-\frac{1}{r}+\frac{\Delta
          L^2}{2r^2}+\frac{Q^2}{2r^2}\nonumber \\
        &=\frac{\Delta L^2}{2r^2}+\frac{\Delta -1}{2} \quad .
\end{align}
We return to the charged test particle in \S \ref{sec:charged}.


\section{Uncharged Particle Orbits in RN Spacetime}
\subsection{Bounds on Q}
For an uncharged particle, the RN geometry is 
qualitatively similar outside the horizon to the Schwarzschild geometry 
\cite{chandra}.   But because 
the RN $V_{\textrm{eff}}$ is larger than the Schwarzschild $V_{\textrm{eff}}$
by a factor of $\frac{Q^2}{2r^2} + \frac{Q^2L^2}{2r^4}$, a thorough 
quantitative analysis is necessary to understand the orbital dynamics.  
Our goal in this section is to 
determine the bounds on $L$ and $Q$ that yield periodic orbits.

Determining bounds on $Q$ is simple: owing to the 
RN geometry's peculiar horizon structure, only certain values of $Q$ are 
realistic \cite{carroll}.  The null hypersurfaces are given by
\begin{equation}
	g^{rr} = \Delta(r) = 1 - \frac{2}{r} + \frac{Q^2}{r^2} = 0,
\end{equation}
which when solved for $r$ gives us the horizons:
\begin{equation}
	r_\pm = 1 \pm \sqrt{1 - Q^2}.
\end{equation}

When $Q^2>1$, the geometry is a naked singularity.  The condition $Q^2=1$ 
describes an extremal RN black hole, which is highly unstable due to the 
fact that adding any mass
at all makes it an undercritically charged black hole.  This
leaves the undercritically charged geometry ($Q^2<1$) as the most 
realistic scenario.  Because $Q$ only appears
in the potential as $Q^2$, the undercritically charged case is 
equivalently given by $-1<Q<1$.

\begin{figure}[htpb]
	\centering
    \begin{minipage}{86mm}
	\vspace{-10pt}
	\includegraphics[width=86mm]{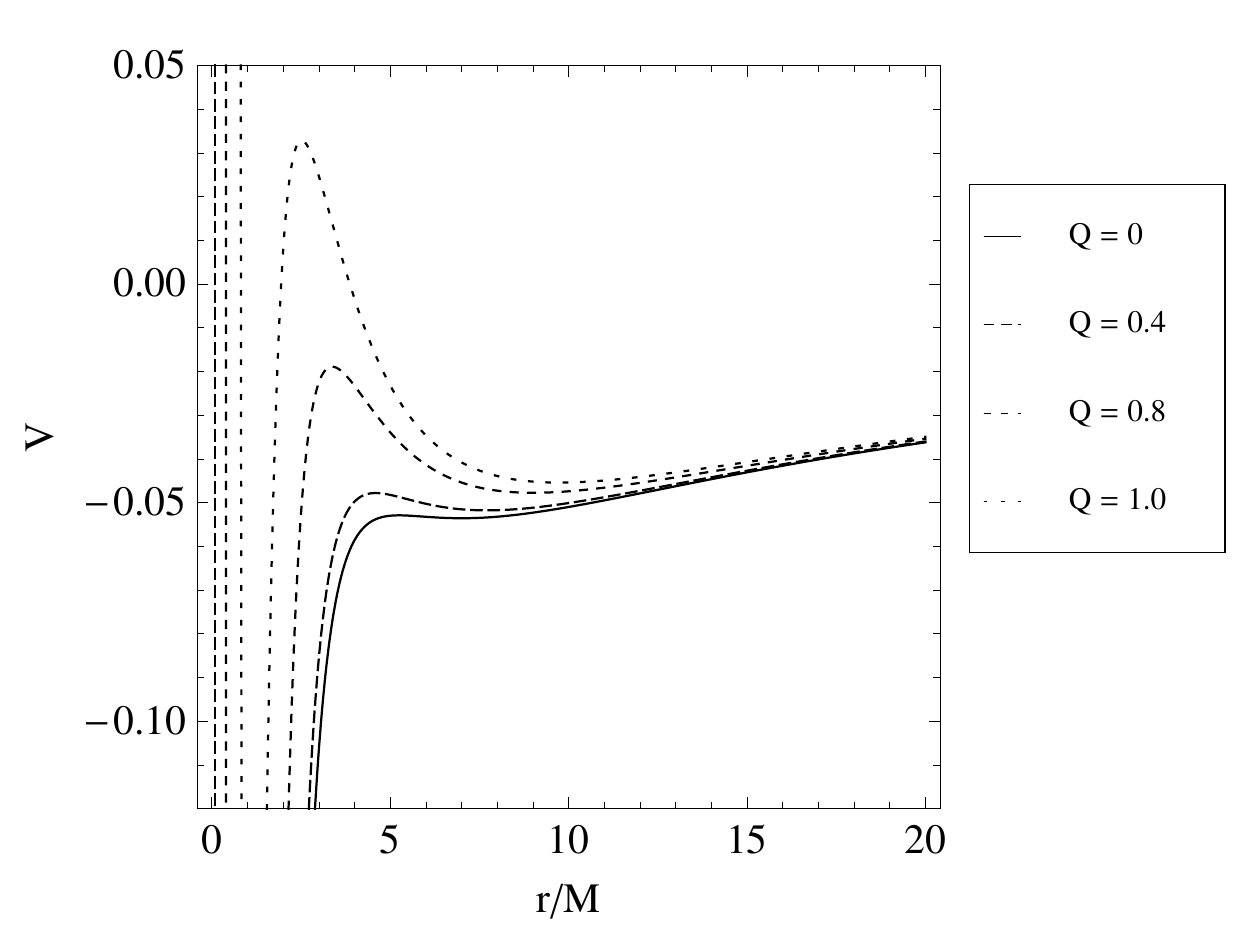}
    \hfill
    \end{minipage}
	\vspace{-20pt}
	\caption{$V_{\textrm{eff}}$ for $Q=0$, $0.4$, $0.8$, and $1$, from bottom to top}
	\label{fig:potentialQ}
\end{figure}

$Q$ affects the shape of the potential as shown
in Figure \ref{fig:potentialQ}. Note that the $Q=0$ case (solid) is 
simply the Schwarzschild effective potential.  
When comparing the RN potential to the Schwarzschild potential, we see that the factor of $+Q^2/2$ in the
$r^{-2}$ term is the reason for the heightened peak and that the $r^{-4}$
term makes the potential blow up at zero. 

\begin{figure}[htpb]
	\centering
    \begin{minipage}{86mm}
	\includegraphics[width=86mm]{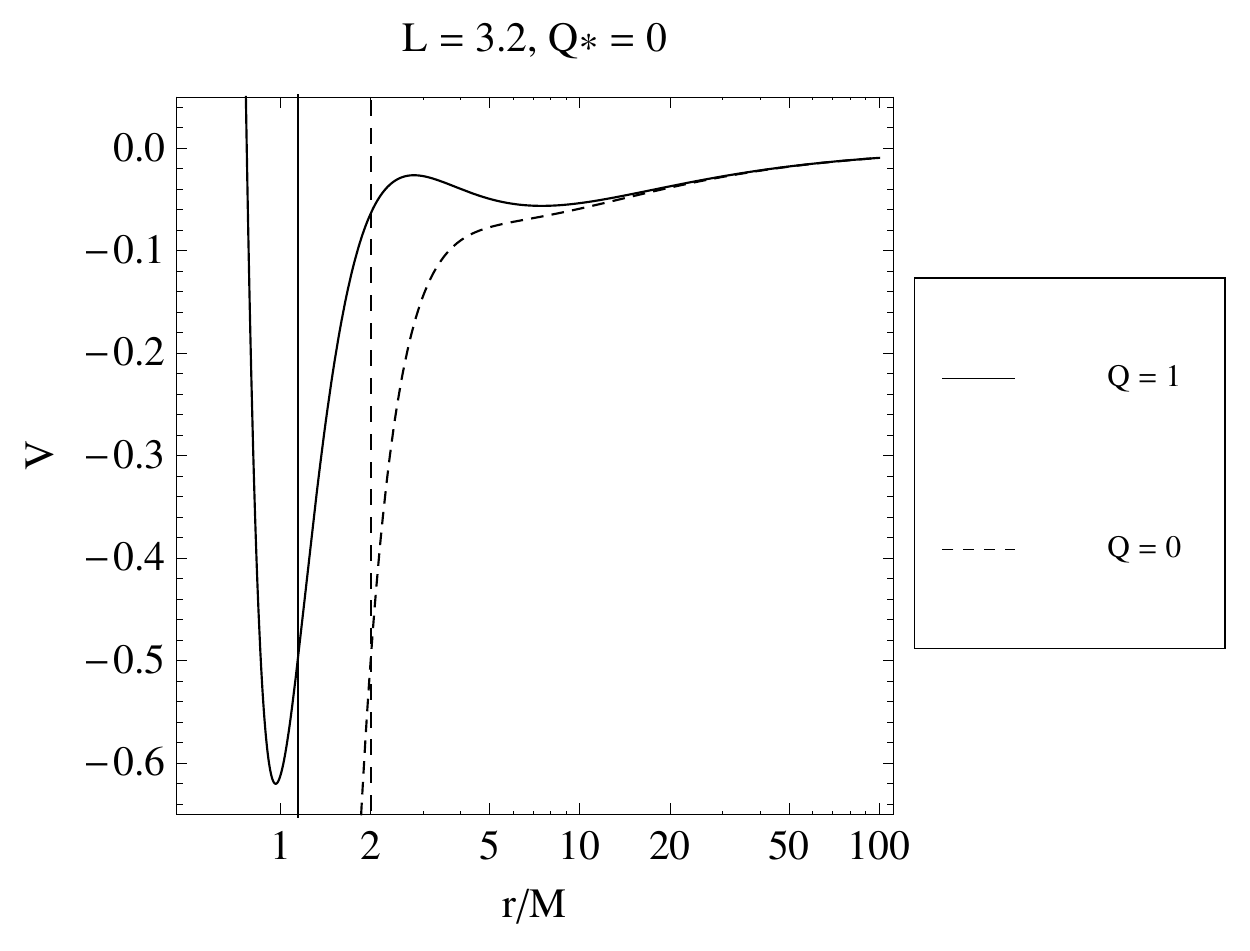}
    \end{minipage}
	\vspace{-15pt}
	\caption{The RN and Schwarzschild potentials with horizons 
	for each and a non-physical internal horizon (dotted) that 
	represents a scenario in which there are multiple stable circular orbits}
	\label{fig:veff3min}
\end{figure}

The Schwarzschild $V_{\textrm{eff}}$ has at most two extrema. 
Because the RN $V_{\textrm{eff}}$ is a quartic, it can have three extrema.
This opens up the possibility of three circular orbits --
two stable and one unstable -- outside the horizon.  The existence of
two stable circular orbits would imply that there are two regions in 
which we can find bounded orbits.  
There would be two stable circular orbits if both minima were to lie outside
the outer horizon, $r_+$.
Figure \ref{fig:veff3min} depicts the Schwarzschild and RN
potentials for $L=3.2$.  The dashed potential is the Schwarzschild
$V_{\textrm{eff}}$ and the Schwarzschild horizon is given by the
dashed vertical line ($r=2$).  The RN $V_{\textrm{eff}}$
and external horizon are solid.  For this
choice of $L$ and $Q$ there are only two extrema outside
the RN horizon, one stable and one unstable.  We want to determine if 
there are any $L$ and
$Q$ for which the external horizon is closer to the singularity than 
the inner minimum of the potential. This would yield two stable circular orbits.

Our first step is to find the radii $r_c$ at which the effective potential has
stationary points, given by
\begin{equation}
	\label{eqn:derivative}
	\frac{dV_{\textrm{eff}}}{dr} = \frac{1}{r^2} - \frac{Q^2+L^2}{r^3}+ 
	\frac{3L^2}{r^4} - \frac{2Q^2L^2}{r^5} = 0.
\end{equation}
We rewrite this as 
\begin{equation}
	\label{eqn:derivative2}
	r_c^3 - (Q^2 + L^2)r_c^2 + 3L^2r_c - 2Q^2L^2=0
\end{equation}
and take its discriminant to get
\begin{widetext}
\begin{equation}
	\label{eqn:discriminant}
	D = -108L^6 + 9L^8 + 126L^6Q^2 - 8L^8Q^2 + 9L^4Q^4 - 24L^6Q^4 - 24L^4Q^6 - 8L^2Q^8.
\end{equation}
\end{widetext}
When $D>0$, $\frac{\partial V_{\textrm{eff}}}{\partial r}$ has 3 distinct, real roots, and
$V_{\textrm{eff}}$ has 3 extrema.  If the condition $D>0$ is ever true when 
the $r_c$ are all $>\, r_+$, the external horizon, we will have two stable
circular orbits.  The solutions of 
Equation (\ref{eqn:derivative2}) for $r_c$ do not provide much insight, so 
we will not reproduce them here.  Instead we represent the solutions graphically. 

\begin{figure}[htpb]
    \centering
    \begin{minipage}{86mm}    
    \vspace{0pt}
    \includegraphics[width=60mm]{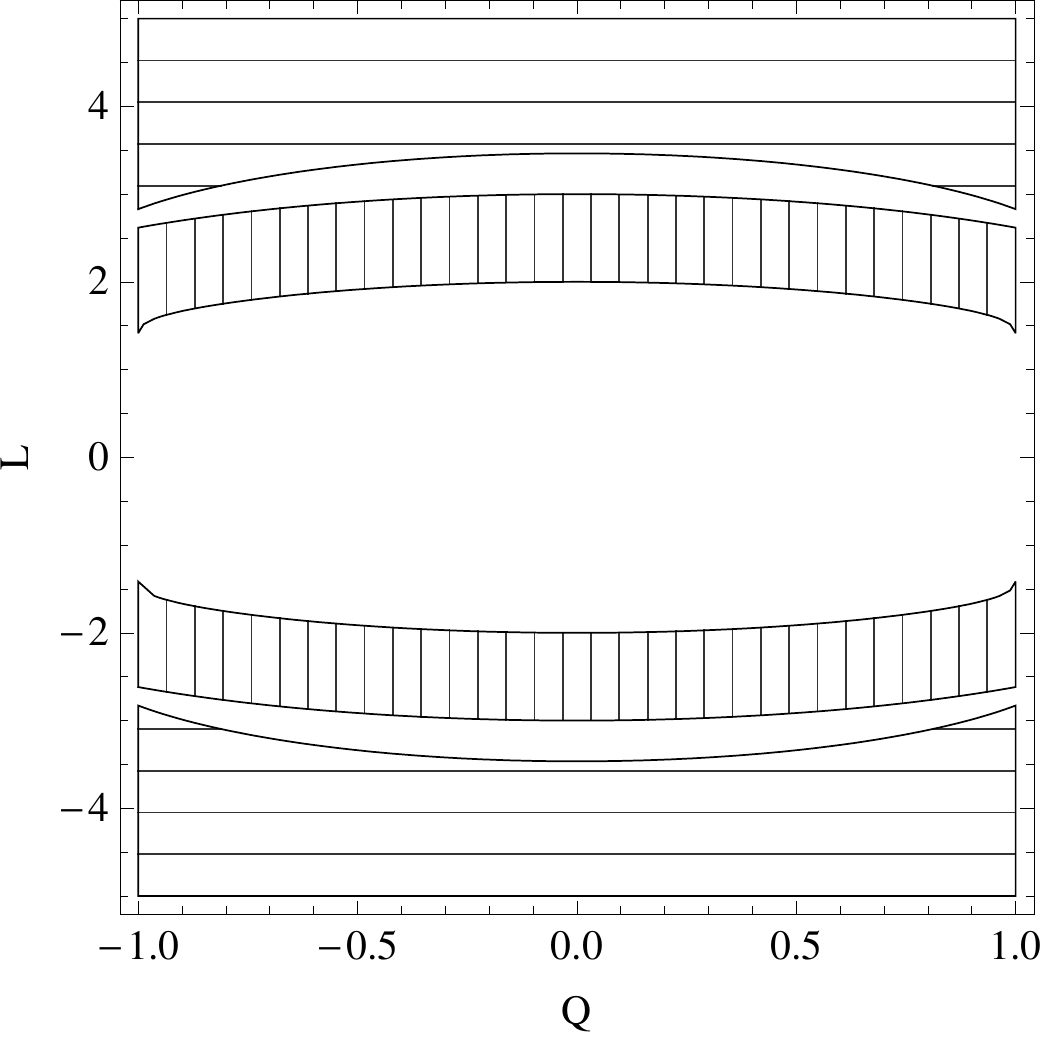}
    \end{minipage}
	\vspace{-10pt}
	\caption{Regions in which the potential has three extrema (horizontal
	hatching) and those in which the smallest extremum is outside the horizon.
	(vertical hatching)}
	\label{fig:discrimTP}
\end{figure}

The disjointedness of the regions in
parameter space that satisfy each of the above conditions is demonstrated
in Figure \ref{fig:discrimTP}.  The regions with horizontal hatching
are those in which $D>0$ and the regions with vertical hatching are where the smallest
circular orbit $r_c > r_+$.  Note that the vertically hatched region does not presume
the existence of three extrema.  There is no overlap between the two regions when
$-1 < Q < 1$ and $Q_*= 0$,
so in these conditions we never see three extrema
outside the event horizon.  Therefore, there is always at most one stable circular
orbit.  This clarifies the region of the effective potential in which we find
periodic orbits and confirms that for $Q_*=0$, there is always at most one stable circular orbit.

\subsection{Bounds on L}
\begin{figure}[htpb]
	\centering
    \begin{minipage}{86mm}
	\includegraphics[width=86mm]{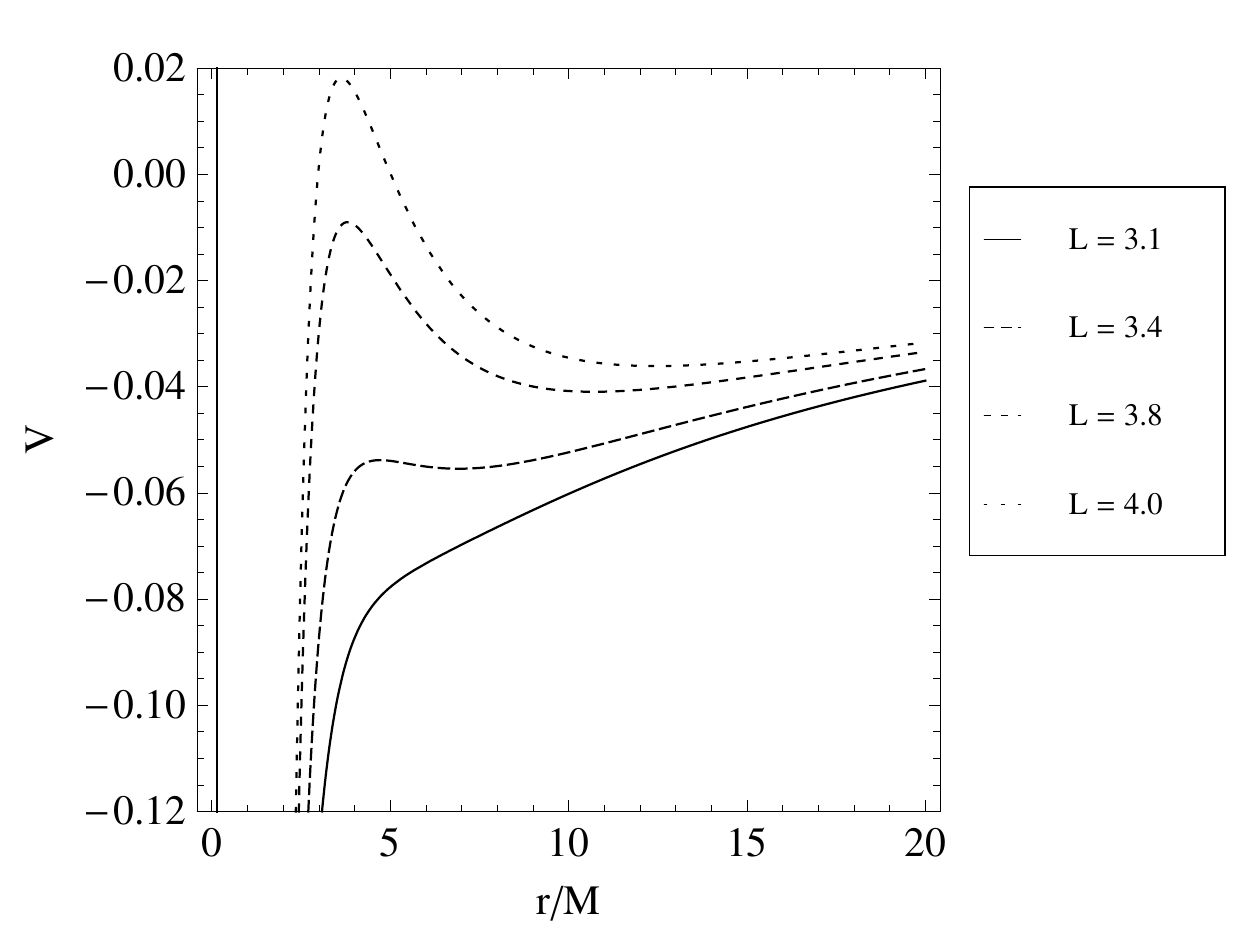}
    \end{minipage}
	\vspace{-15pt}
	\caption{$V_{\textrm{eff}}$ for $L=3.1$, $3.4$, $3.8$, and $4.0$, from bottom to top, with
	$Q$ fixed.}
	\label{fig:potentialL}
\end{figure}

We have established that in the undercritically-charged RN
geometry for a neutral particle, for any L there is
at most one stable circular orbit.  Figure \ref{fig:potentialL} 
depicts the potential for various $L$ with $Q$
fixed. 
We will now determine the bounds in
the undercritically-charged RN geometry that specify the region in
which we find most zoom-whirl behavior \cite{levin},
\begin{equation}
	L_{\rm ISCO}<L<L_{\rm IBCO},
\end{equation}
where ISCO stands for ``Innermost Stable Circular Orbit"
and IBCO for ``Innermost Bound Circular Orbit".  
$L_{\textrm{ISCO}}$ is the lowest value of $L$ for which the potential has a local
minimum, and therefore marks the first appearance of a stable circular orbit.  
For $L<L_{\textrm{ISCO}}$, all orbits will plunge into
the black hole, so $L_{\textrm{ISCO}}$ sets the lower limit on bound orbits. $L_{\textrm{IBCO}}$
marks the first appearance of an unstable circular orbit that is
energetically bound. It sets the upper limit only in the sense that we
expect to see the most zoom-whirl behavior in the strong-field when
such an unstable bound orbit comes into play \cite{levin}. Orbits 
will whirl more as they
roll up the potential towards the unstable bound circular orbit. 

To derive expressions for $L_{\textrm{ISCO}}$ and $L_{\textrm{IBCO}}$ we start with 
the conditions for circular orbits.
Equation (\ref{eqn:rdotpot}), written with $V_{\textrm{eff}}$ as it 
appears in Equation (\ref{eqn:potential2}), is
\begin{equation}
	\frac{1}{2}\dot{r}^2 + \frac{\Delta L^2}{2 r^2} + \frac{\Delta - 1}{2}
	=\mathcal{E}_{\textrm{eff}},
\end{equation}
The two conditions for circular orbits are $\dot r = 0$ and $\ddot r = 0$, which imply that 
\begin{eqnarray}
	V_{\rm eff} & = & \frac{E^2-1}{2} \nonumber \\
	\frac{\partial V_{\rm eff}}{\partial r} & = & 0.
	\label{eqn:circconditions}
\end{eqnarray}
The first condition, when solved for $L^2$, gives us
\begin{equation}
	\label{eqn:L2no1}
	L^2=\left(\frac{E^2}{\Delta}-1\right)r^2.
\end{equation}
We may rewrite $\frac{\partial V}{\partial r}$ as
\begin{eqnarray}
	\frac{\partial V_{\textrm{eff}}}{\partial r} & = & \Delta\left(\frac{-L^2}{r^3}+\frac{\Delta'}{\Delta}\left(\frac{L^2}{2r^2}+\frac{1}{2}\right)\right) \\
	& = & \Delta\left(\frac{-L^2}{r^3}+
		\frac{\Delta'}{\Delta^2}\left(V_{\textrm{eff}}+\frac{1}{2}\right)\right)
	\label{eqn:dvdr2}
\end{eqnarray}
where 
\begin{equation}
	\Delta' = \frac{d\Delta}{dr} = \frac{2}{r^2} - \frac{2Q^2}{r^3}.
\end{equation}
Applying the conditions for circular orbits in Equation (\ref{eqn:circconditions}) gives us
\begin{eqnarray}
	\label{eqn:altpotentialderiv}
	\frac{\partial V_{\textrm{eff}}}{\partial r} & = & \Delta\left(\frac{-L^2}{r^3}+\frac{\Delta'}{\Delta^2}\left(\frac{E^2}{2}\right)\right)
        = 0 \quad
\end{eqnarray}
Solving for $L^2$ gives us
\begin{equation}
	\label{eqn:L2no2}
	L^2=r^3\frac{\Delta'}{\Delta^2}\left(\frac{E^2}{2}\right).
\end{equation}
Equating Equations (\ref{eqn:L2no1}) and (\ref{eqn:L2no2}) and solving for $E^2$ gives
\begin{equation}
	\label{eqn:Ecirc}
	E_c^2(r_c,Q)=\frac{2\Delta^2}{2\Delta-r\Delta'}.
\end{equation}
Putting Equation (\ref{eqn:Ecirc}) into Equation (\ref{eqn:L2no1}) and solving for $L^2$ yields
\begin{equation}
	\label{eqn:Lcirc}
	L_c^2(r_c,Q)=r^2\left(\frac{r \Delta'}{2\Delta -
          r\Delta'}\right).
\end{equation}

\begin{figure}[htpb]
	\centering
	\vspace{-15pt}
    \begin{minipage}{86mm}
	\includegraphics[width=86mm]{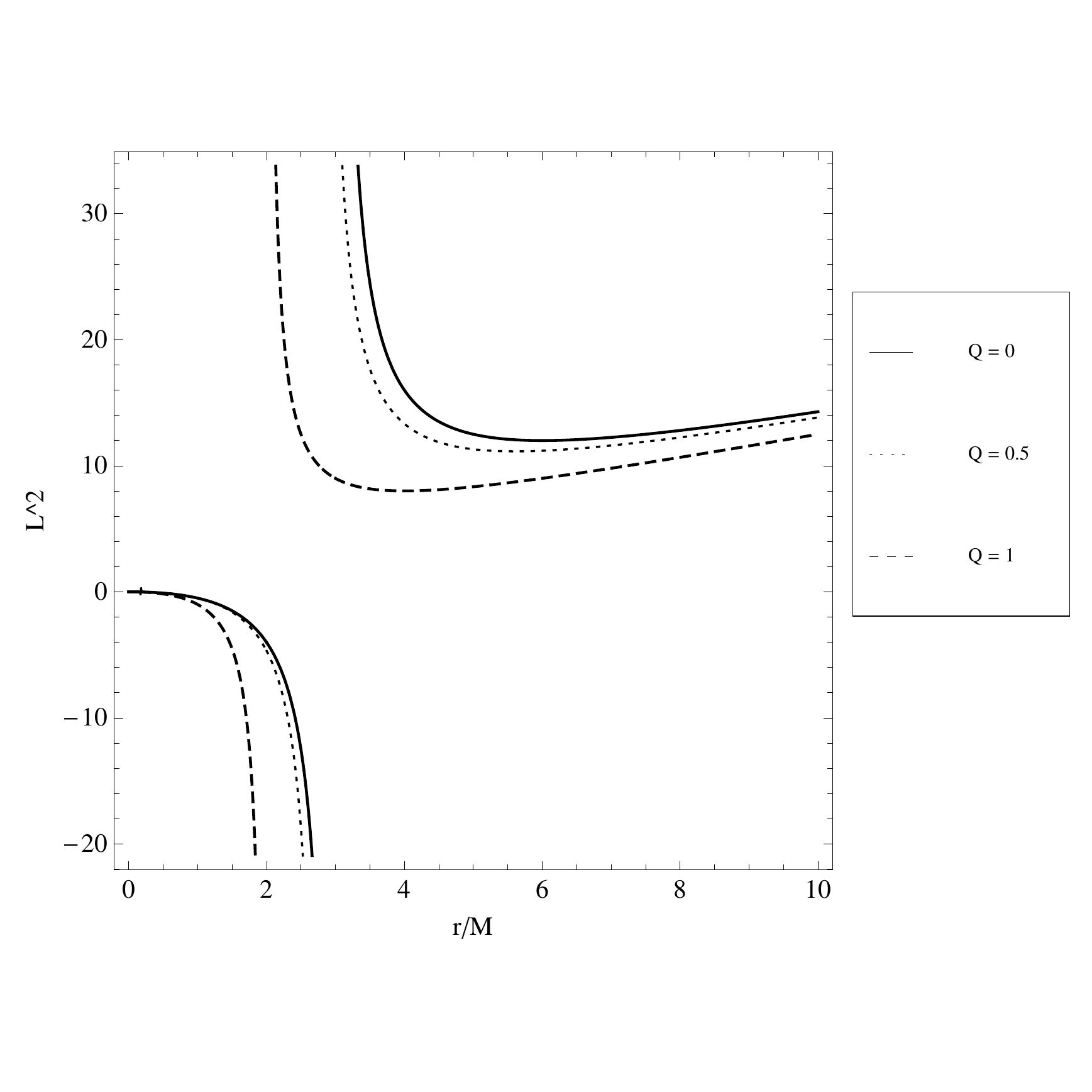} \\
	\vspace{-35pt}
	\includegraphics[width=84mm]{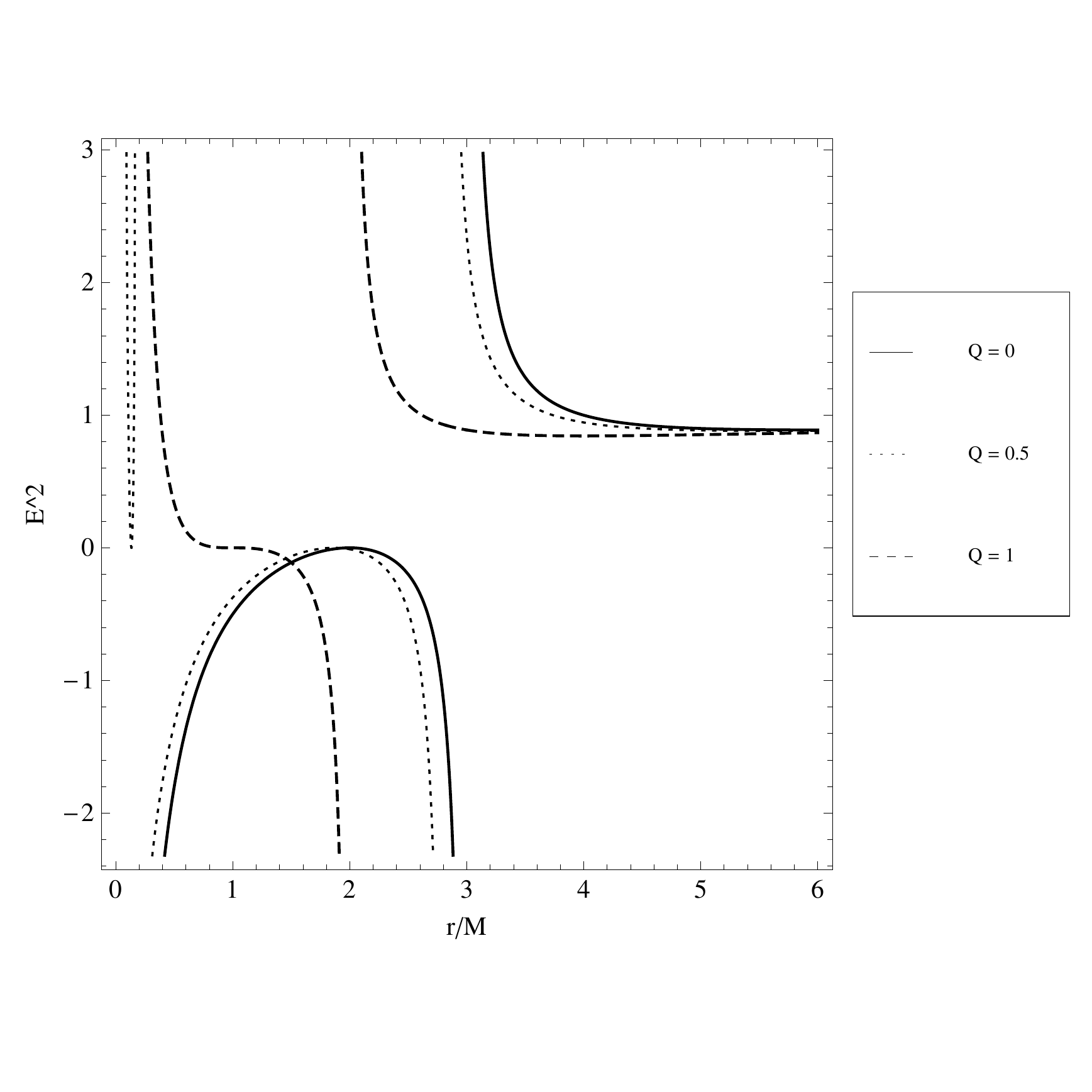}
    \end{minipage}
	\vspace{-40pt}
	\caption{$L_c^2$ and $E_c^2$ as functions of $r$ for $Q=1$, $Q=0.5$, and $Q=0$.}
	\label{fig:LEplot}
\end{figure}

\begin{figure}[htpb]
	\centering
    \begin{minipage}{86mm}
	\includegraphics[width=86mm]{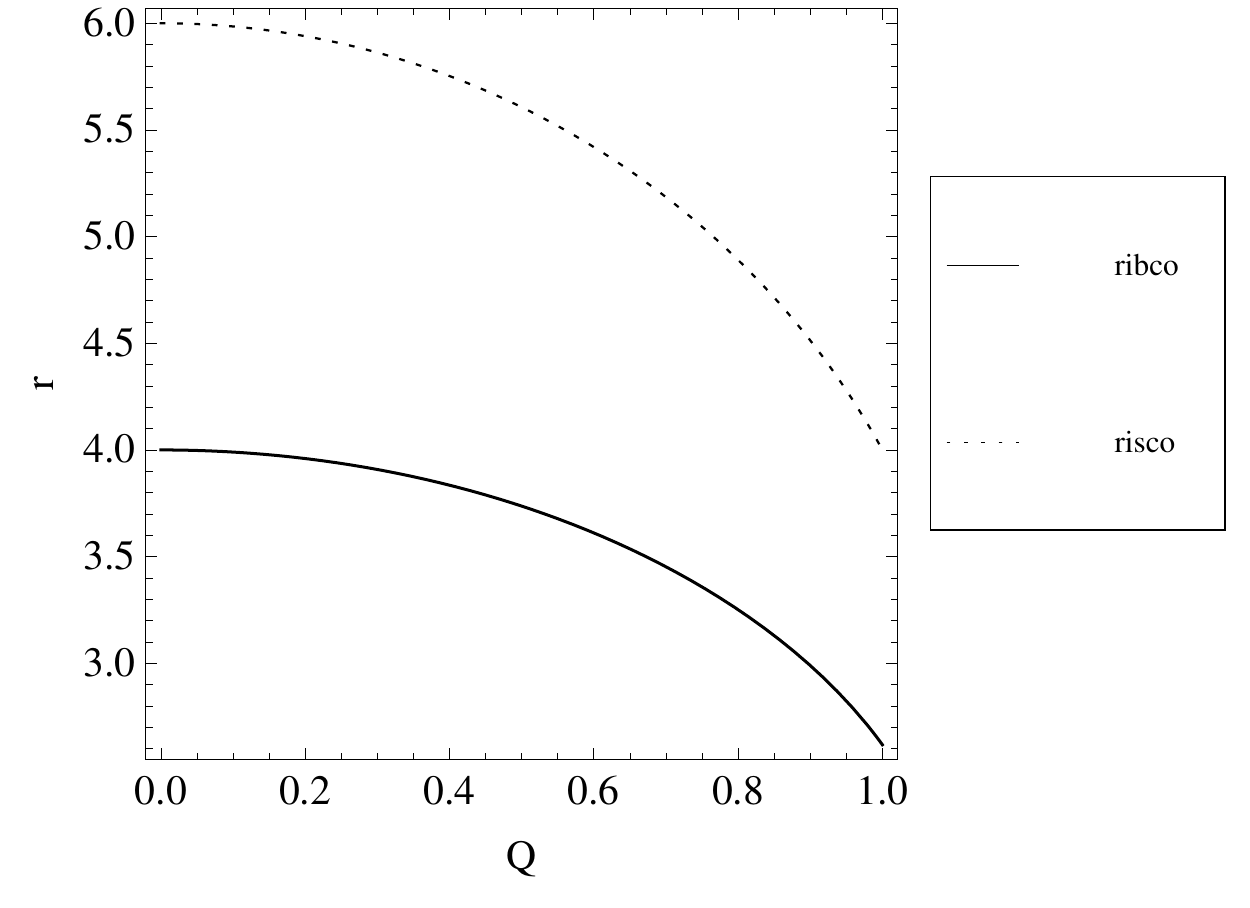}
    \end{minipage}
	\vspace{-20pt}
	\caption{$r_{\textrm{IBCO}}$ (solid) and $r_{\textrm{ISCO}}$ (dotted) as a function of $Q$}
	\label{fig:ribcorisco}
\end{figure}

With $L_c$ and $E_c$ in hand, each of which is depicted
in Figure~\ref{fig:LEplot}, our next task is to
calculate $r_{\textrm{IBCO}}$ and $r_{\textrm{ISCO}}$.  Because at infinity
the potential is $0$, $r_{\textrm{IBCO}}$ is simply the radius of unstable
circular orbit $r_c$ such that $V_{\textrm{eff}}(r_c)=0$.
We first solve
\begin{equation}
	V_{\textrm{eff}}=0=\frac{\Delta L^2}{2 r^2}+ \frac{\Delta-1}{2} \\
\end{equation}
for $L^2$ to obtain
\begin{equation}
	L_{V=0}^2=\frac{1-\Delta}{\Delta}r^2.
\end{equation}
Setting this equal to $L_c^2$ gives us an expression we may use to find
$r_{\textrm{IBCO}}$:
\begin{eqnarray}
	L_c^2 & = & L_{V=0}^2 \\
	r^2\left(\frac{r \Delta'}{2\Delta - r\Delta'}\right) & = &
        \left (\frac{1-\Delta}{\Delta}\right )r^2 \\
	0 & = & 2\Delta - r\Delta' - 2\Delta^2.
\end{eqnarray}
Solving this for r gives us
\begin{eqnarray}
	\label{eqn:ribco}
	r_{\textrm{IBCO}} & = & \frac{1}{6}\left(8-\frac{8\cdot2^{1/3}(-4+3Q^2)}{G(Q)}+2^{2/3}G(Q)\right) \\
	G(Q) & = & \sqrt[3]{128-144Q^2+27Q^4+3\sqrt{-96Q^6+81Q^8}}. \nonumber
\end{eqnarray}
$L_{\textrm{IBCO}}$ is then $L_c$ with $r_{\textrm{IBCO}}$ in place of $r$:
\begin{equation}
    L_{\textrm{IBCO}}^2 = \frac{-Q^2 r_{\textrm{IBCO}}^2+r_{\textrm{IBCO}}^3}{2Q^2 - 3r_{\textrm{IBCO}}
    + r_{\textrm{IBCO}}^2}.
\end{equation}
We determine $r_{\textrm{ISCO}}$ by taking
$V_{\textrm{eff}}''=0$ (differentiated with respect to $r$), or
\begin{equation}
	\frac{\partial^2 V_{\textrm{eff}}}{\partial r^2} = 
	\frac{3 L^2 \Delta}{r^4} - \frac{2 L^2 \Delta'}{r^3} + \frac{(L^2 + r^2)\Delta''}{2 r^2} = 0,
\end{equation}
where
\begin{equation}
	\Delta''=-\frac{4}{r^3}+\frac{6Q^2}{r^4}.
\end{equation}
Solving for $L^2$ gives us
\begin{equation}
	L_{\textrm{V}''\textrm{=0}}^2 = \frac{r^4 \Delta''}{4\Delta'r - 6\Delta - \Delta'' r^2}
\end{equation}
Setting $L_{\textrm{V}''\textrm{=0}}^2$ equal to $L_c^2$ and solving for $r$ gives us
\begin{eqnarray}
	\label{eqn:risco}
	r_{\textrm{ISCO}} & = & 2+\frac{4-3Q^2}{H(Q)}+H(Q) \\
	H(Q) & = & \sqrt[3]{8-9Q^2+2Q^4+\sqrt{5Q^4-9Q^6+4Q^8}}, \nonumber
\end{eqnarray}
which when plugged into $L_c$ gives us
\begin{equation}
    L_{\textrm{ISCO}}^2 = \frac{-Q^2 r_{\textrm{ISCO}}^2+r_{\textrm{ISCO}}^3}{2Q^2 - 3r_{\textrm{ISCO}}
    + r_{\textrm{ISCO}}^2}.
\end{equation}
$r_{\textrm{IBCO}}$ and $r_{\textrm{ISCO}}$ are plotted in Figure~~\ref{fig:ribcorisco}.

\begin{figure}[htpb]
    \centering
    \includegraphics[width=7cm]{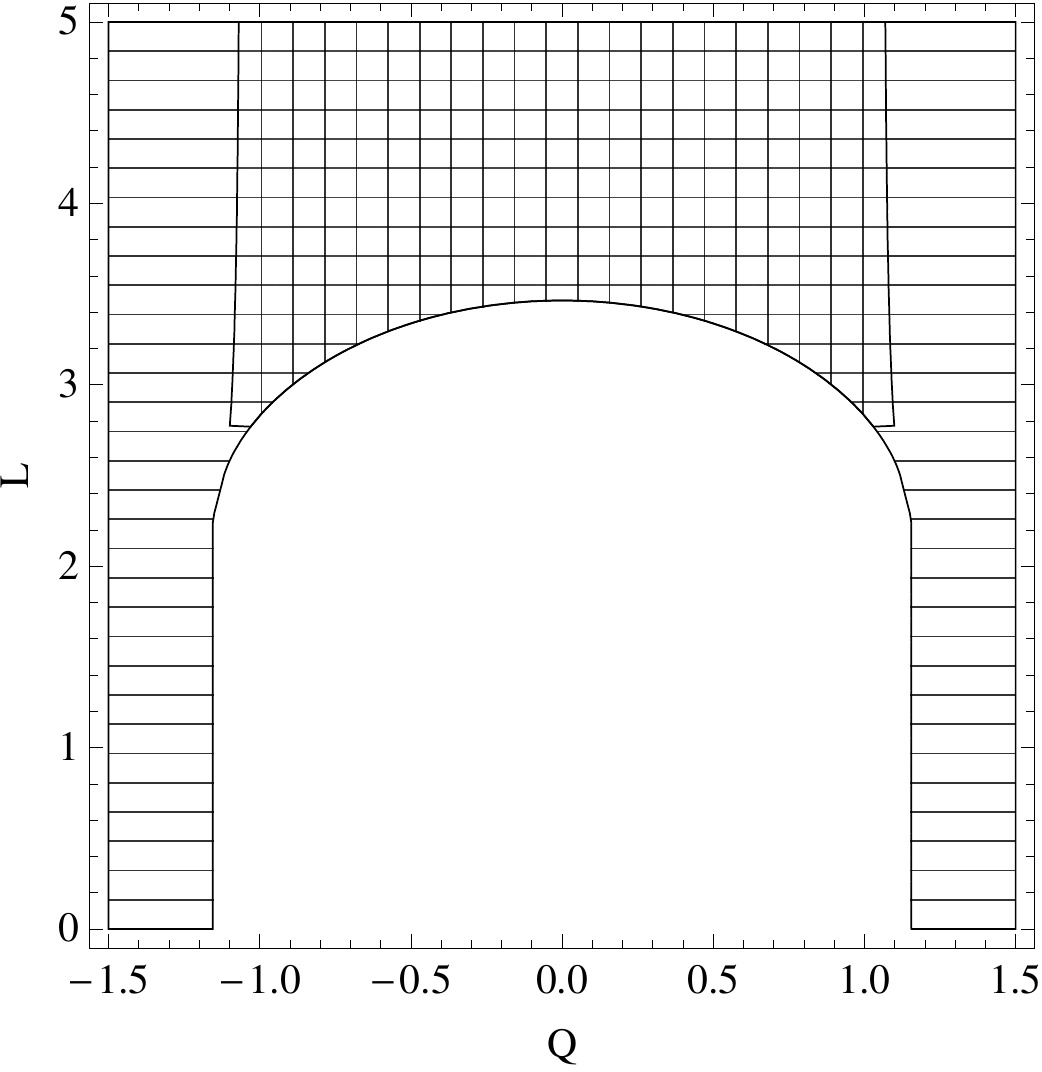}
	\caption{Regions in which $L^2>L_{\textrm{ISCO}}^2$ (horizontal hatching)
	and where the discriminant $D>0$ (vertical hatching).  When we impose the 
	bound $-1<Q<1$ the two regions are equivalent, which demonstrates that only
	when $L^2>L_{\textrm{ISCO}}^2$ does the potential have three extrema.}
	\label{fig:regionplot1}
\end{figure}

We can confirm that the presence of three stationary points in the potential is
tied to $L_{\textrm{ISCO}}$ by checking whether the regions in parameter space $D > 0$ 
and $L^2 > L_{\textrm{ISCO}}^2$ coincide.  Figure \ref{fig:regionplot1}
shows that this is the case within our bounds for $Q$.

Having established bounds on $Q$ and $L$ for the $Q_*=0$ case, we can taxonomize all 
zoom-whirl behavior in the strong field bounded by
$L_{\textrm{ISCO}}<L<L_{\textrm{IBCO}}$, as we show in \S \ref{sec:tables}.  First, we briefly
discuss homoclinic orbits and then the $Q_*\ne 0$ case.

\subsection{Homoclinic Orbits}
\begin{figure}[htpb]
    \centering
    \begin{minipage}{\columnwidth}
    \includegraphics[width=.8\textwidth]{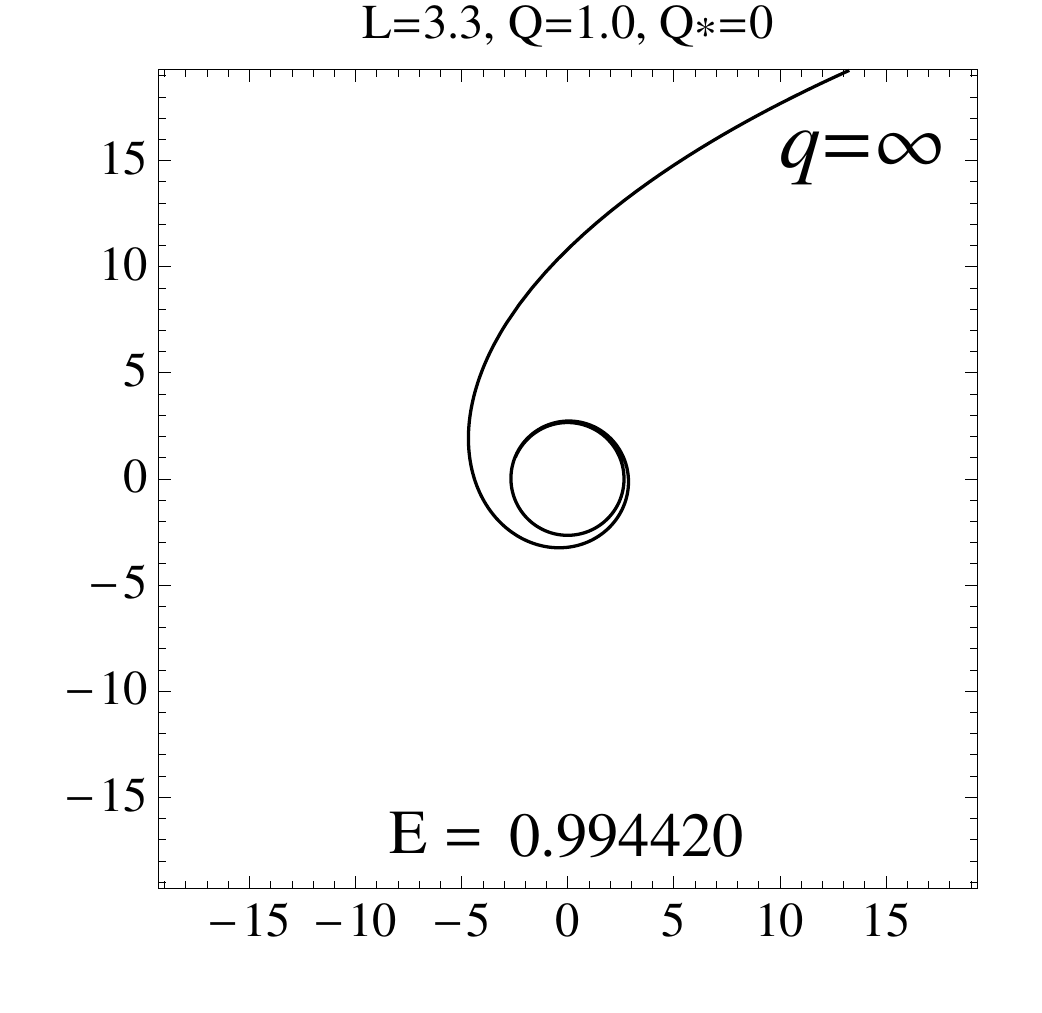}
    \caption{A homoclinic orbit for $Q = 1.0$, $Q_* = 0$, $L = 3.3$}
    \label{fig:homoclinic}
    \end{minipage}
\end{figure}

For any $L$ such that $L_{\textrm{ISCO}} < L < L_{\textrm{IBCO}}$, there is an 
unstable circular orbit called a homoclinic orbit.  The homoclinic orbit is 
the separatrix between orbits that plunge to the horizon
and those that do not \cite{levin_h1}.  It provides us with the infinite-whirl
limit and is therefore an important landmark in the orbital landscape.
Because homoclinic orbits are central
to strong-field dynamics \cite{levin, levin_h1, levin_h2, levin_d2},
we include the orbital plot of a homoclinic orbit for an uncharged
particle in RN spacetime in Figure~\ref{fig:homoclinic}.


\section{Charged Particle Orbits in RN Spacetime}
\label{sec:charged}

Having discussed the uncharged test particle scenario, we will now move on
to orbits in which the particle has nonzero charge. While the $Q_*\neq0$
effective potential (Equation (\ref{eqn:potential}))
depends on both $L$ and $E$, it still only depends on one dynamical variable, $r$. Because there is
no dependence on $\theta$, $\varphi$ or $t$, every bound orbit has fixed
apastra and periastra that are functions of $Q$, $L$, and $E$, so we still find periodicity.
As such, we can
make use of this potential to study orbits for charged particles.  
However, the method for finding periodic orbits is more complicated than in the $Q_*=0$ case,
because given a particular effective potential we may no longer choose $\mathcal{E}_{\textrm{eff}}$ 
so as to produce a periodic orbit, because the potential is coupled to $E$. 
\begin{figure}[htpb]
	\centering
    \begin{minipage}{86mm}
	\vspace{-30pt}
    \includegraphics[width=86mm]{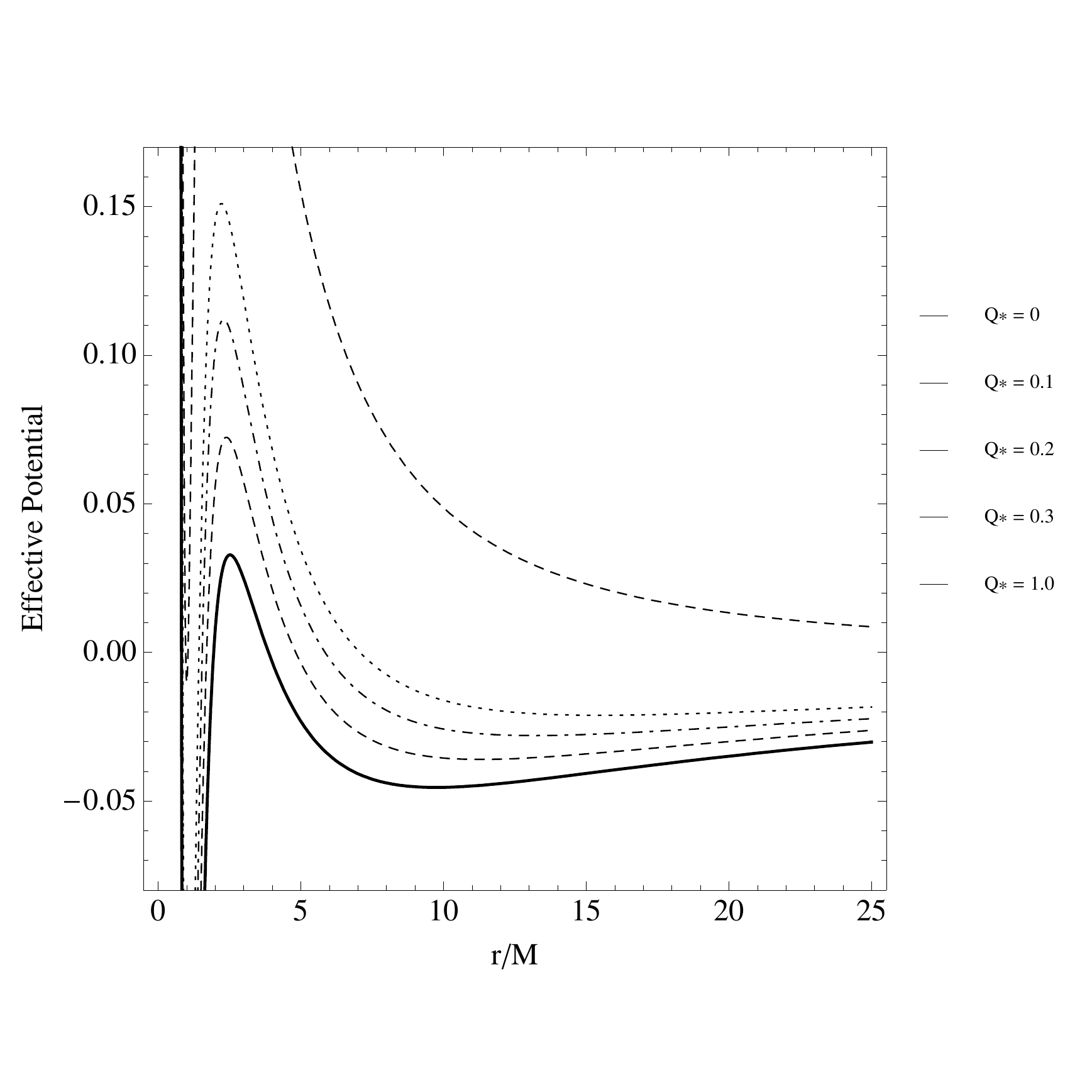}
    \vspace{-35pt}
	\caption{$V_{\textrm{eff}}$ for $Q_*=\{0, 0.1, 0.2, 0.3, 1.0\}$ from bottom to top, with
	$\mathcal{E}_{\textrm{eff}}=\frac{-1+E^2}{2} =0.99$, $Q = 1$, $L = 3.5$}
    \end{minipage}
	\label{fig:veffqvar}
\end{figure}

\subsection{Identifying Regions with Bounded Orbits} 
\begin{figure}
    \begin{minipage}{\columnwidth}
    \centering
    \includegraphics[width=50mm]{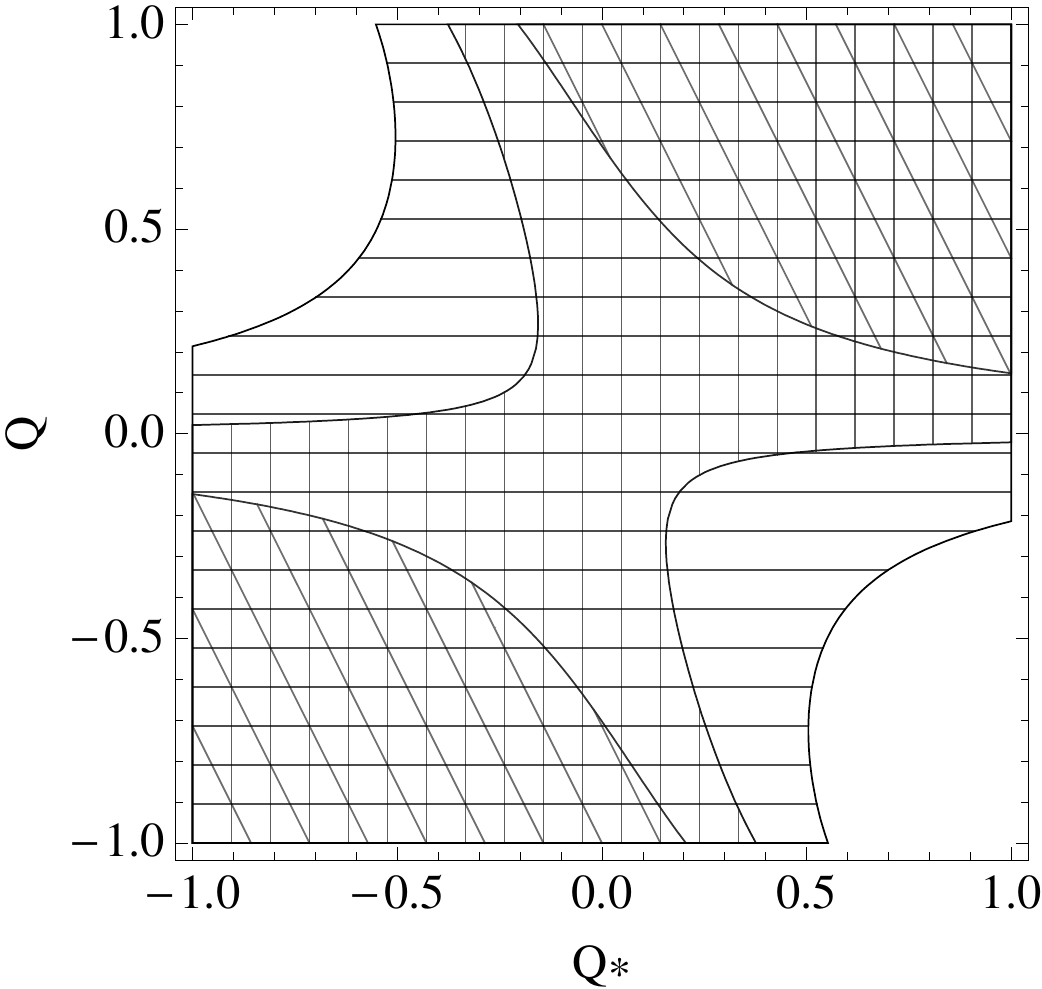}
    \caption{Regions of positive discriminant for $L=\{3.2, 3.5, 3.8\}$; 
	hatching: (diagonal, vertical, horizontal); $E = 0.98$}
    \label{fig:discrimQq}
    \end{minipage}
\end{figure}

\begin{figure}
    \begin{minipage}{\columnwidth}
    \centering
    \vspace{-1cm}
    \includegraphics[width=55mm]{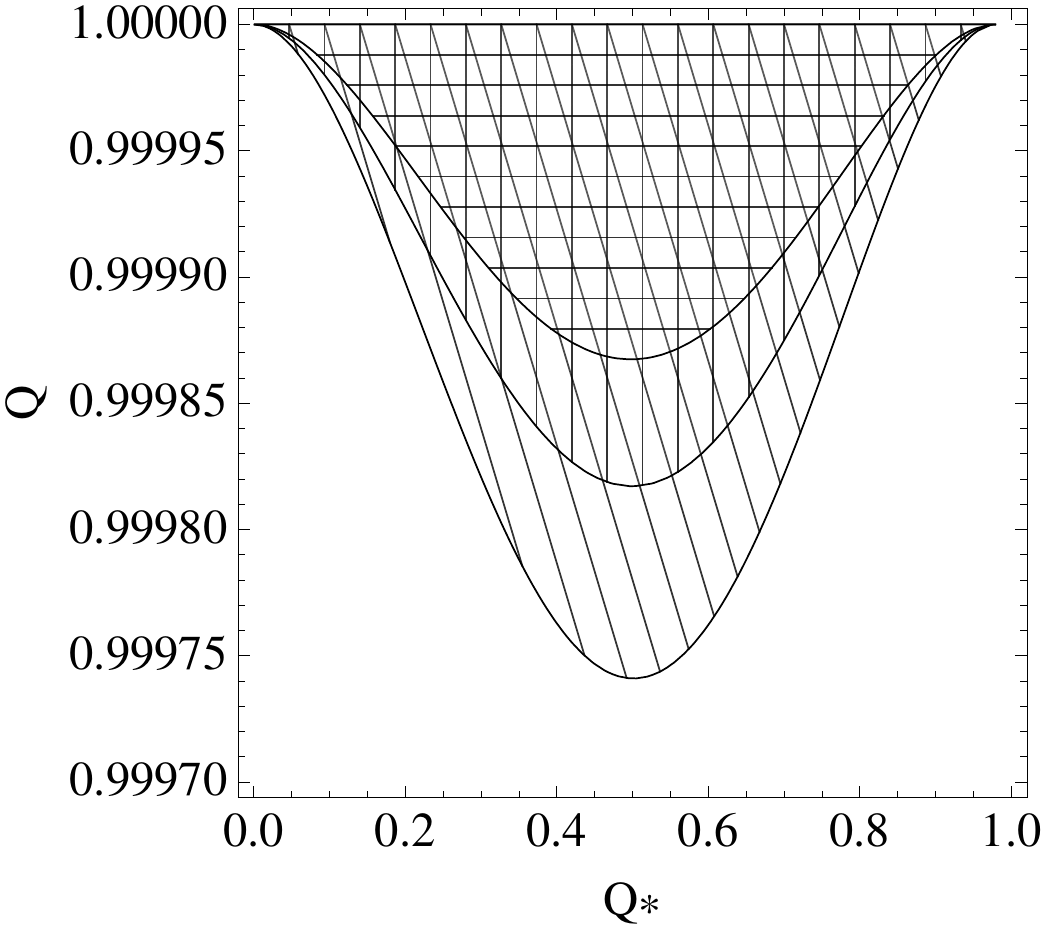}
    \includegraphics[width=55mm]{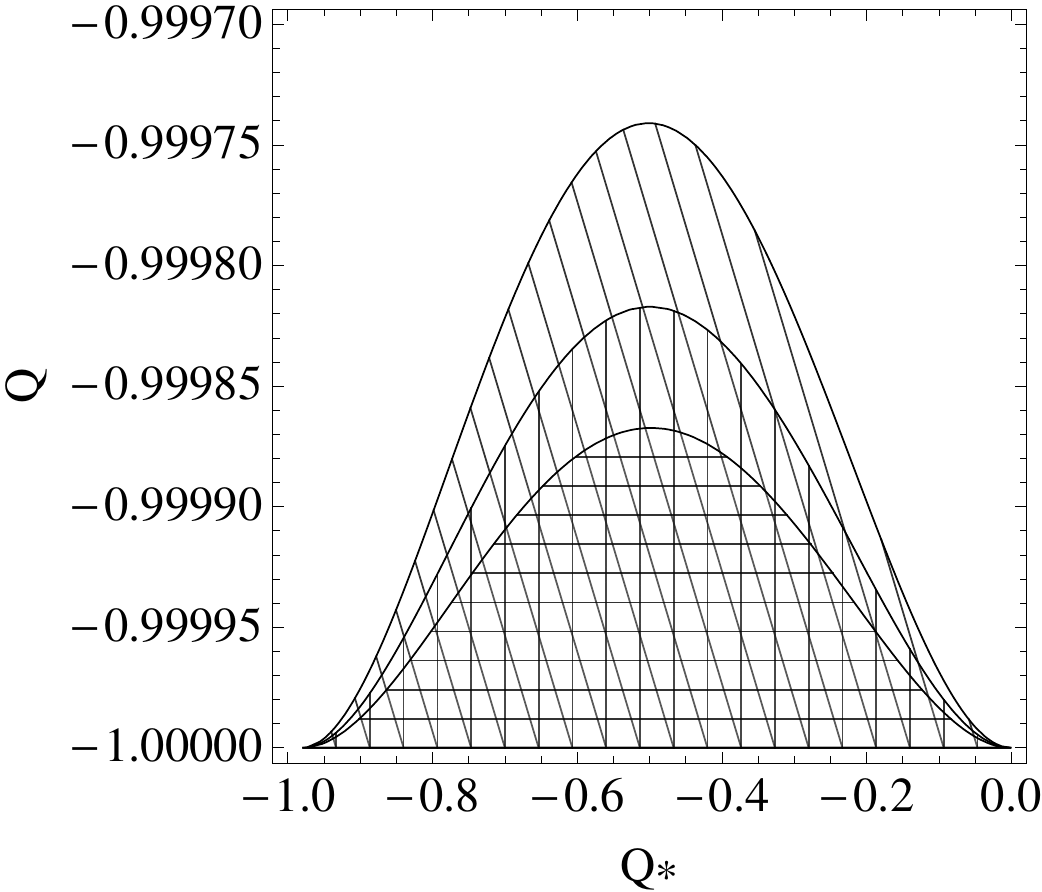}
    \caption{Regions in which the potential has three extrema that are all 
	    $>r_+$, for $L=\{3.2, 3.5, 3.8\}$;
        hatching: (diagonal, vertical, horizontal); $E = 0.98$} 
    \label{fig:tp1horizon}
    \end{minipage}
\end{figure}

Because we are considering only test particles we assume that the
particle mass $m$ is $<< M$, which implies that $Q_*< (m/M) < 1$. 
As before, we first want to determine whether the potential ever has three extrema
outside the horizon, which we do by determining if it is ever simultaneously the case
that the discriminant of
$\frac{\partial V_{\textrm{eff}}}{\partial r}$ is positive -- which means that there are three extrema --
and that the smallest of the extrema is at a radius $>r_+$.
The region of parameter space in which each condition is true
is depicted in Figures \ref{fig:discrimQq} and \ref{fig:tp1horizon}, respectively. 
The discriminant here is not the same as in Equation 
(\ref{eqn:discriminant}), but rather
\begin{eqnarray}
    D & = & -108L^6 + 9L^8 + 108EL^6Q_*Q + 126L^6Q^2  \\ \nonumber
    & & - 8L^8Q^2 - 18L^6Q_*^2Q^2 - 108EL^6Q_*Q^3   \\ \nonumber
    & & + 9L^4Q^4 - 24L^6Q^4 - 126L^4Q_*^2Q^4  \\ \nonumber
    & & + 24L^6Q_*^2Q^4 + 9L^4Q_*^4Q^4 + 108EL^4Q_*Q^5  \\ \nonumber
    & & + 108EL^4Q_*^3Q^5 - 24L^4Q^6 + 48L^4Q_*^2Q^6  \\ \nonumber
    & & - 108E^2L^4Q_*^2Q^6 - 24L^4Q_*^4Q^6 - 8L^2Q^8 \\ \nonumber
    & & + 24L^2Q_*^2Q^8 - 24L^2Q_*^4Q^8 + 8L^2Q_*^6Q^8 \ \ .\nonumber
    \label{eqn:discriminant2}
\end{eqnarray}

Each region in Figure \ref{fig:tp1horizon} is a subset
of the corresponding region in Figure \ref{fig:discrimQq}, which means that
for these values of $L$, there are in fact regions in which the potential has
three extrema -- two minima -- outside the horizon.  
Note that this only occurs for near-extremal geometries and
only when $Q$ and $Q_*$ have like sign, but that for all $Q_*$ there exists 
a $Q$ for which there are three external extrema.  


This is not the complete picture.  The existence of two minima outside the horizon
does not imply that there are multiple stable circular orbits, because unlike in the
$Q_*=0$ case, $\mathcal{E}_{\textrm{eff}}$ is fixed by the $E$ we chose for plotting the potential.
For there to be two stable circular orbits such that an energetic
particle could ``roll" from one stable circular orbit over the local maximum and
into the stable circular orbit, the following condition must be satisfied:
\begin{equation}
    V_{\textrm{eff}}(r_1, L, Q, q, E) = V_{\textrm{eff}}(r_2, L, Q, q, E),
\end{equation}
where $r_1$ and $r_2$ are the radii of stable circular orbits.  

\begin{figure}[htpb]
	\centering
	\vspace{15pt}
    \begin{minipage}{86mm}
	\includegraphics[width=60mm]{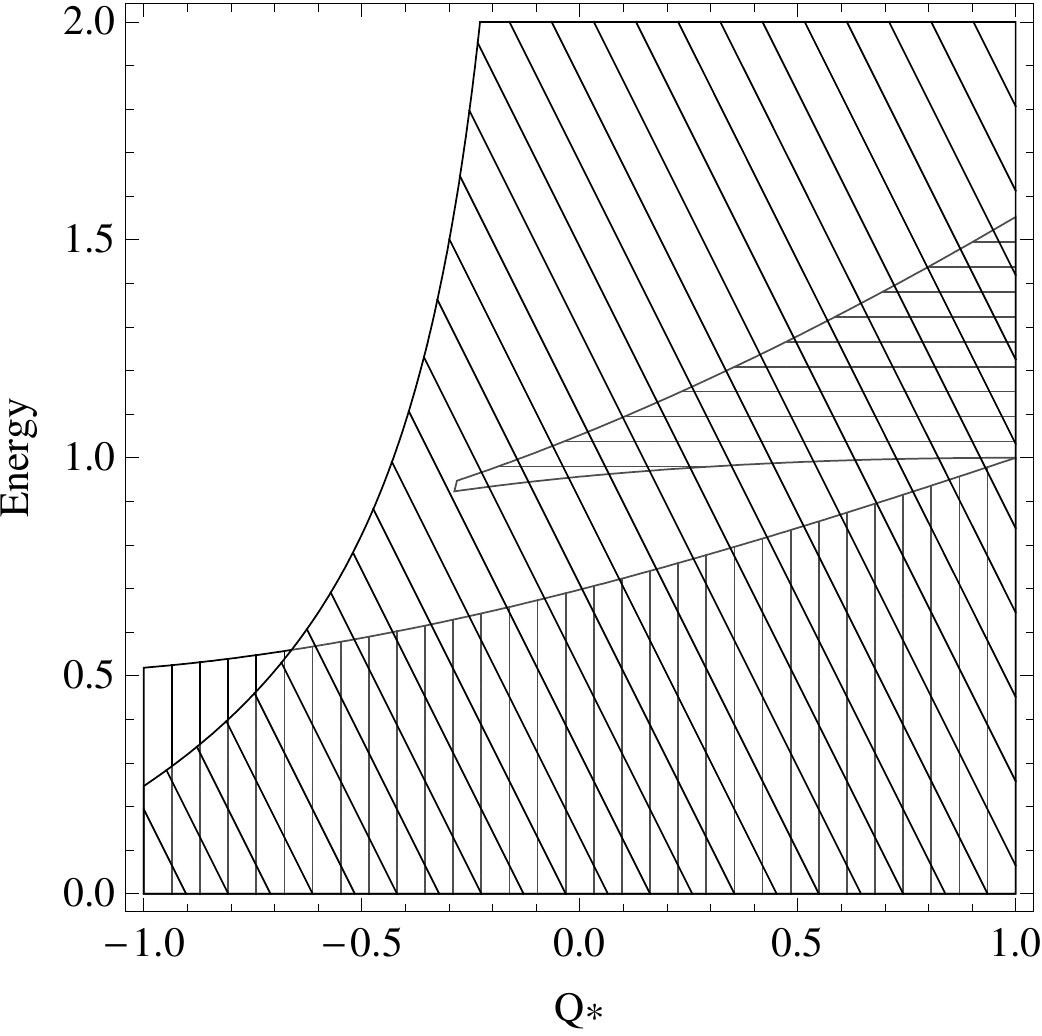}
	\caption{$Q = 1.0$; $L = 3.6$.  Each point in the parameter plot 
	defines an effective potential and $\mathcal{E}_{\textrm{eff}}$.  
	The regions shown are those where the resulting orbits' periastra 
	are at a radius outside the horizon (horizontal hatching), 
	where the periastra are inside the potential's smaller minimimum 
	(vertical hatching), and where the discriminant is positive 
	(diagonal hatching).  The coincidence of the three regions 
	gives us the subset of parameter space in which we can find bounded 
	orbits in the first of the potential's two minima}
	\label{fig:chargedregions}
    \end{minipage}
\end{figure}

This is a special case of when
the effective potential has two minima and $\mathcal{E}_{\textrm{eff}}$ cuts through the
more central of them in such a way that the periastron of the orbit it defines is outside the horizon.
If the general case never occurs, the special case is impossible as well.  We check for the
general case using a parameter plot, shown in Figure \ref{fig:chargedregions}.  Were the three
regions shown in the plot ever to overlap, we would have parameters for which the potential has
two minima such that $\mathcal{E}_{\textrm{eff}}$ cuts across the first minimum to yield periodic orbits.
The regions in this plot do not overlap for $Q = 1.0$, $L = 3.6$. Regenerating the plot
for various $Q$ and $L$ reveals the effect of changing each on the shape of each region 
and makes it clear that the three regions never overlap.  We cannot expect to find
any choice of parameters that yields a second region of bounded orbits in the 
potential's more central minimum.
This outcome does not preclude the existence of bounded orbits in the second minimum, where there
are no obstacles to applying the taxonomy.

\subsection{Bounds on L}
The final step in characterizing the charged particle RN geometry is to calculate
$L_{\textrm{ISCO}}$ and $L_{\textrm{IBCO}}$.  
We first write $V'_{\textrm{eff}}$ and $V''_{\textrm{eff}}$:
\begin{eqnarray}
    V'_{\textrm{eff}}  & = & \frac{1 - E Q_* Q}{r^2} + \frac{Q^2Q_*^2 - L^2 - Q^2}{r^3} \\ \nonumber
	& &+ \frac{3 L^2}{r^4} - \frac{2L^2Q^2}{r^5}
\end{eqnarray}

\begin{eqnarray}
    V''_{\textrm{eff}} & = & \frac{2 E Q_* Q - 2}{r^3} + \frac{3L^2 + 3Q^2 - 3Q_*Q^2}{r^4} \\ \nonumber
	& & -\frac{12 L^2}{r^5} + \frac{10L^2 Q^2}{r^6}
\end{eqnarray}
The condition for circular orbits is $V'_{\textrm{eff}} = 0$; solving this for $L^2$
and equating it with the result of solving $V_{\textrm{eff}}=0$ for $L^2$ yields an expression
we may solve for $E_c^2$, given in Equation~(\ref{eqn:ELcirc}).  Plugging $E_c^2$ into 
$L_{V'=0}^2$ and simplifying yields $L_c^2$, also given below. 
\begin{widetext}
\begin{eqnarray}
    E_c^2(r, Q, Q_*) & = & \frac{\left(3Q_*Q^3 -4rQQ_* + r^2QQ_* - 
    B (Q^2 -2r +r^2)\right)^2}
    {4r^2(r^2 - 3r + 2Q^2)^2} \nonumber \\
    L_c^2(r, Q, Q_*) & = & \frac{r^2 \left(-4 Q^4+Q_*^2 Q^4-6 r^2+2 r^3+Q^2 r \left(10-2 Q_*^2-2 r+Q_*^2 r\right) + B \left(Q_*Q^3+Q_*Qr^2-2Q_*Qr\right)\right)}{2\left(2Q^2 - 3r + r^2\right)^2} \nonumber \\
    B & = & \sqrt{8Q^2 + Q^2Q_*^2 -12r +4r^2} \,.
    \label{eqn:ELcirc}
\end{eqnarray} 
\end{widetext}
The ISCO exists at the inflexion point, given by $V''(r) = 0$.  So the solution of
$L^2_{V'=0} = L^2_{V''=0}$ for $r$ gives us $r_{\textrm{ISCO}}$.  Likewise, the solution
of $L^2_{V'=0} = L^2_{V=0}$ for $r$ gives us $r_{\textrm{IBCO}}$.  Plugging each of these
into $L_c^2$ then yields $L_{\textrm{IBCO}}$ and $L_{\textrm{ISCO}}$.  Because these solutions
do not provide any additional insight, we do not reproduce them here.

Having defined the bounds in which we find periodic orbits, we can move on to applying the
taxonomy to orbits in RN spacetime.  The taxonomy can 
be used to compare the parameters that yield orbits of a given $q$ in the 
$Q\neq0$, $Q_*=0$; $Q\neq0$, $Q_*\neq0$; and $Q=0$ cases, as we will now show.


\section{Periodic Tables of Orbits in RN Spacetime}
\label{sec:tables}

\subsection{Overview of the Taxonomy}
Before discussing the taxonomy as it applies to the RN spacetime, we will
summarize its salient features, which were presented in an earlier paper \cite{levin}.
As mentioned, orbits to which we can apply this taxonomy appear within
the $L_{\textrm{IBCO}}$ and $L_{\textrm{ISCO}}$ of the geometry, which define the
region of parameter space in which the potential accommodates bounded orbits.

The taxonomy assigns to each orbit a distinct rational number using a
scheme that takes advantage of the fact that in the strong-field
regime, every periodic orbit exhibits certain clearly visible 
topological characteristics.  In this section we will discuss how to
assign to each orbit a rational number based on the orbit's
topological features.  

\begin{figure*}[p]
	\centering
    \begin{minipage}{.9\textwidth}
	\includegraphics[width=.23\textwidth]{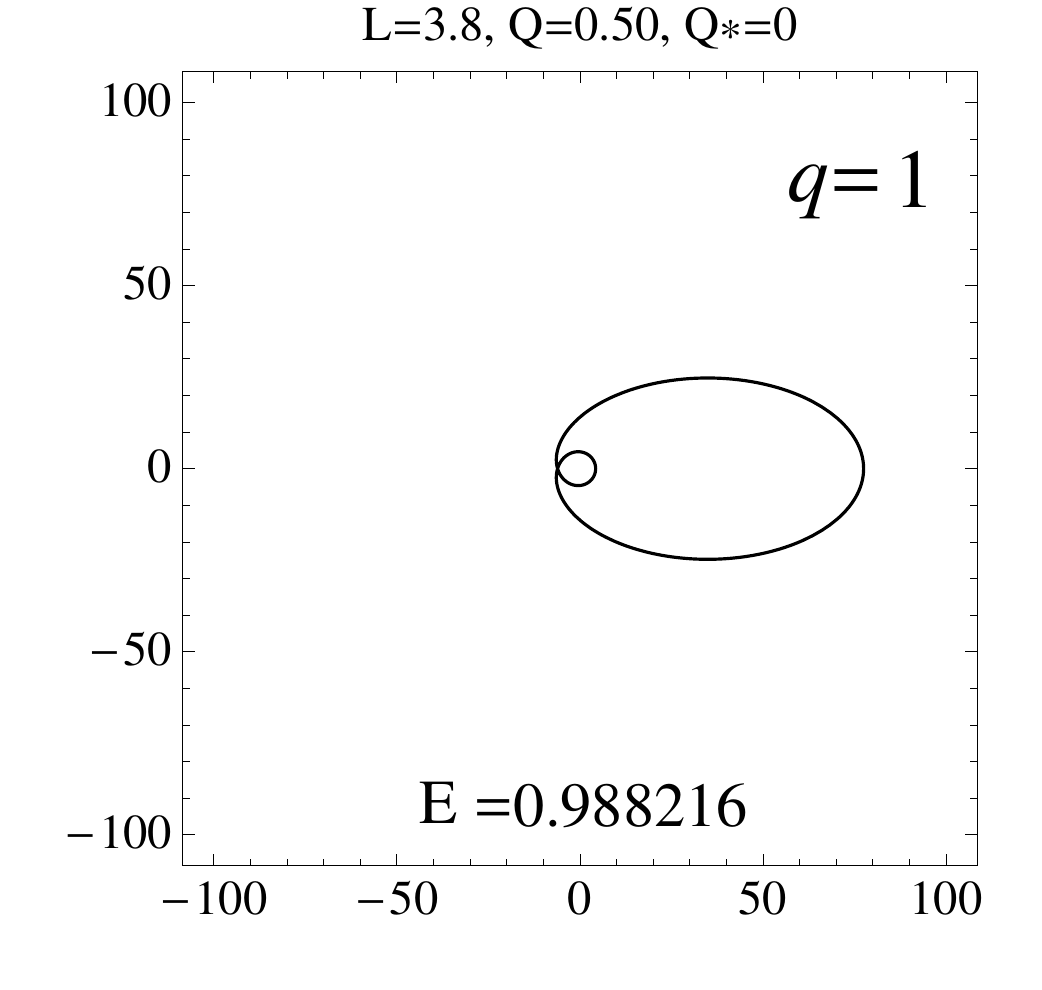}
	\hspace{-15pt}
	\includegraphics[width=.23\textwidth]{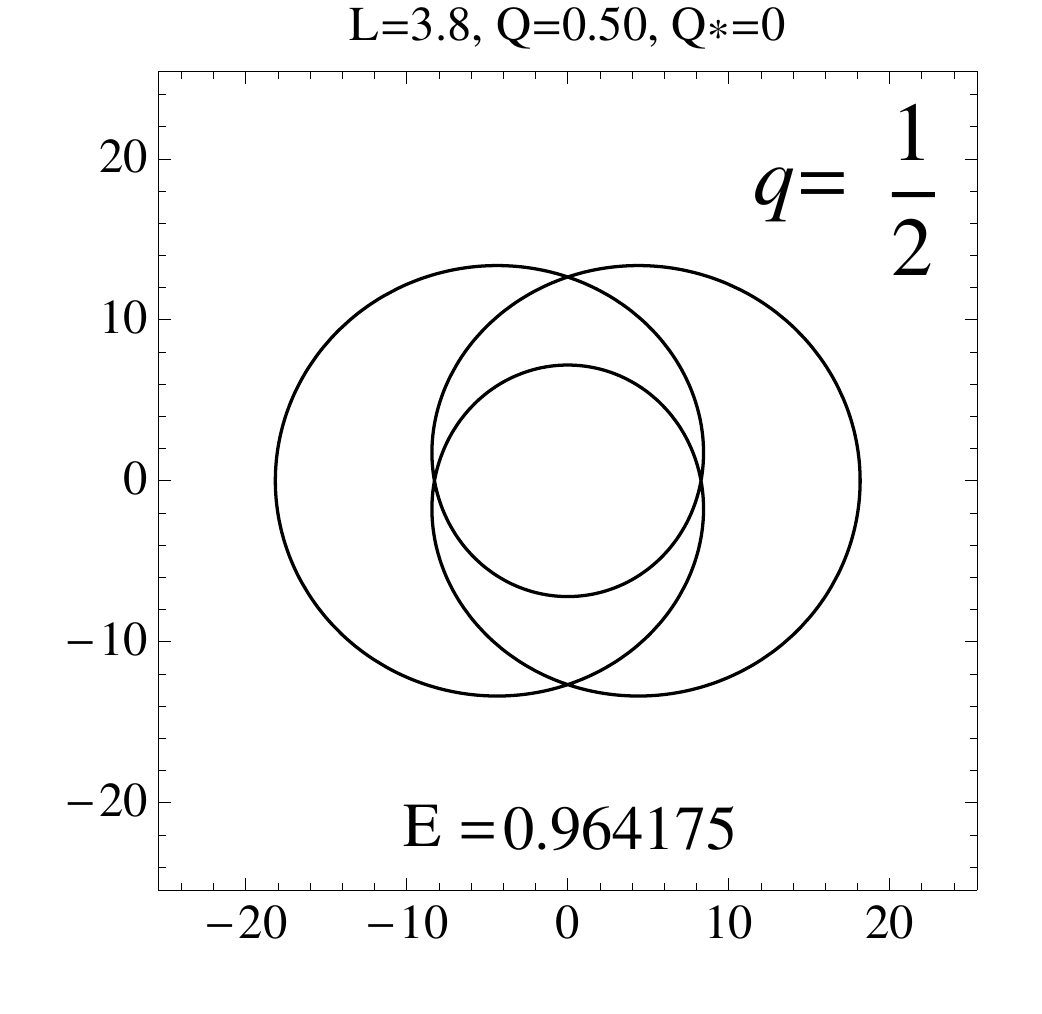}
    \hspace{-15pt}
    \includegraphics[width=.23\textwidth]{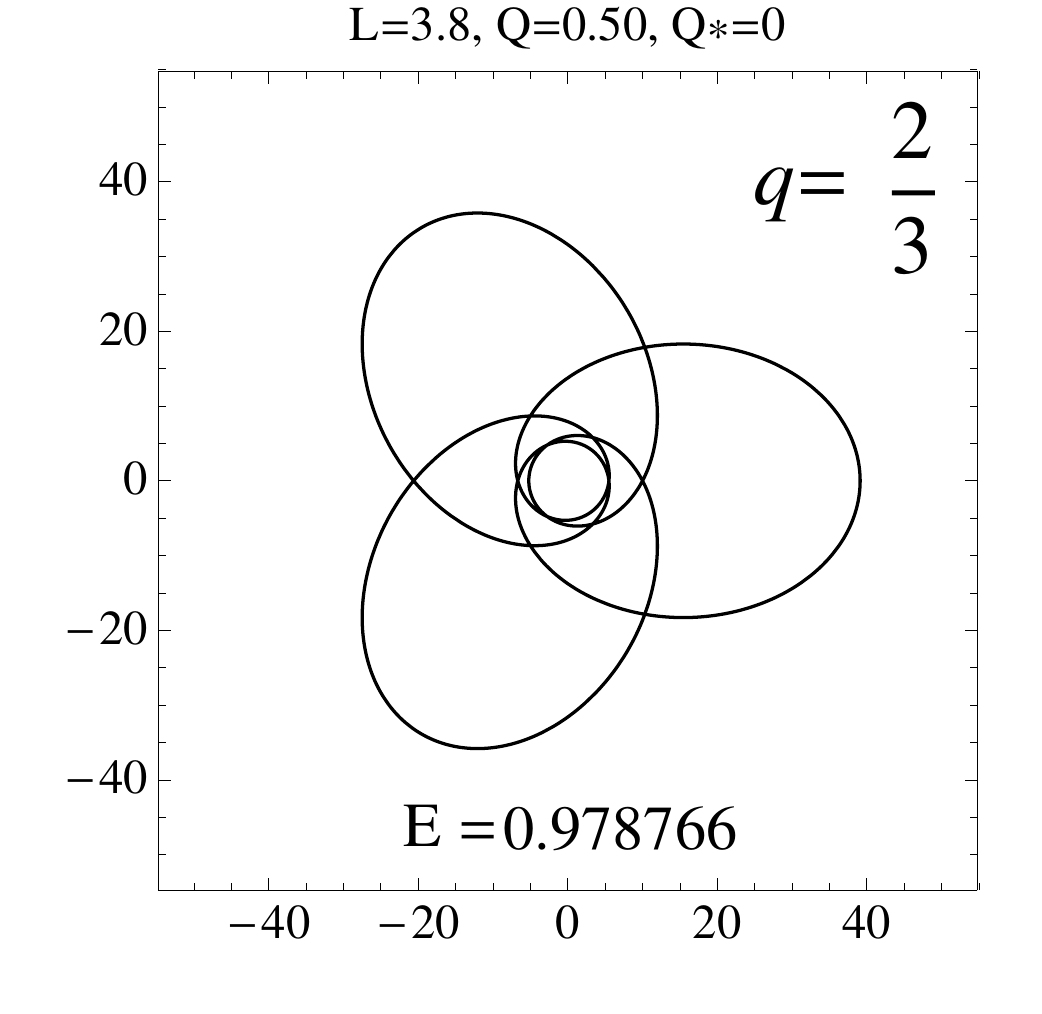}
    \hspace{-15pt}
	\includegraphics[width=.23\textwidth]{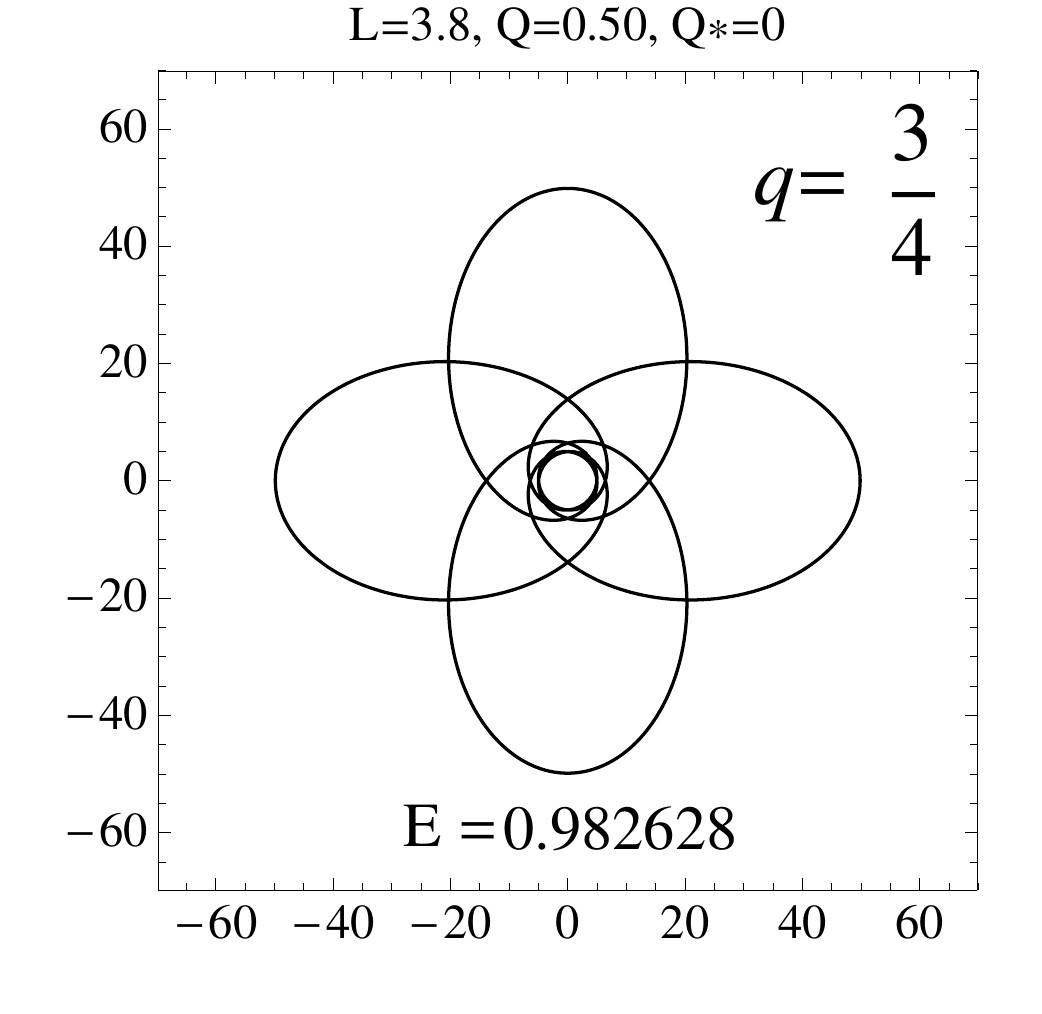} \hfill \\
	\caption{Periodic orbits with $z$ = 1, 2, 3, and 4 for $L=3.8$, $Q=0.5$, $Q_{*}=0$
	    \label{fig:leaves}}
    \vspace{15pt}
	\includegraphics[width=.23\textwidth]{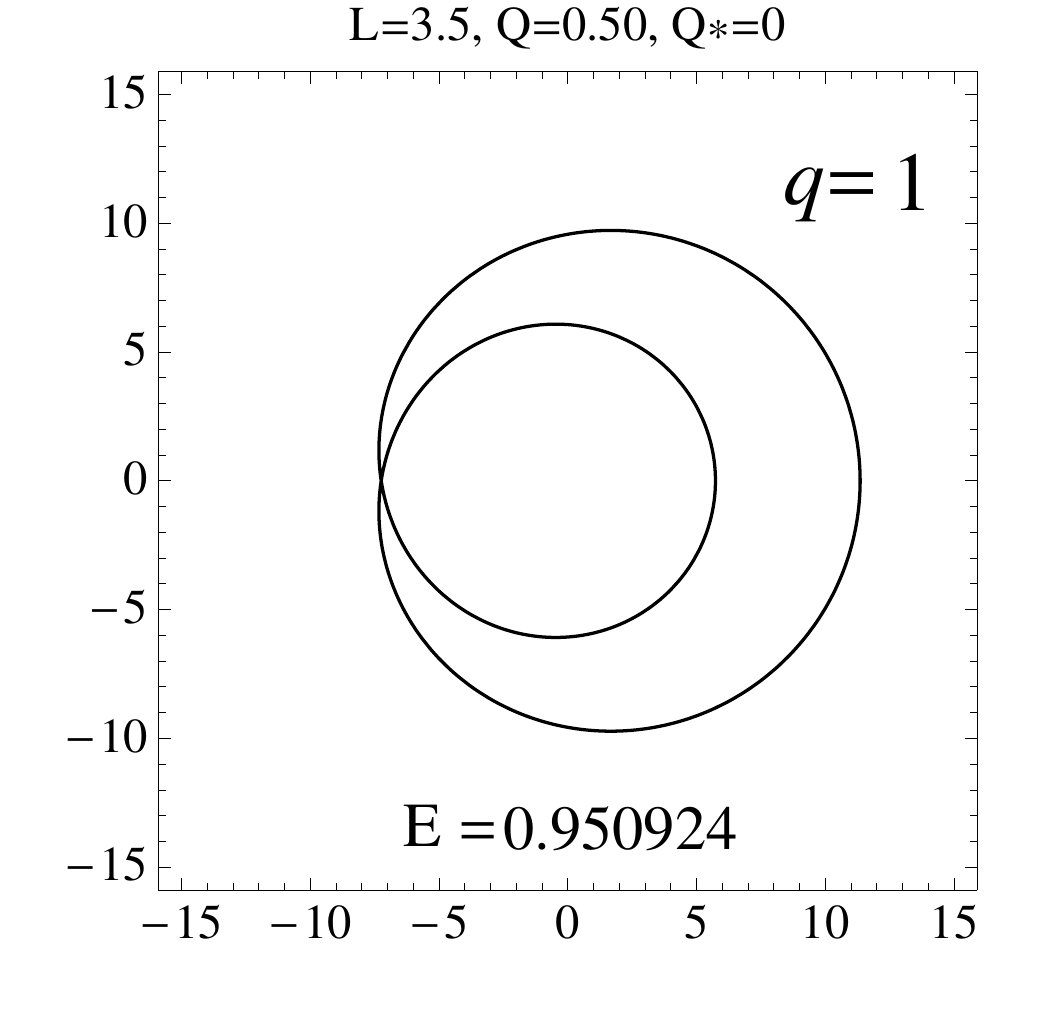}
	\includegraphics[width=.23\textwidth]{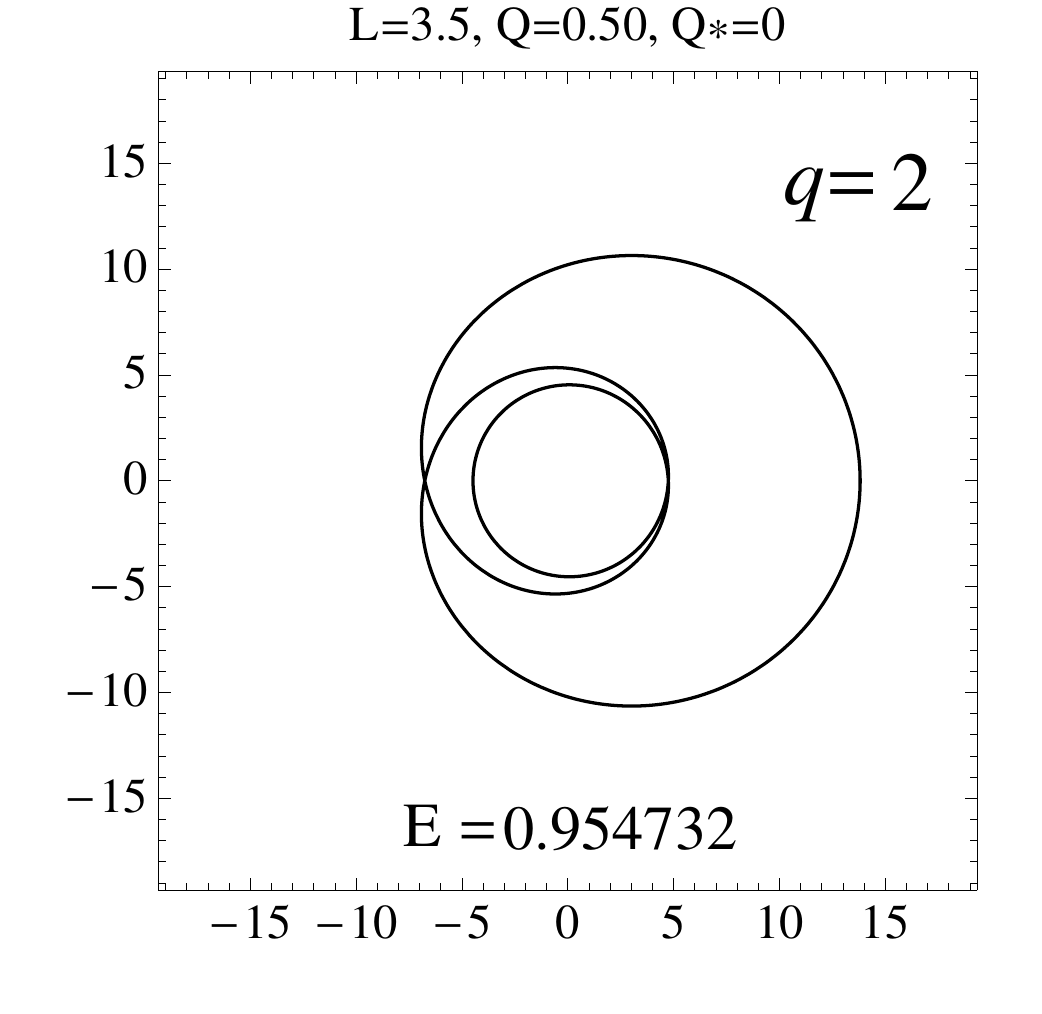}
	\includegraphics[width=.23\textwidth]{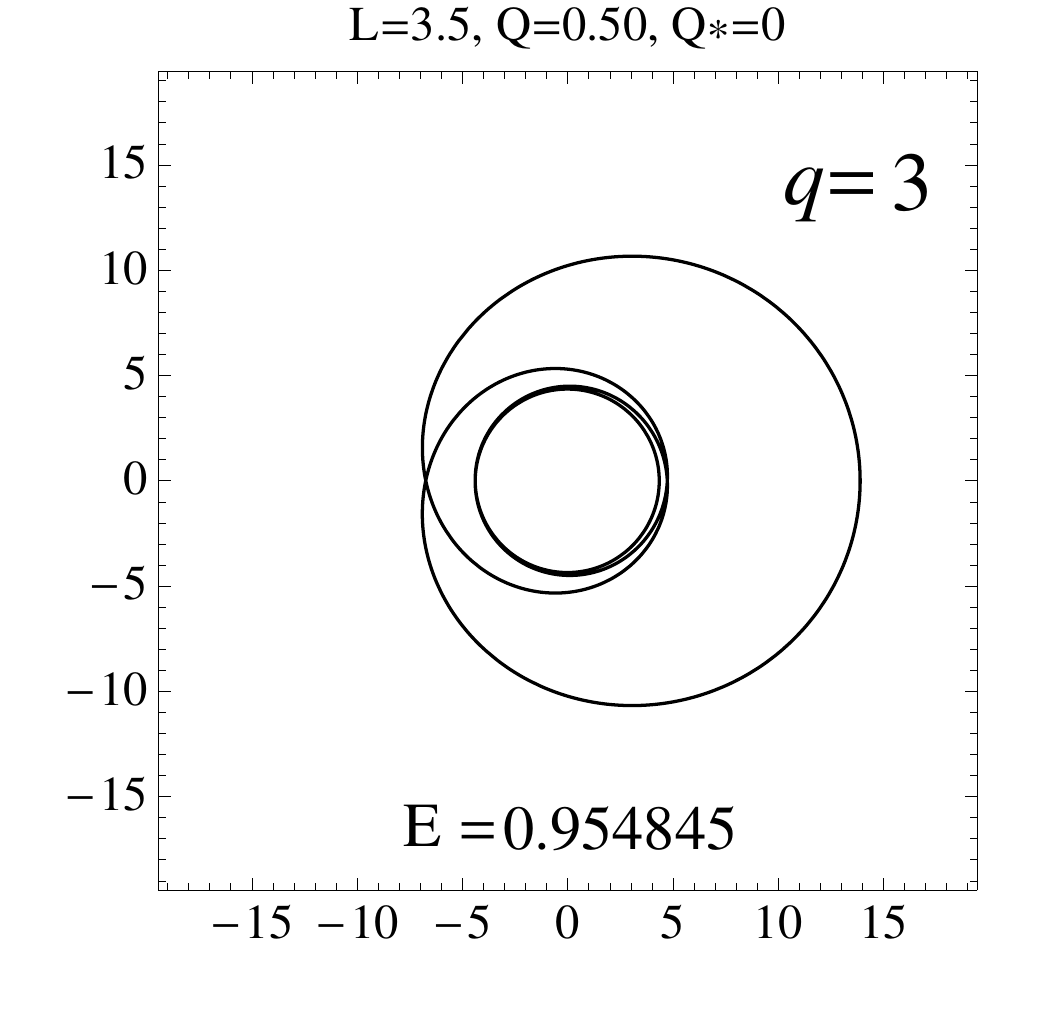}
	\caption{The $(w,v,z)=(w,0,1)$ orbits for $L=3.5$, 
	    $Q=0.5$.  From left to right, these are $(0,0,1)$, $(1,0,1)$,
	    and $(2,0,1)$.
	    Orbits with $w>3$ are indistinguishable at this
	    scale because the additional ``whirls" are densely packed in $r$.
	    \label{fig:w01orbits}}
    \vspace{15pt}
	\includegraphics[width=.23\textwidth]{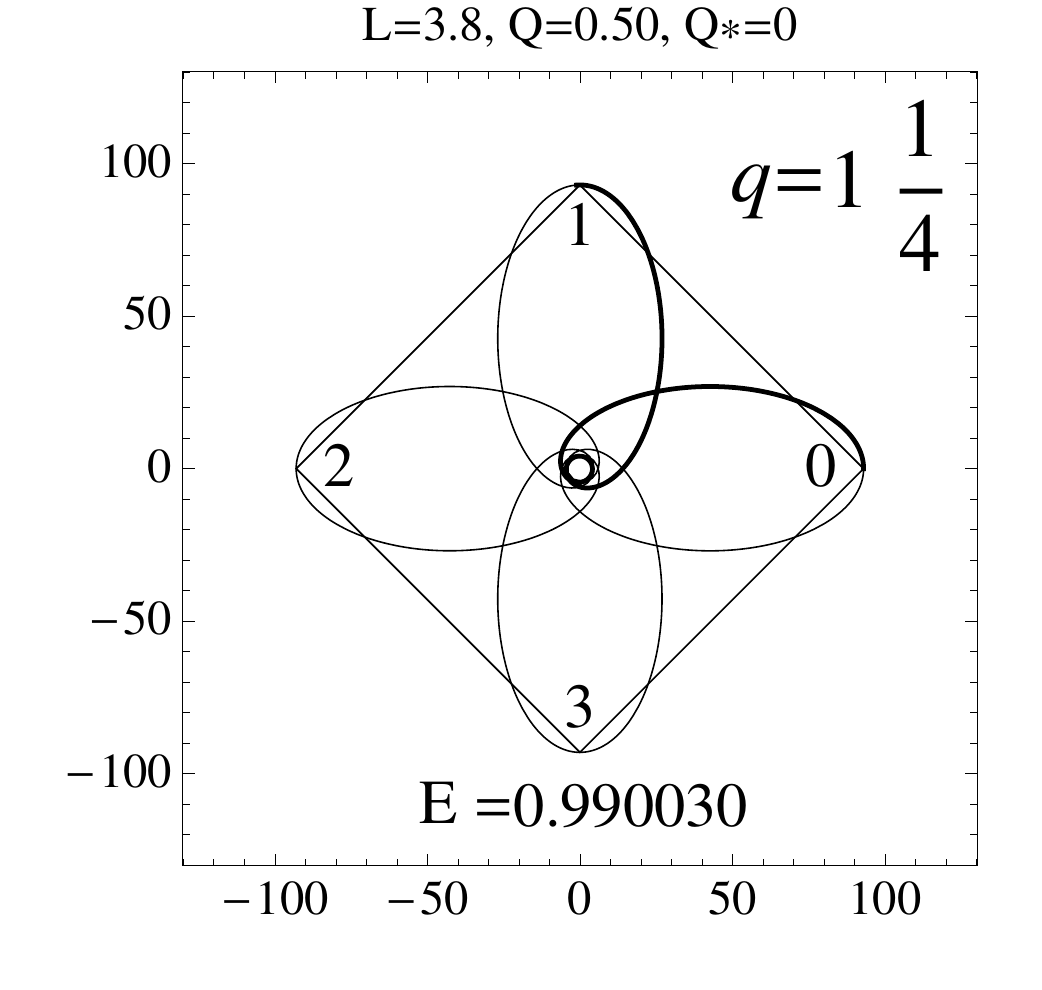}
	\hspace{-13pt}
	\includegraphics[width=.23\textwidth]{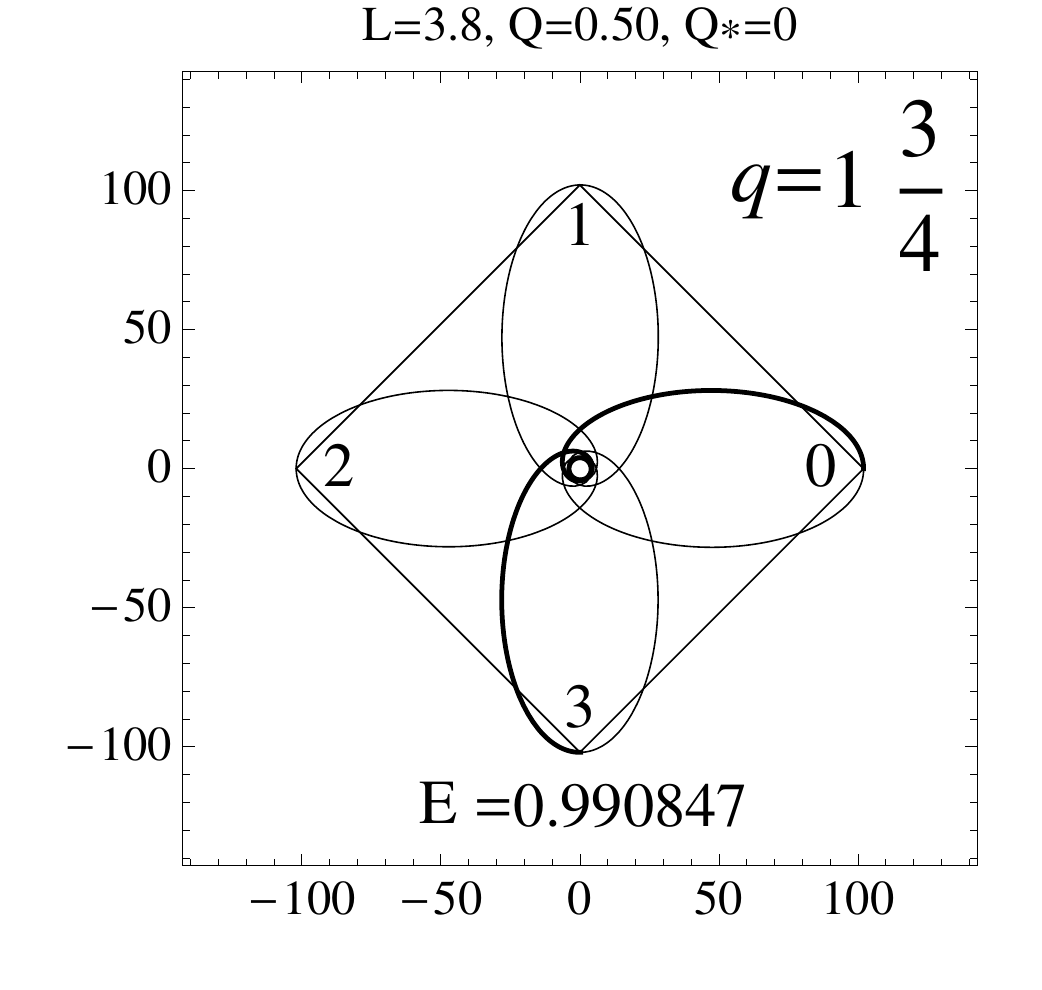}
	\caption{Two $z=4$, $w=1$ orbits with different $v$ values.\label{fig:vorbits}}
    \end{minipage}
\end{figure*}

Each periodic orbit is associated with a rational number $q$, defined as
\begin{equation}
\label{eqn:q}
q=w+\frac{v}{z},
\end{equation}
where $w$, $v$, and $z$ are integers.  Each of these integers corresponds
to a specific topological characteristic of a given periodic orbit.  
The most easily visualized is ``z," the number of leaves, or ``zooms" 
in the particle's orbit.  In its path around an RN black hole, 
our test particle will trace out a number of 
leaves before closing.  Figure \ref{fig:leaves} depicts orbits with various $z$ values.

The integer $w$ defines the number of ``whirls" the particle makes in
its path from apastron to periastron to the subsequent apastron.  To
understand this, note that every object travels at
least a full $2\pi$ around the central black hole. The number of
whirls is defined as the additional integer number of $2\pi$ executed
beyond this. In other words, the number of
extra turns around the center of the geometry gives us the value of $w$.   
Figure \ref{fig:w01orbits} shows orbits with various $w$ values.

We require a third number
$v$, the ``vertex" number, to distinguish between orbits that have
equal $z$ and $w$ but are geometrically different
nonetheless.  This is easily seen in Figure \ref{fig:vorbits}, in
which both orbits have $z=4$, $w=1$, but where we see that the
particle can skip leaves in its motion from apastron to apastron.  

We label successive apastra of a periodic orbit with integers, 
counting the starting apastron of the orbit as $0$ and increasing in the 
same rotational sense as the orbit (counterclockwise for prograde orbits, 
clockwise for retrograde orbits), as shown in Figure \ref{fig:vorbits}.  
In general, any periodic orbit with $z>2$ can skip any number of vertices 
less than $z$ when moving between apastra.  We define $v$ to be the index 
of the first vertex hit by the orbit after $v=0$ (the bounds on $v$ are therefore 
$1\leq v \leq z-1$).  When $z=1$, we define $v=0$, which is the only 
sensible choice for $v$ because it implies that that successive apastra for 
single-leaf ($z = 1$) orbits are actually the 
same single apastron (see Figure \ref{fig:leaves}).

Finally, we must address the degeneracy that arises when the quotient $v/z$ 
is a reducible fraction.  For a given $w$, there are multiple 
choices of $z$ and $v$ that describe the same orbit; for instance, when we 
have $q = 1\frac{2}{4}$ the particle skips every other apastron and never 
hits vertices 1 or 3, which means it closes after only two leaves.  This is 
equivalent to a $q = 1\frac{1}{2}$ orbit.  To avoid this issue, we require 
that $v$ and $z$ be relatively prime.  The bounds on v are therefore
\begin{eqnarray}
	1 \leq v \leq z-1 & & \textrm{if } z > 1 \;\text{and}\;	z, v\; \text{are relatively prime} \nonumber \\
	v=0 & & \textrm{if } z=1.
\end{eqnarray}
The rational $q$ defines all of the topological features of any
closed equatorial orbit and corresponds to the precession of the orbit
beyond a Keplerian ellipse.

\begin{figure*}[p]
    \begin{minipage}{.6\textwidth}
	\centering
	\includegraphics[width=0.32\textwidth]{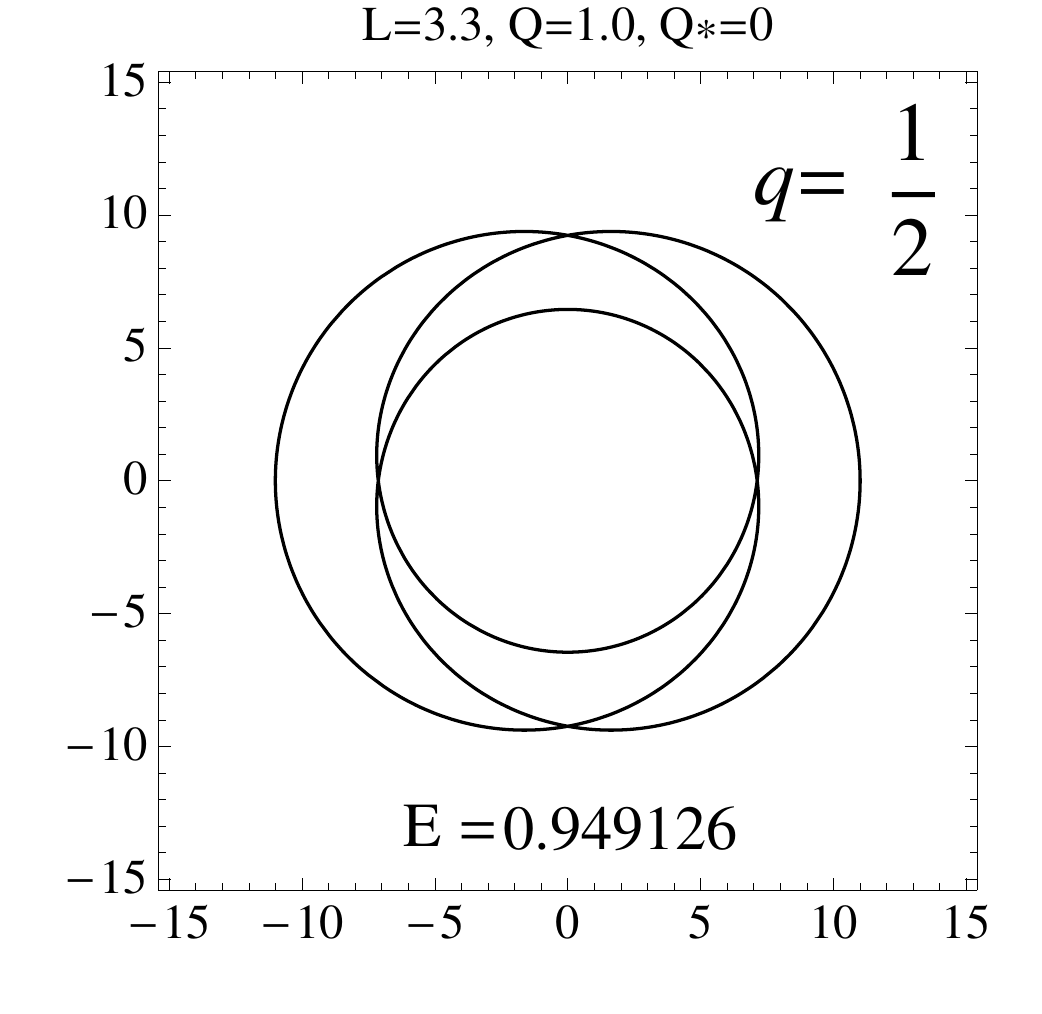}
	\hspace{-10pt}
	\includegraphics[width=0.32\textwidth]{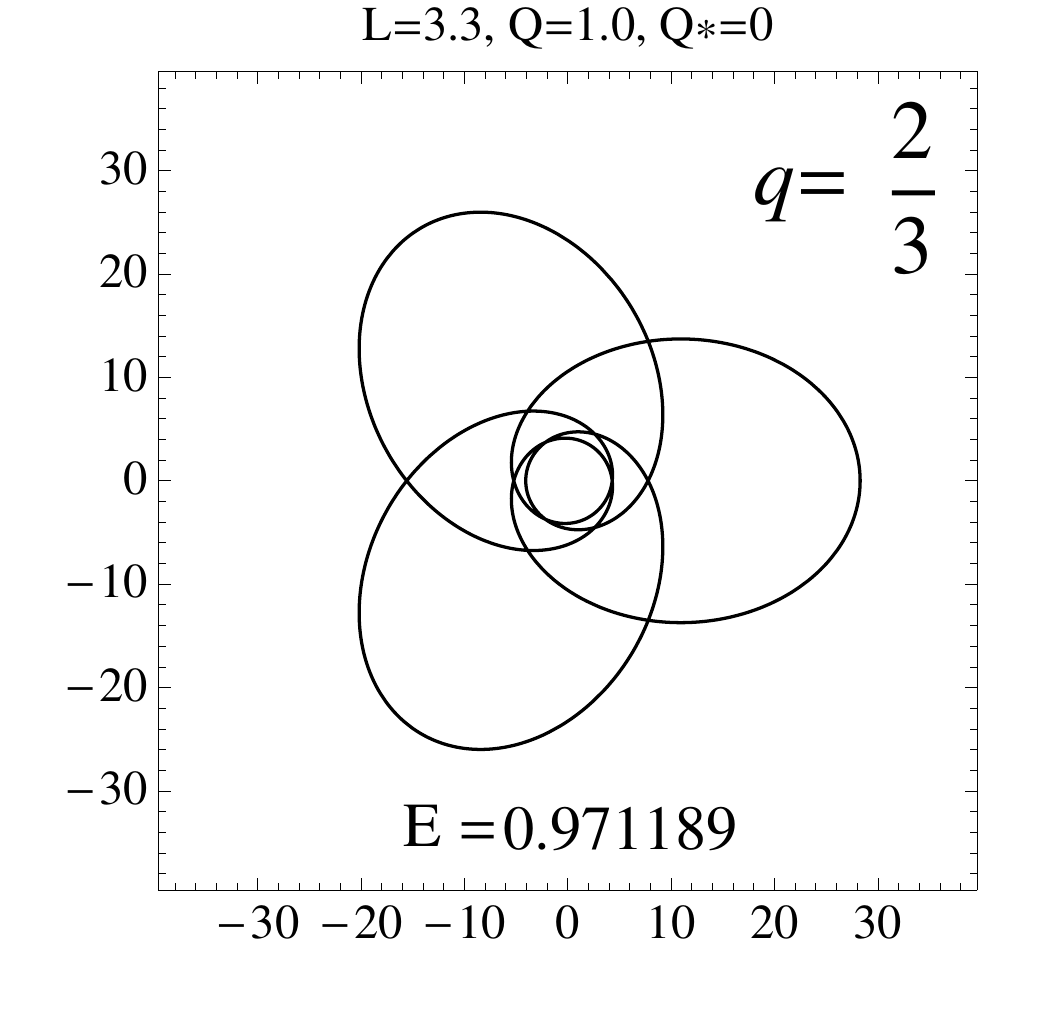}
	\hspace{-10pt}
	\includegraphics[width=0.32\textwidth]{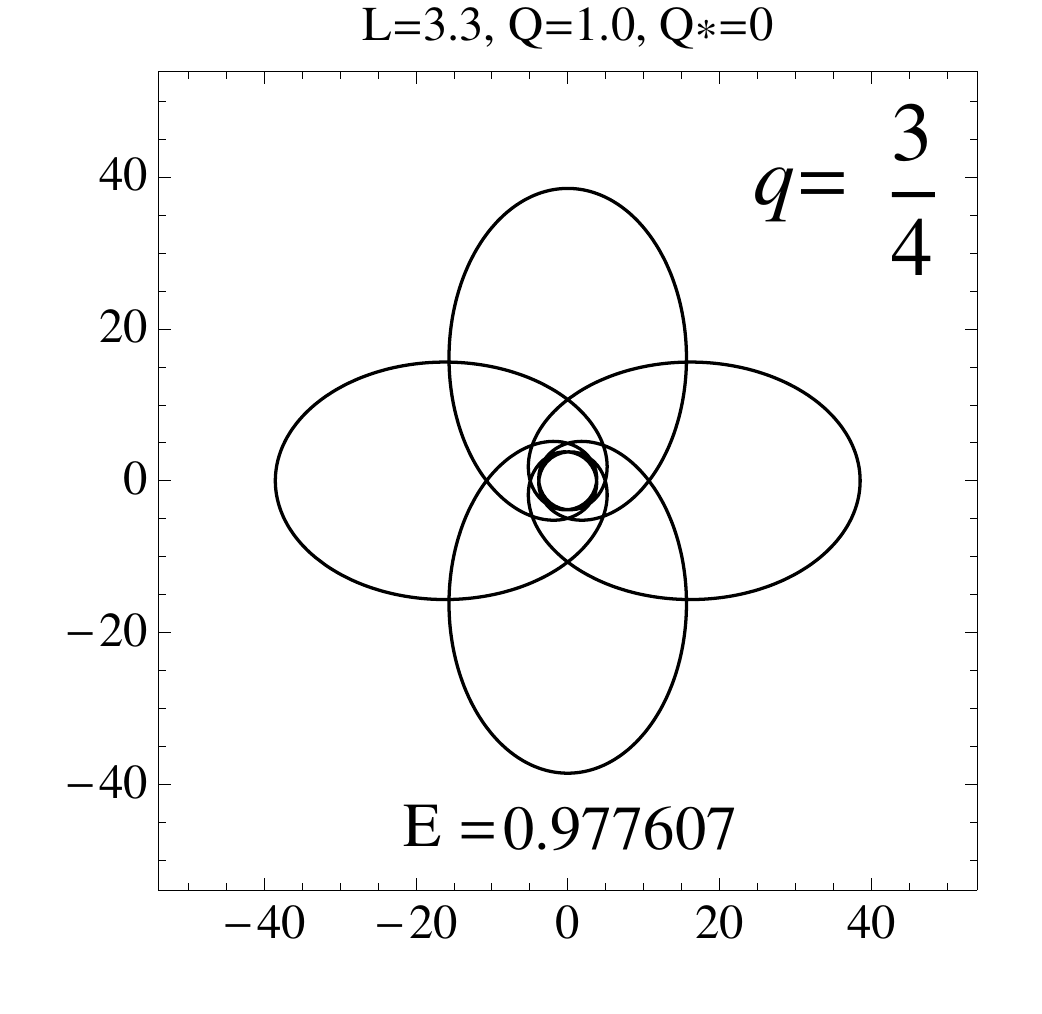} \hfill \\
    \hspace{10pt}
	\includegraphics[width=0.32\textwidth]{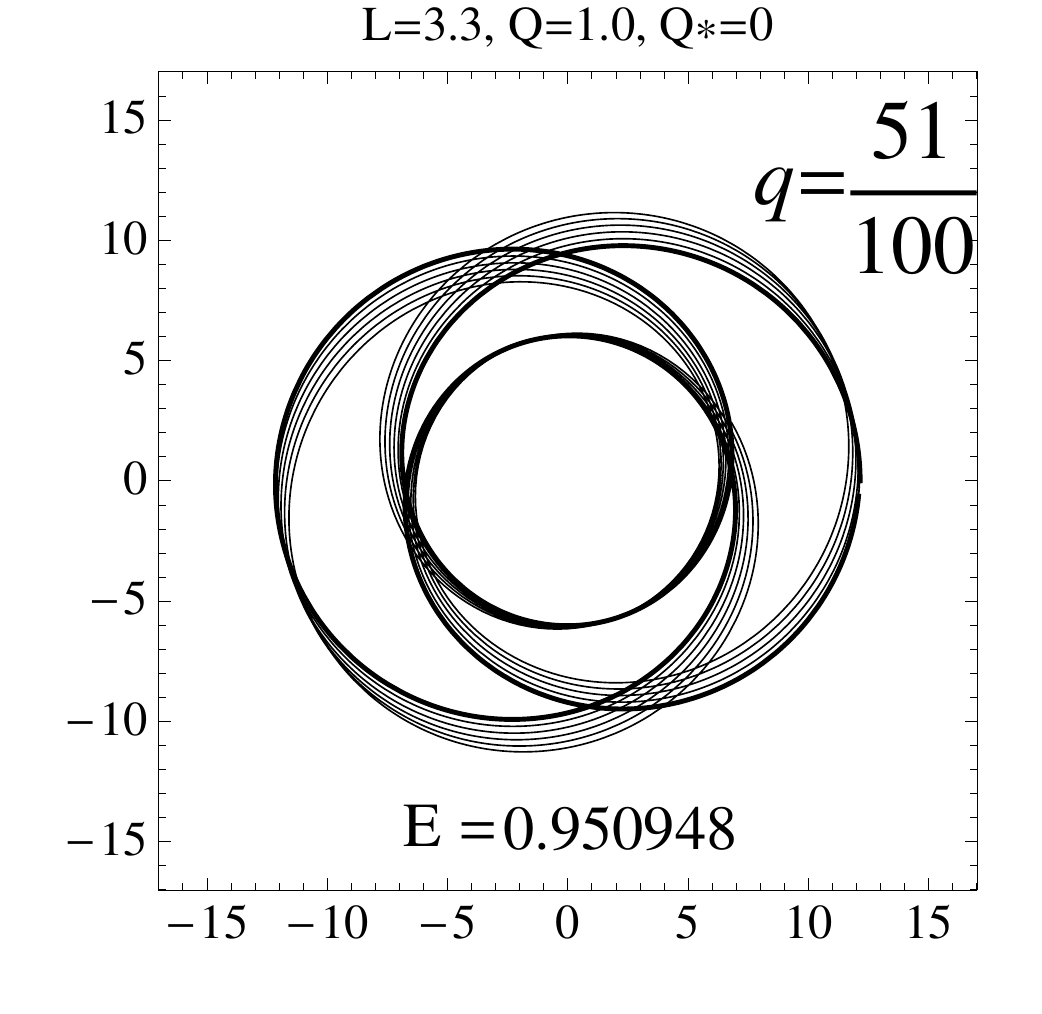}
	\hspace{-10pt}
	\includegraphics[width=0.32\textwidth]{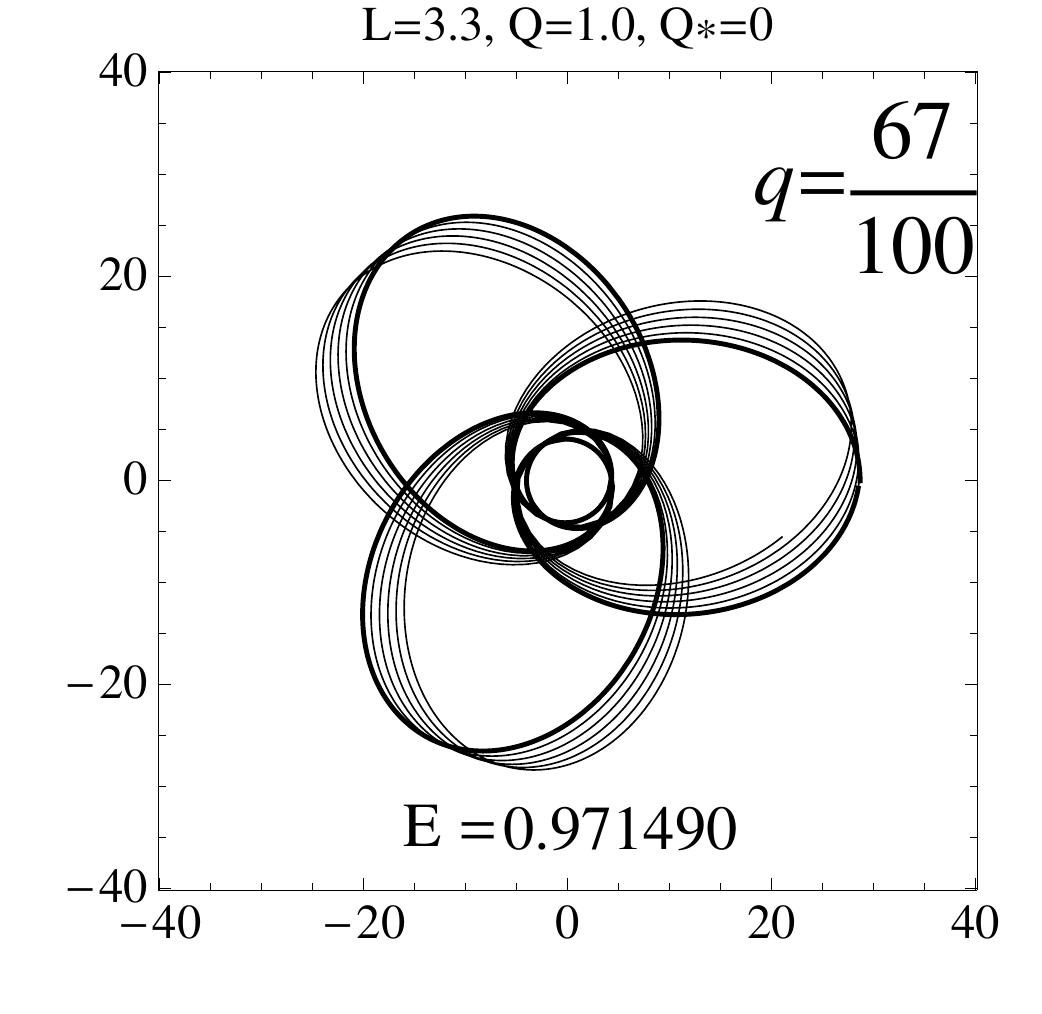}
	\hspace{-10pt}
	\includegraphics[width=0.32\textwidth]{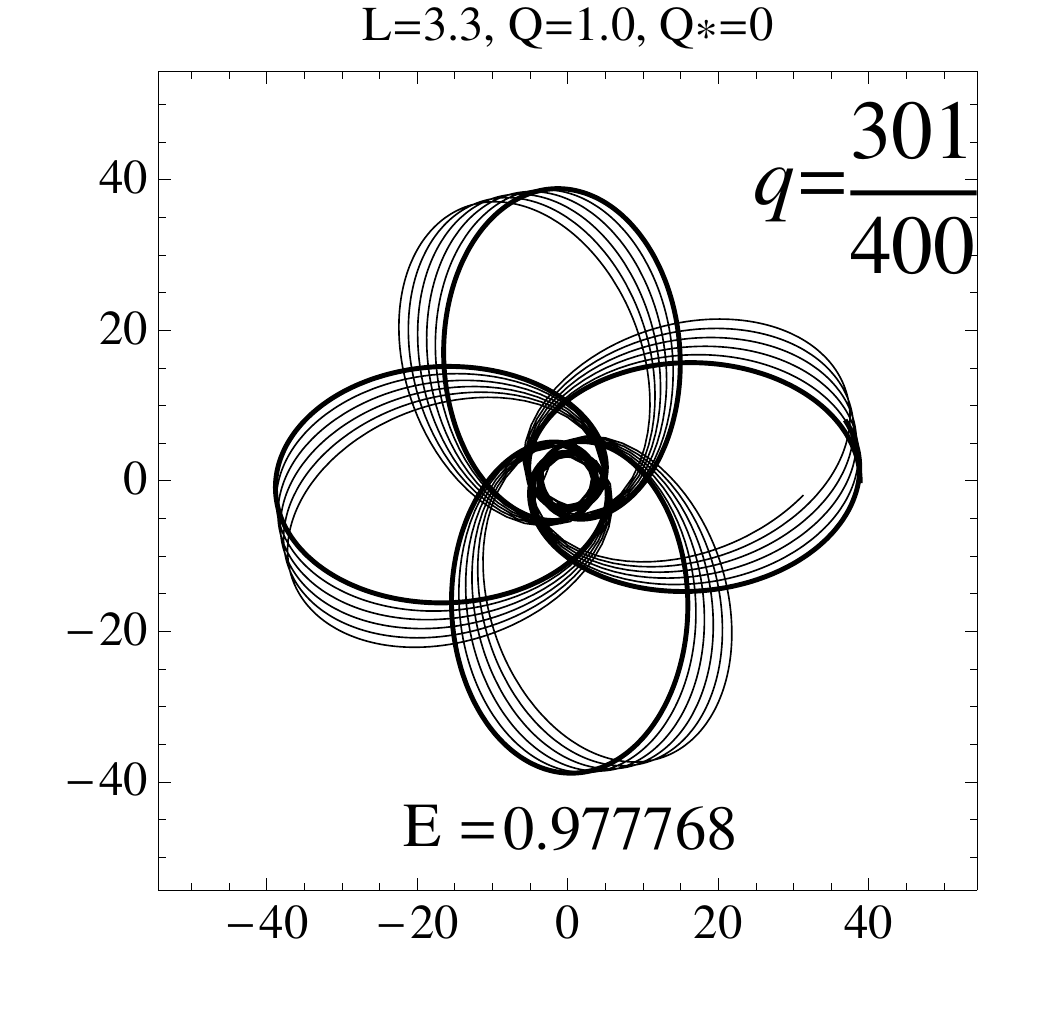} \hfill
	\vspace{-10pt}
	\caption{A periodic table of orbits; in each column the upper orbit may be approximated by the corresponding orbit in the second row. 
	\label{fig:pt1}}
    \end{minipage}
\end{figure*}

The utility of the taxonomy becomes apparent when we compare pairs of orbits 
with nearby $q$ values.  Figure \ref{fig:pt1} depicts several pairs 
of orbits; each orbit in the center column may be approximated with the one 
in the left column.

\subsection{Comparisons Between RN and Schwarzschild}

\begin{figure*}[p]
    \begin{minipage}{.8\textwidth}
	\centering
	\includegraphics[width=0.32\textwidth]{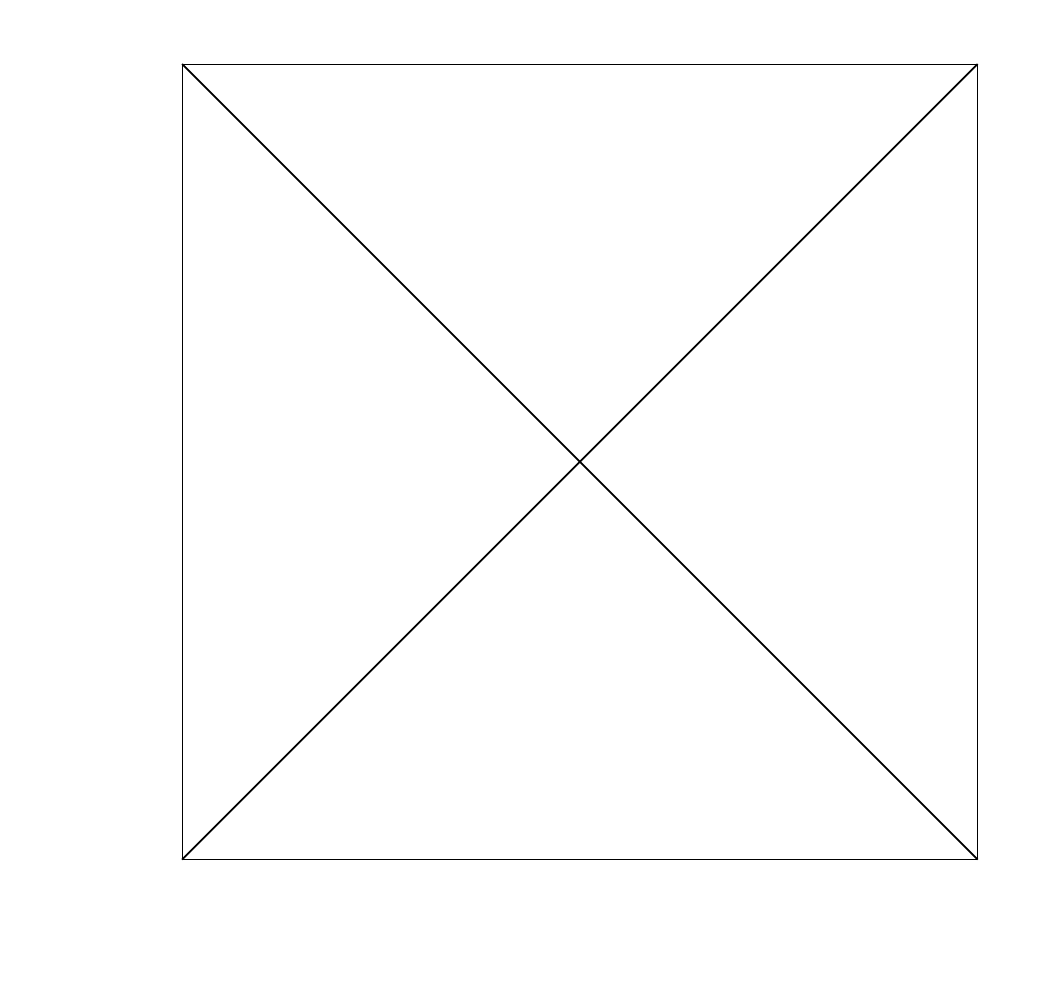}
	\hspace{-10pt}
	\includegraphics[width=0.32\textwidth]{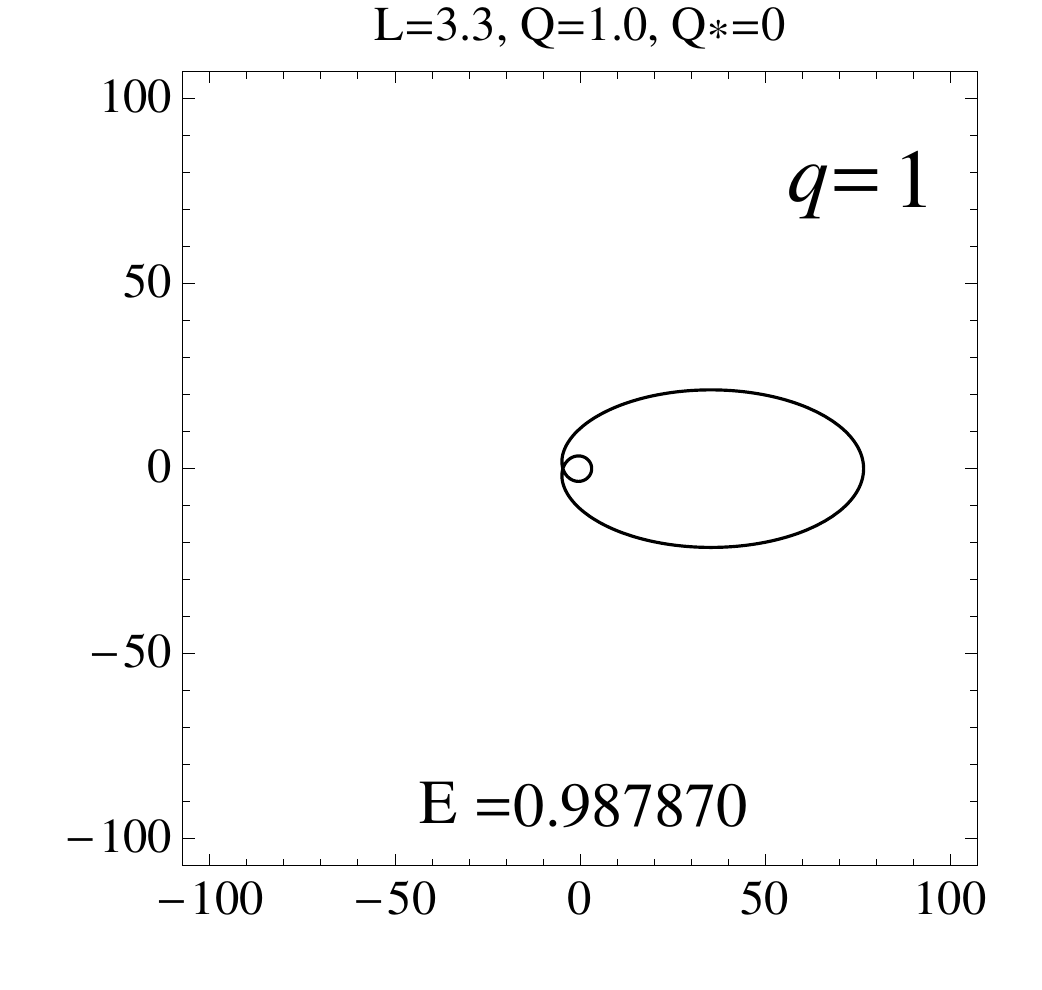}
	\hspace{-10pt}
	\includegraphics[width=0.32\textwidth]{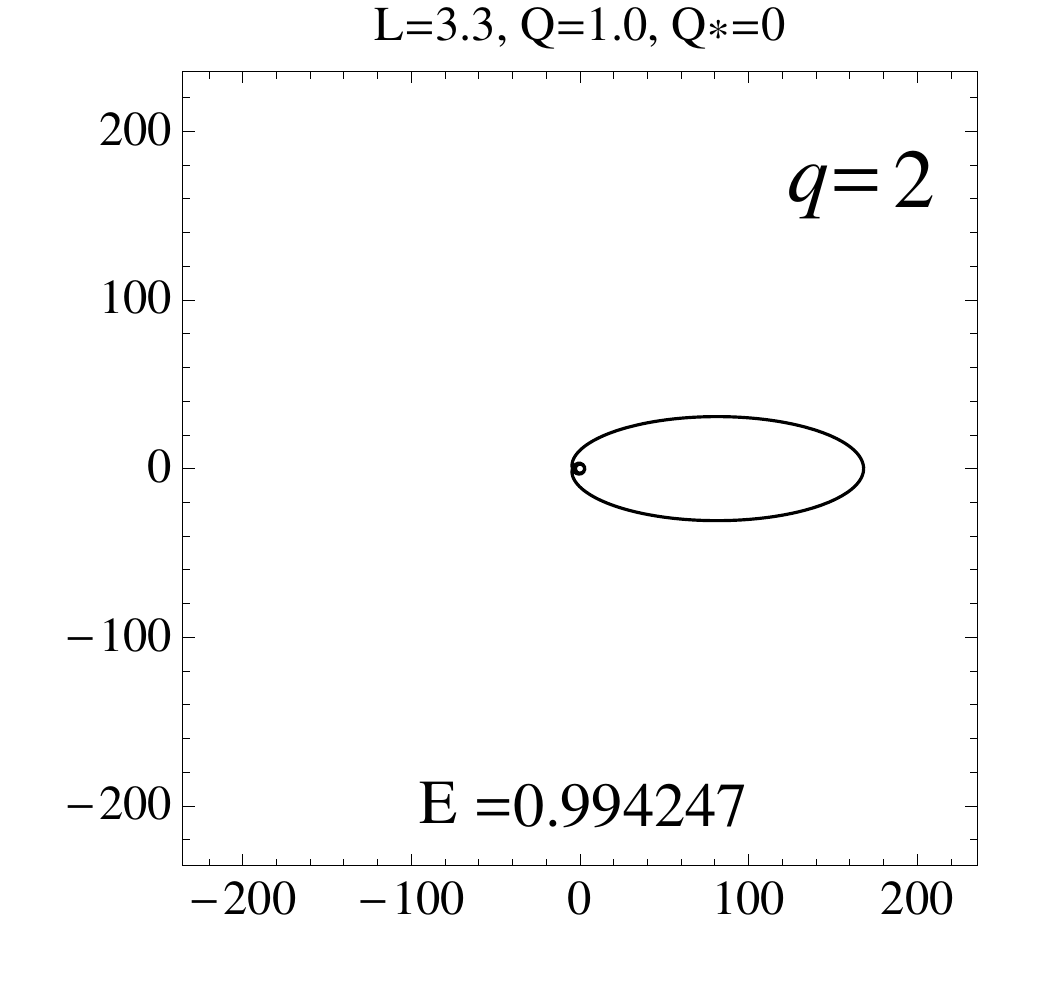} \hfill \\
	\includegraphics[width=0.32\textwidth]{plots/blank.pdf}
	\hspace{-10pt}
	\includegraphics[width=0.32\textwidth]{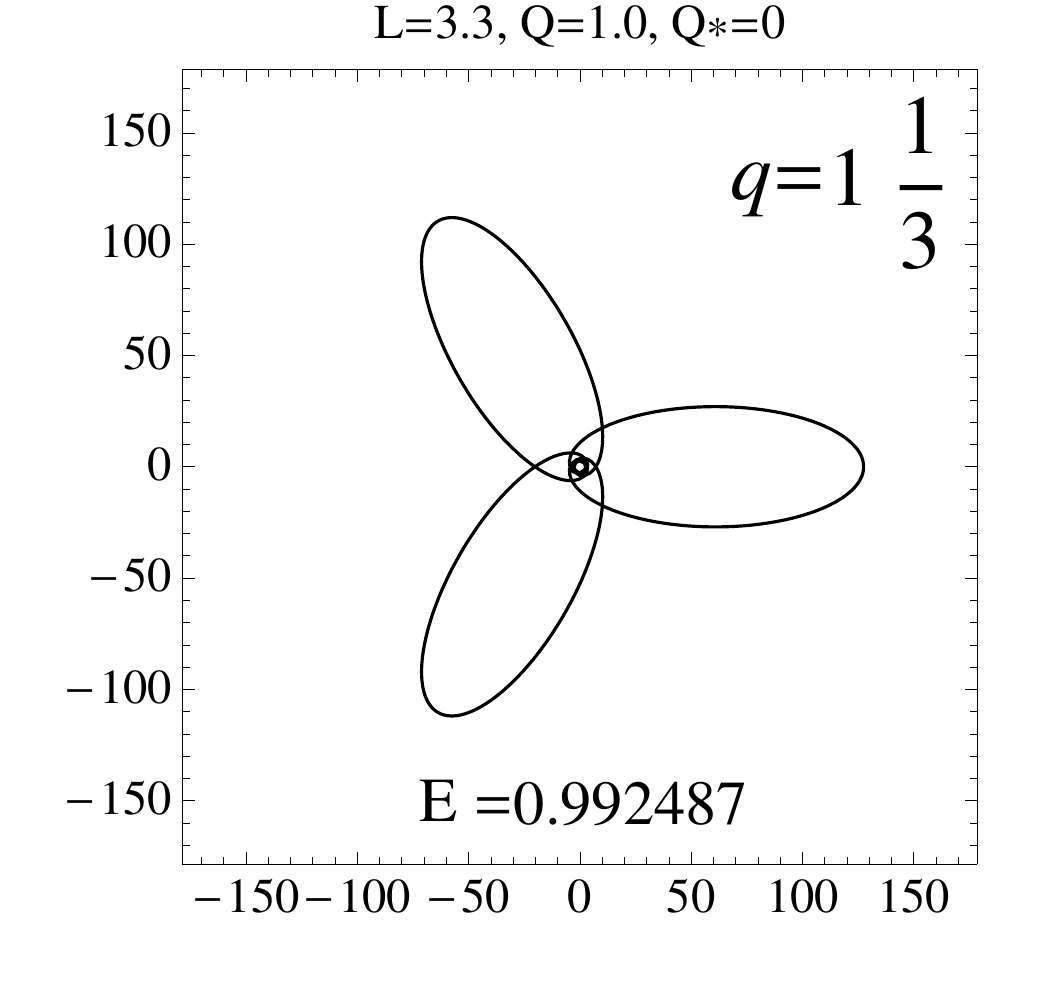}
	\hspace{-10pt}
	\includegraphics[width=0.32\textwidth]{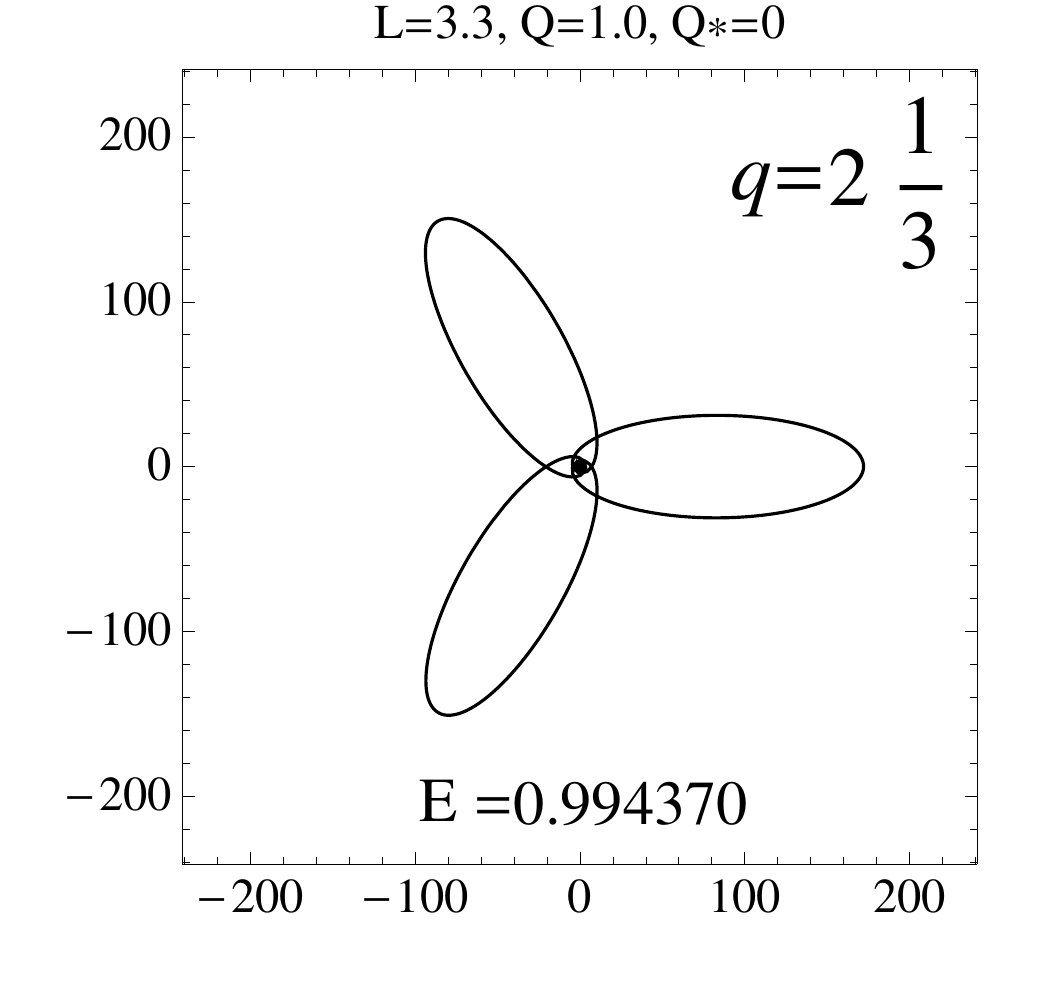} \hfill \\
	\includegraphics[width=0.32\textwidth]{plots/orbit_L33_Q10_q0_012.pdf}
	\hspace{-10pt}
	\includegraphics[width=0.32\textwidth]{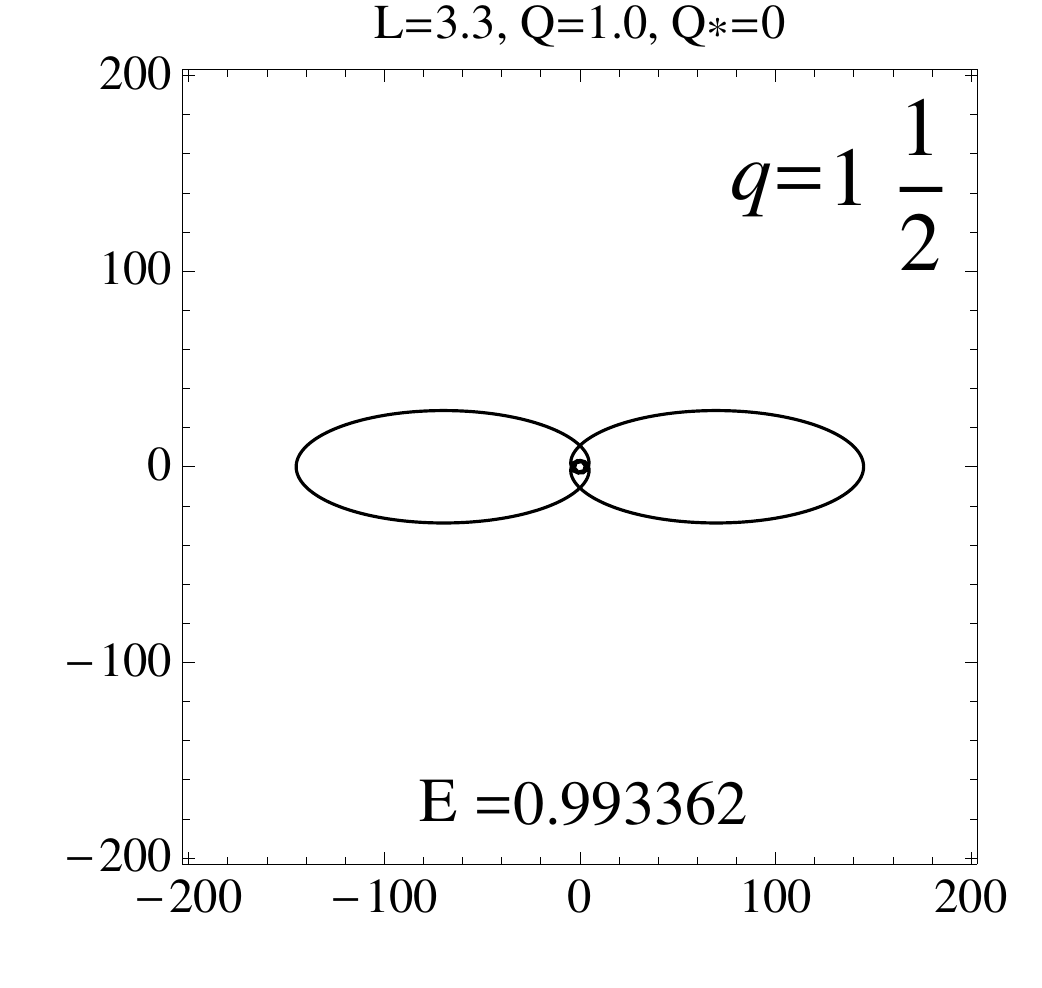}
	\hspace{-10pt}
	\includegraphics[width=0.32\textwidth]{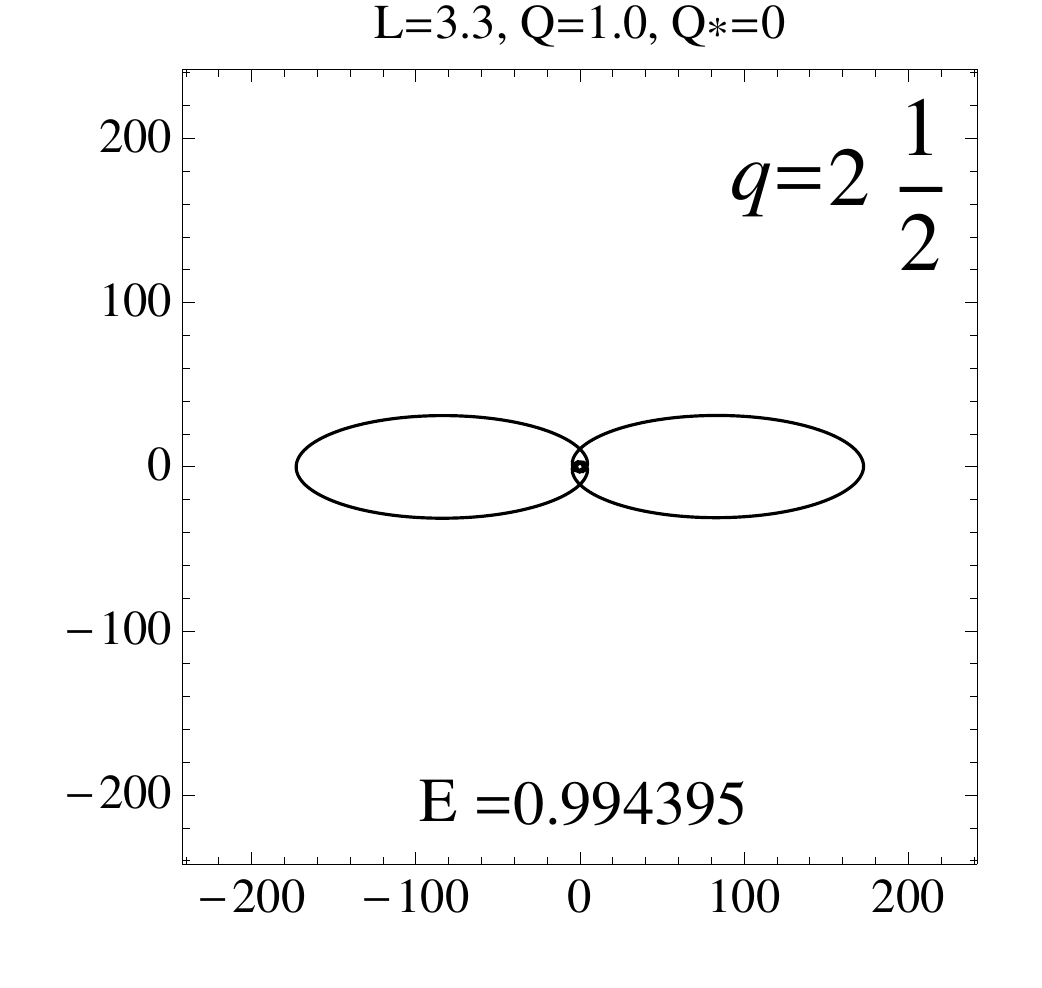} \hfill \\
	\includegraphics[width=0.32\textwidth]{plots/orbit_L33_Q10_q0_023.pdf}
	\hspace{-10pt}
	\includegraphics[width=0.32\textwidth]{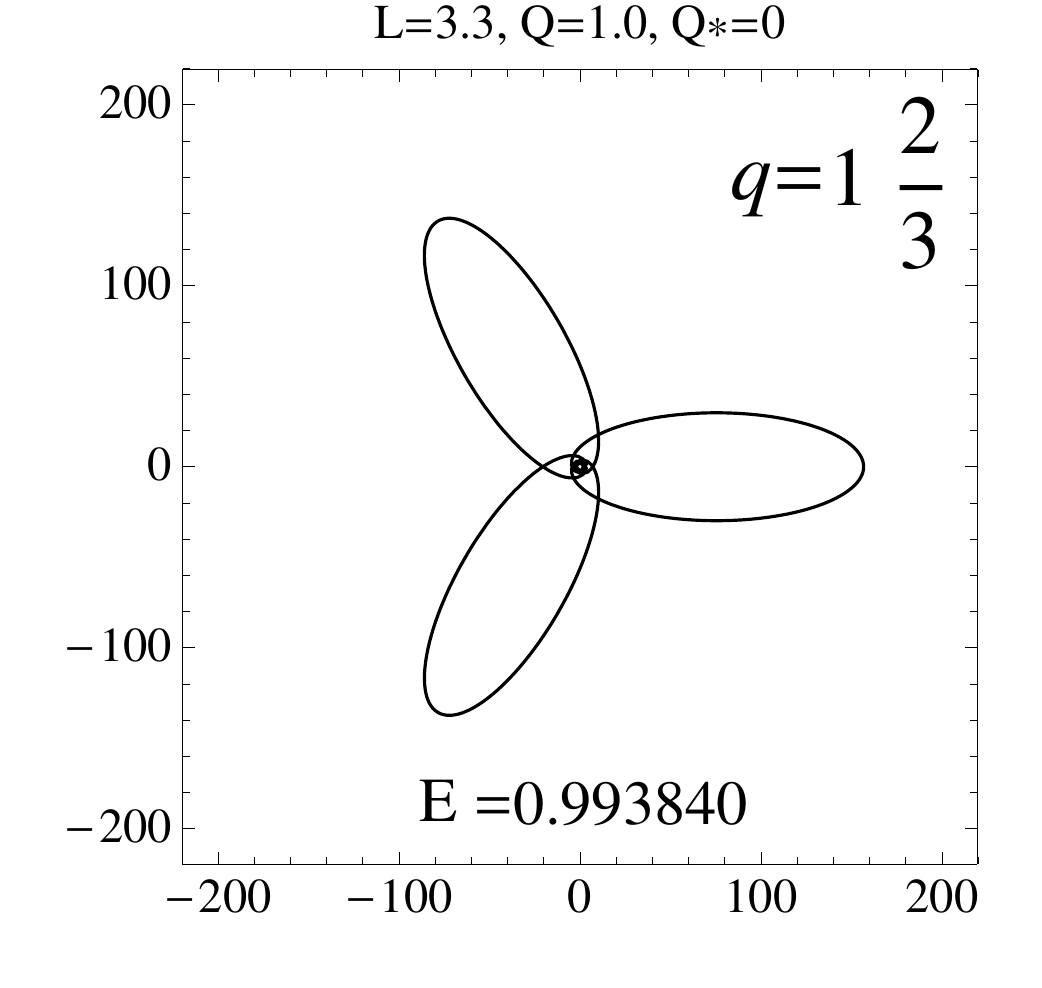}
	\hspace{-10pt}
	\includegraphics[width=0.32\textwidth]{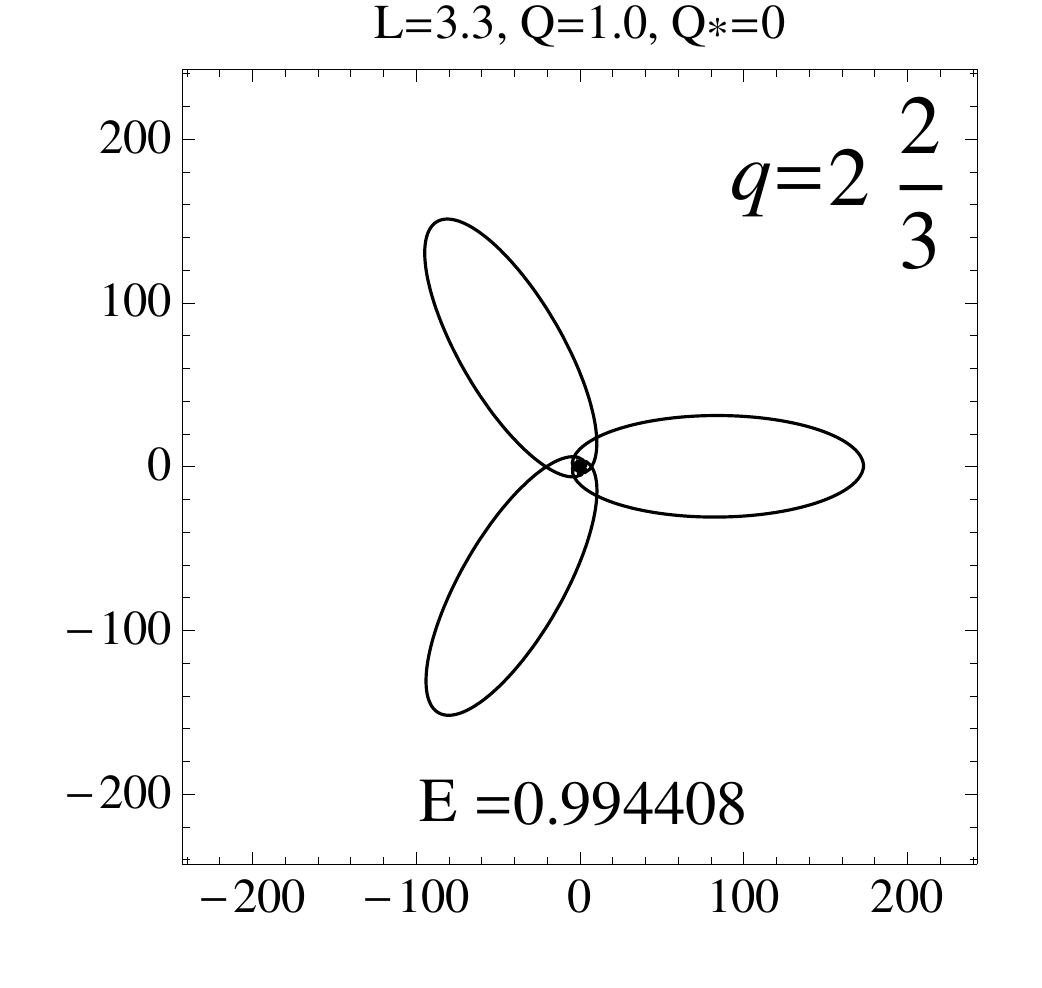} \hfill \\
	\vspace{-10pt}
	\caption{All extremal RN $z = $ 1, 2, 3 orbits with $w = $ 0 for the first column, $w = $ 1 
    for the middle column, and $w = $ 2 for the last column. For all orbits, $L = 3.3$ and $Q = 1.0$.
    Orbits increase in energy from top to bottom and left to right. 
    Note that the first and second entries in the first column are blank because the
    $q = 0+\frac{0}{1}$ and $q = 0+\frac{1}{3}$ orbits are inaccessible.
	\label{fig:pt}}
    \end{minipage}
\end{figure*}

\begin{figure*}[p]
    \begin{minipage}{.8\textwidth}
	\centering
	\includegraphics[width=0.32\textwidth]{plots/blank.pdf}
	\hspace{-10pt}
	\includegraphics[width=0.32\textwidth]{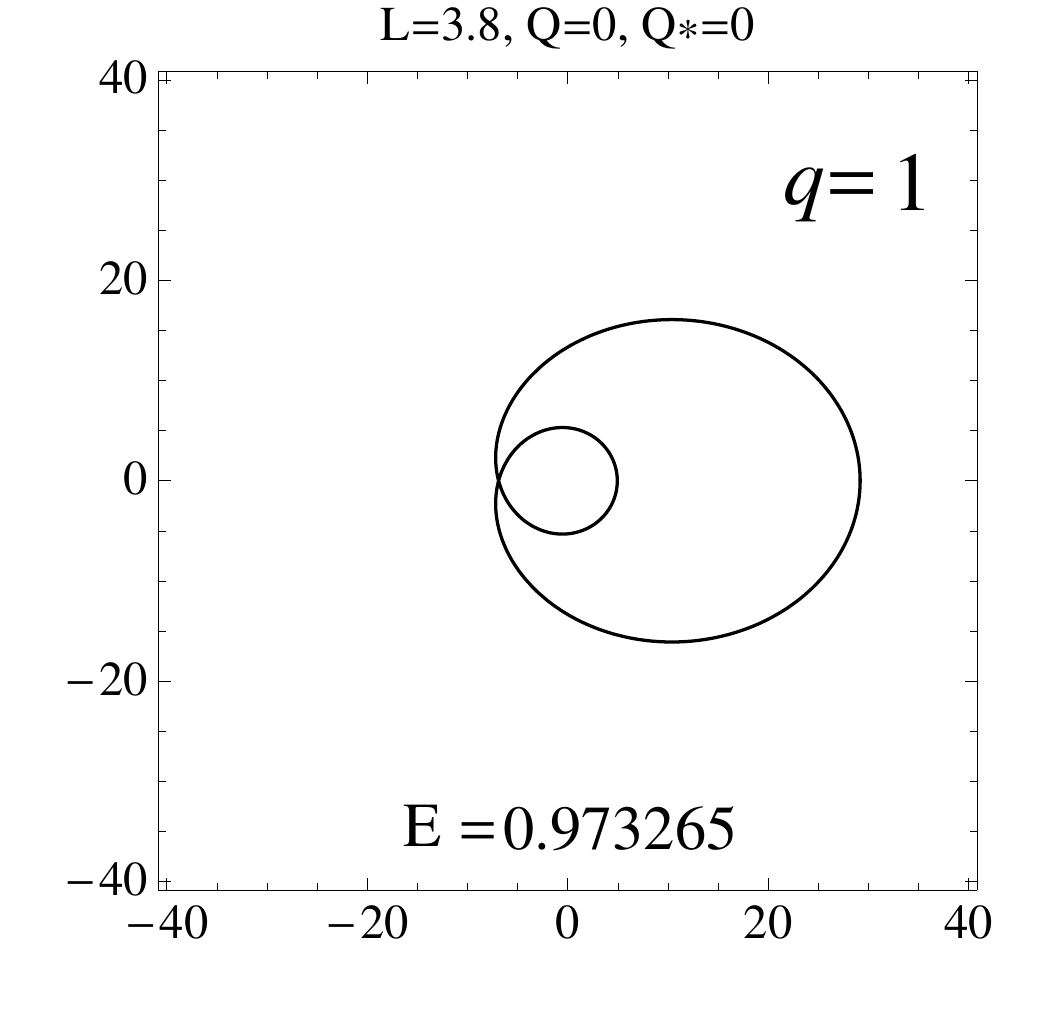}
	\hspace{-10pt}
	\includegraphics[width=0.32\textwidth]{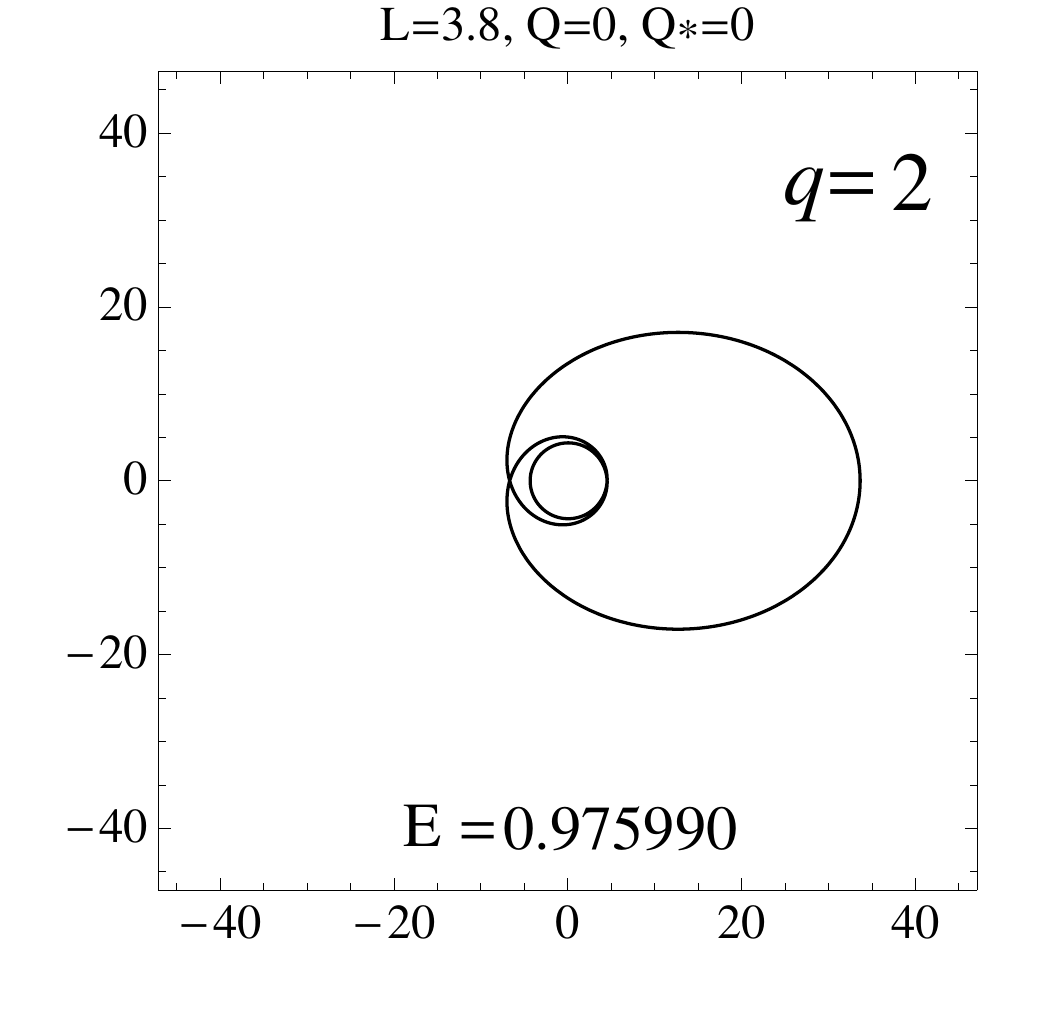} \hfill \\
	\includegraphics[width=0.32\textwidth]{plots/blank.pdf}
	\hspace{-10pt}
	\includegraphics[width=0.32\textwidth]{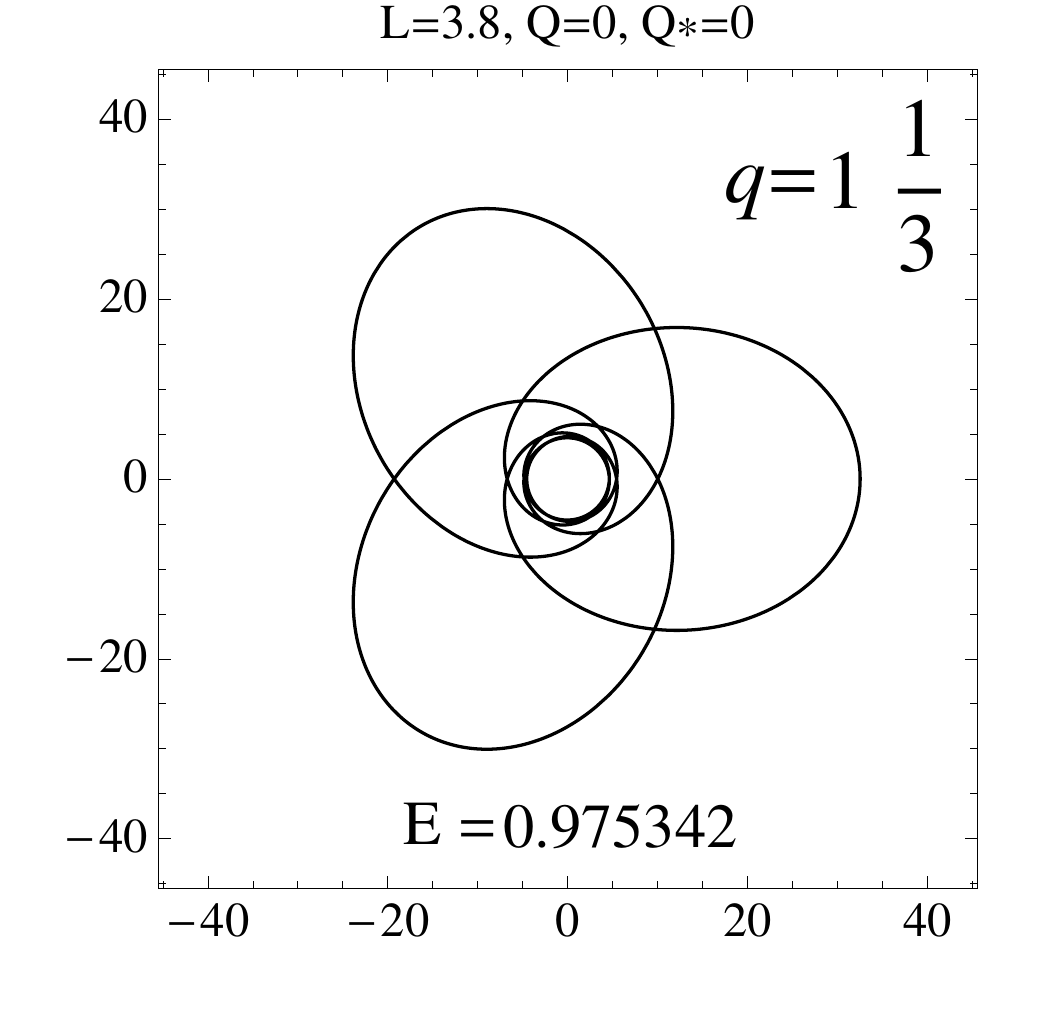}
	\hspace{-10pt}
	\includegraphics[width=0.32\textwidth]{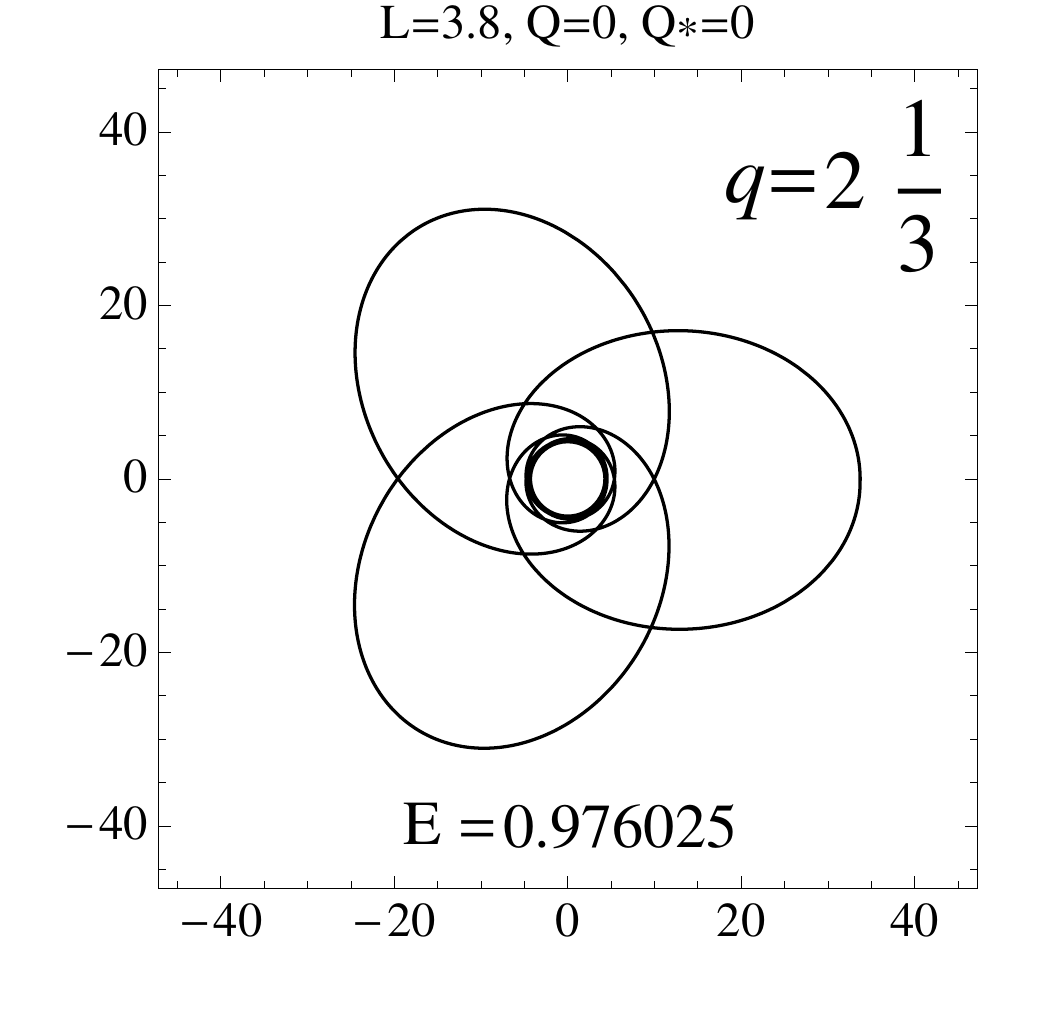} \hfill \\
	\includegraphics[width=0.32\textwidth]{plots/blank.pdf}
	\hspace{-10pt}
	\includegraphics[width=0.32\textwidth]{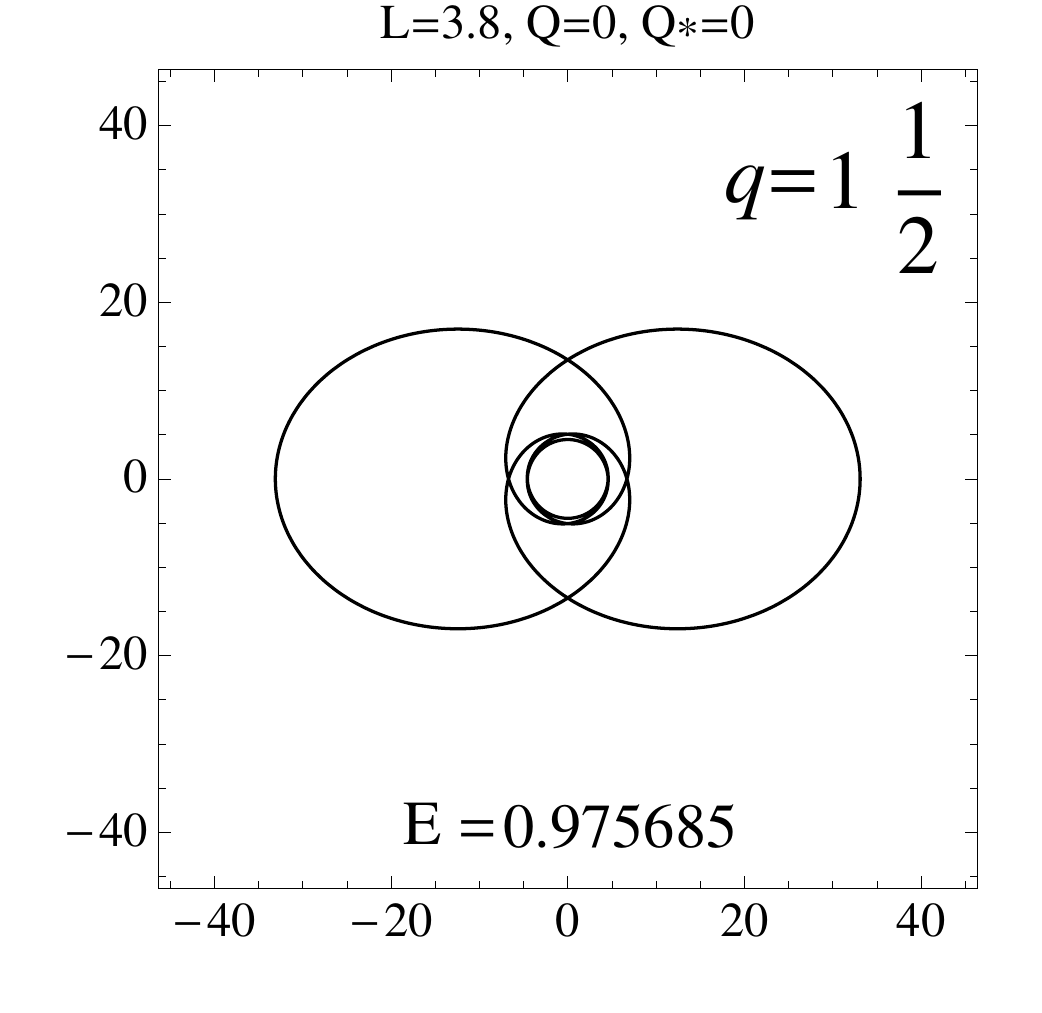}
	\hspace{-10pt}
	\includegraphics[width=0.32\textwidth]{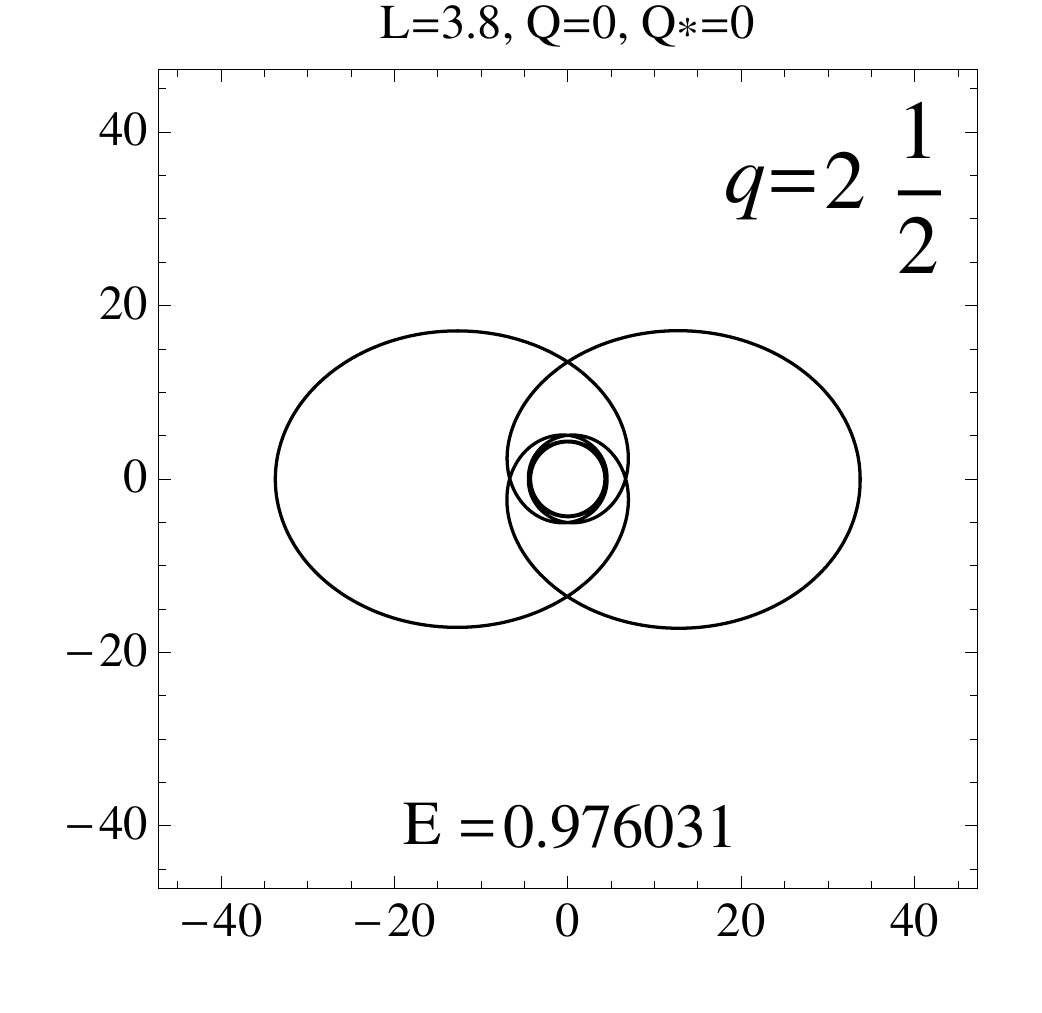} \hfill \\
	\includegraphics[width=0.32\textwidth]{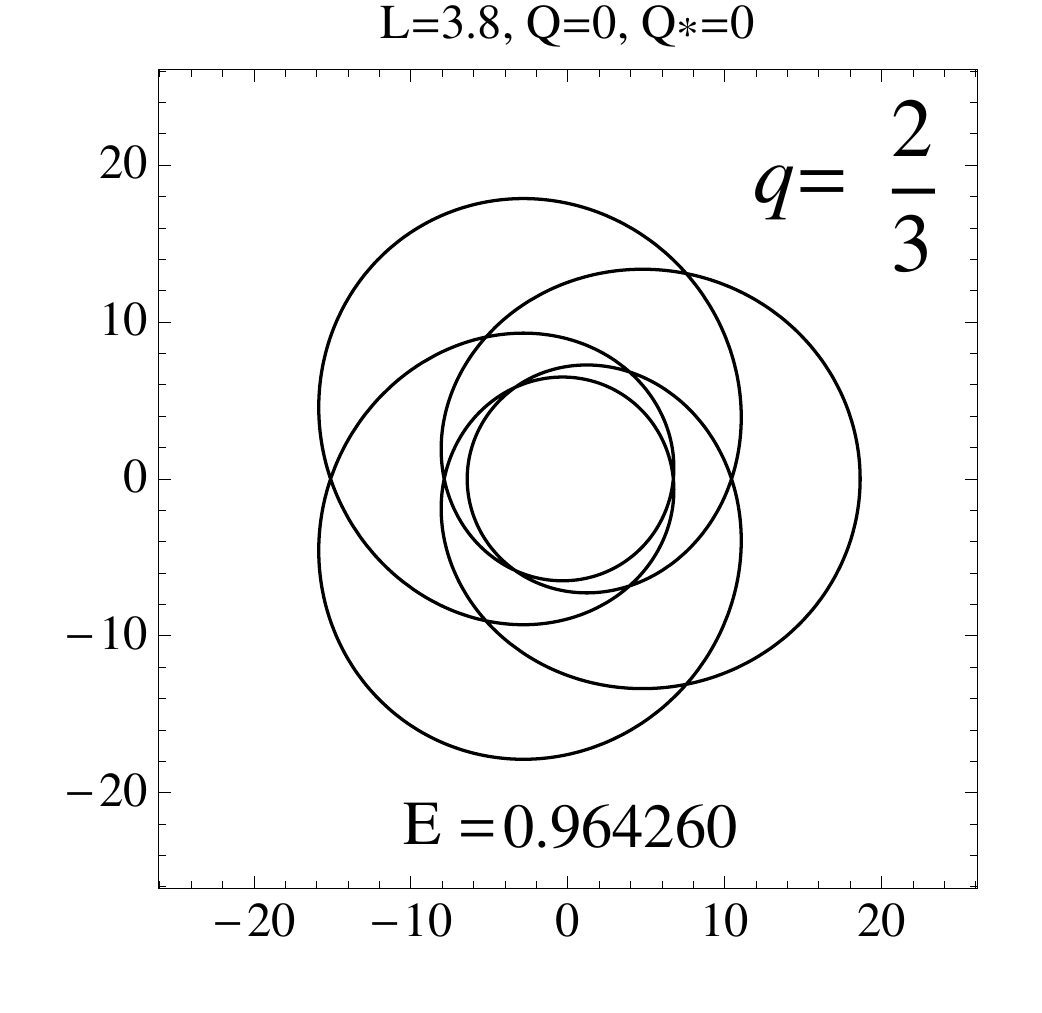}
	\hspace{-10pt}
	\includegraphics[width=0.32\textwidth]{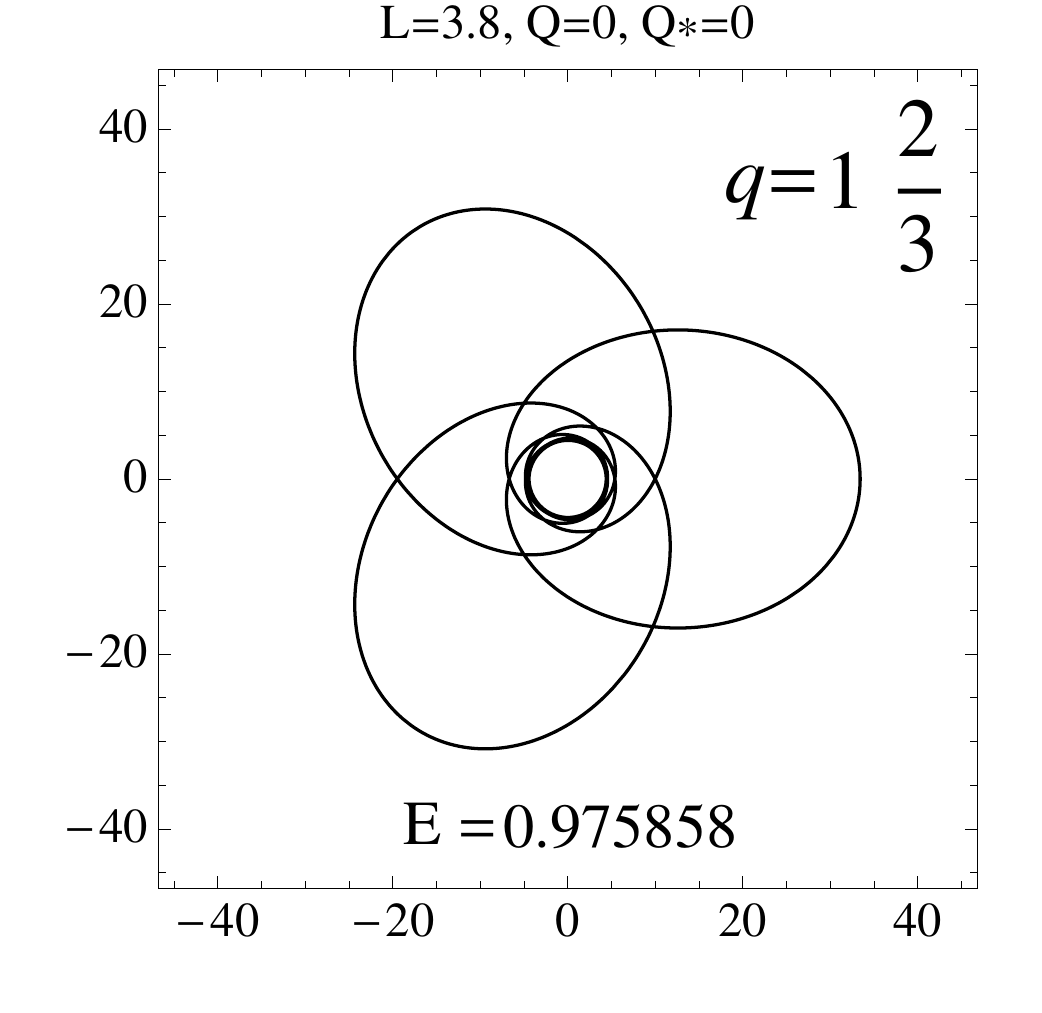}
	\hspace{-10pt}
	\includegraphics[width=0.32\textwidth]{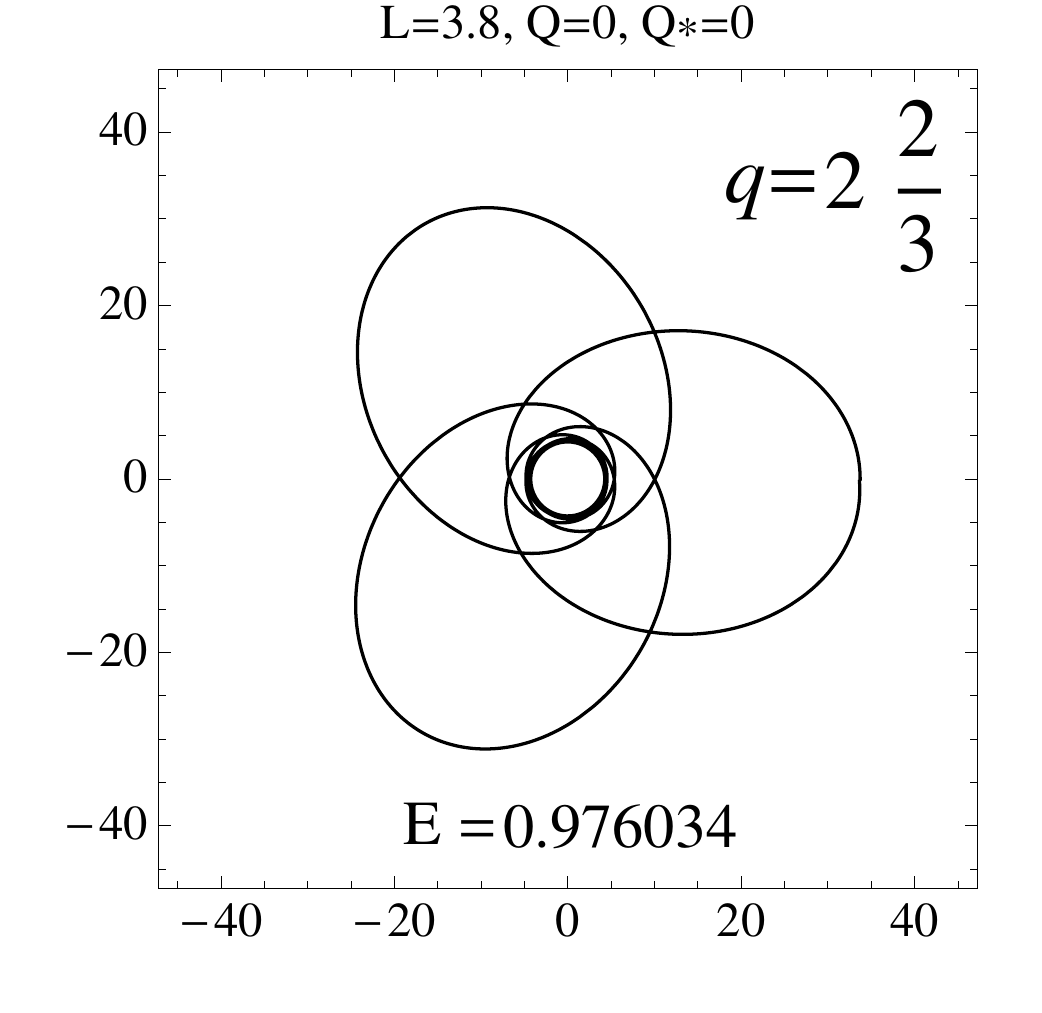} \hfill \\
	\vspace{-10pt}
	\caption{All Schwarzschild $z = $ 1, 2, 3 orbits with $w = $ 0 for the first column, $w = $ 1 
    for the middle column, and $w = $ 2 for the last column.
    Orbits increase in energy from top to bottom and left to right.
    \label{fig:ptschwarzschild}}
    \end{minipage}
\end{figure*}

\begin{figure*}[p]
    \begin{minipage}{.8\textwidth}
	\centering
	\includegraphics[width=0.32\textwidth]{plots/blank.pdf}
	\hspace{-10pt}
	\includegraphics[width=0.32\textwidth]{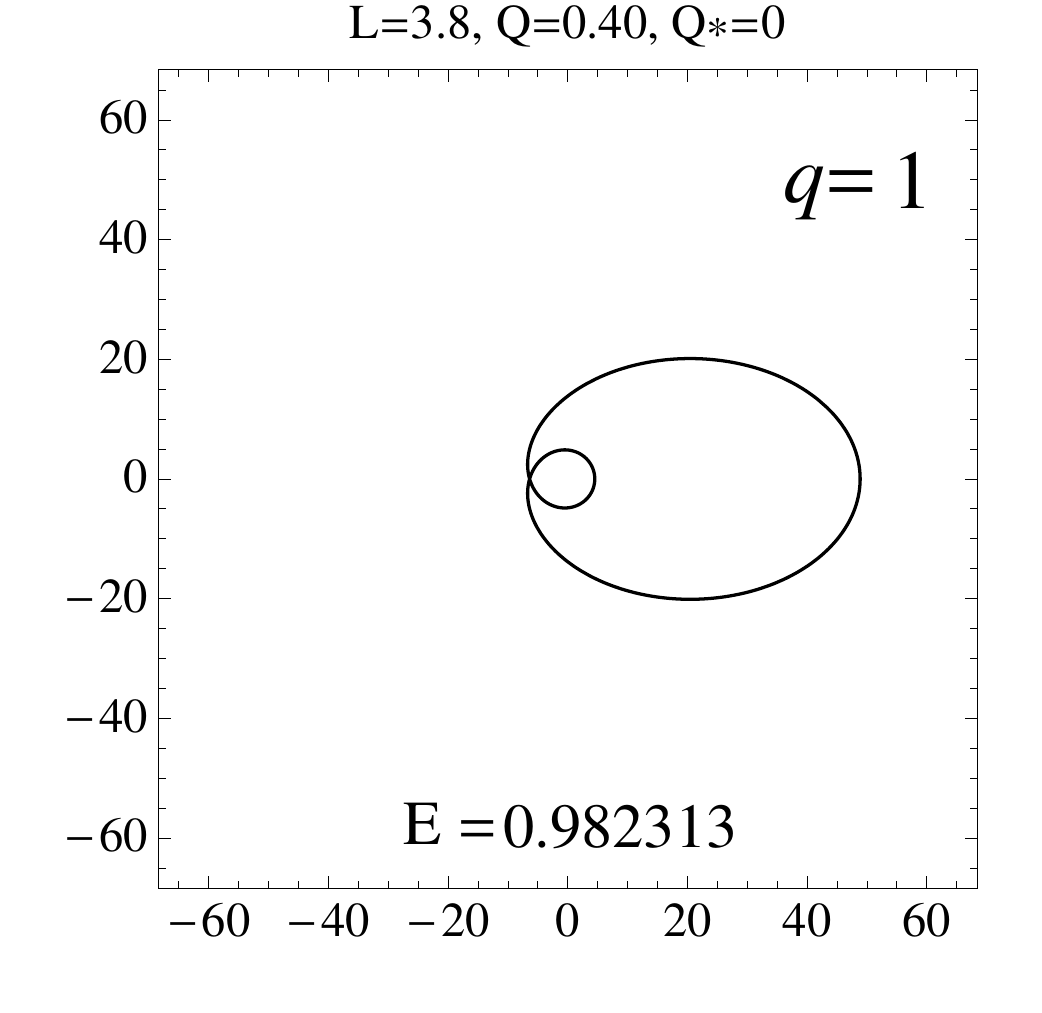}
	\hspace{-10pt}
	\includegraphics[width=0.32\textwidth]{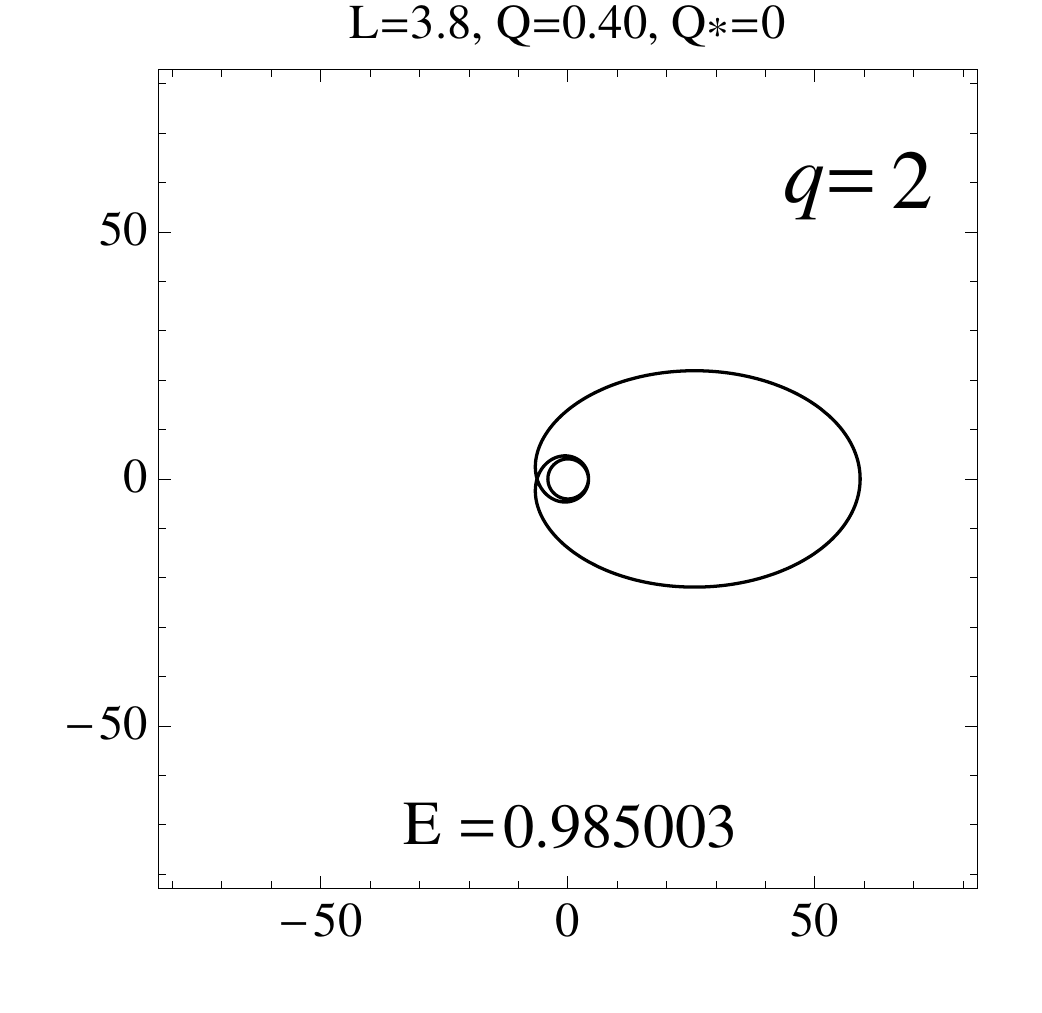} \hfill \\
	\includegraphics[width=0.32\textwidth]{plots/blank.pdf}
	\hspace{-10pt}
	\includegraphics[width=0.32\textwidth]{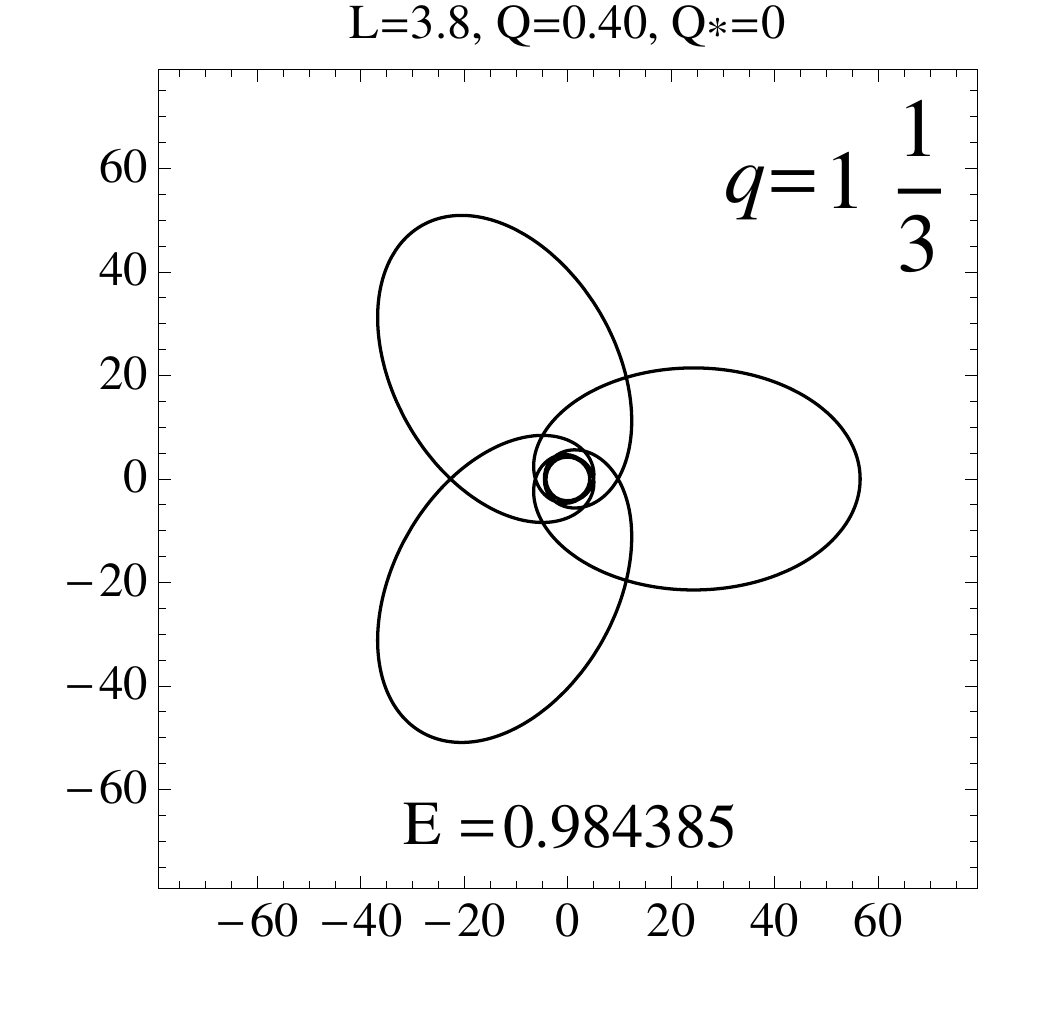}
	\hspace{-10pt}
	\includegraphics[width=0.32\textwidth]{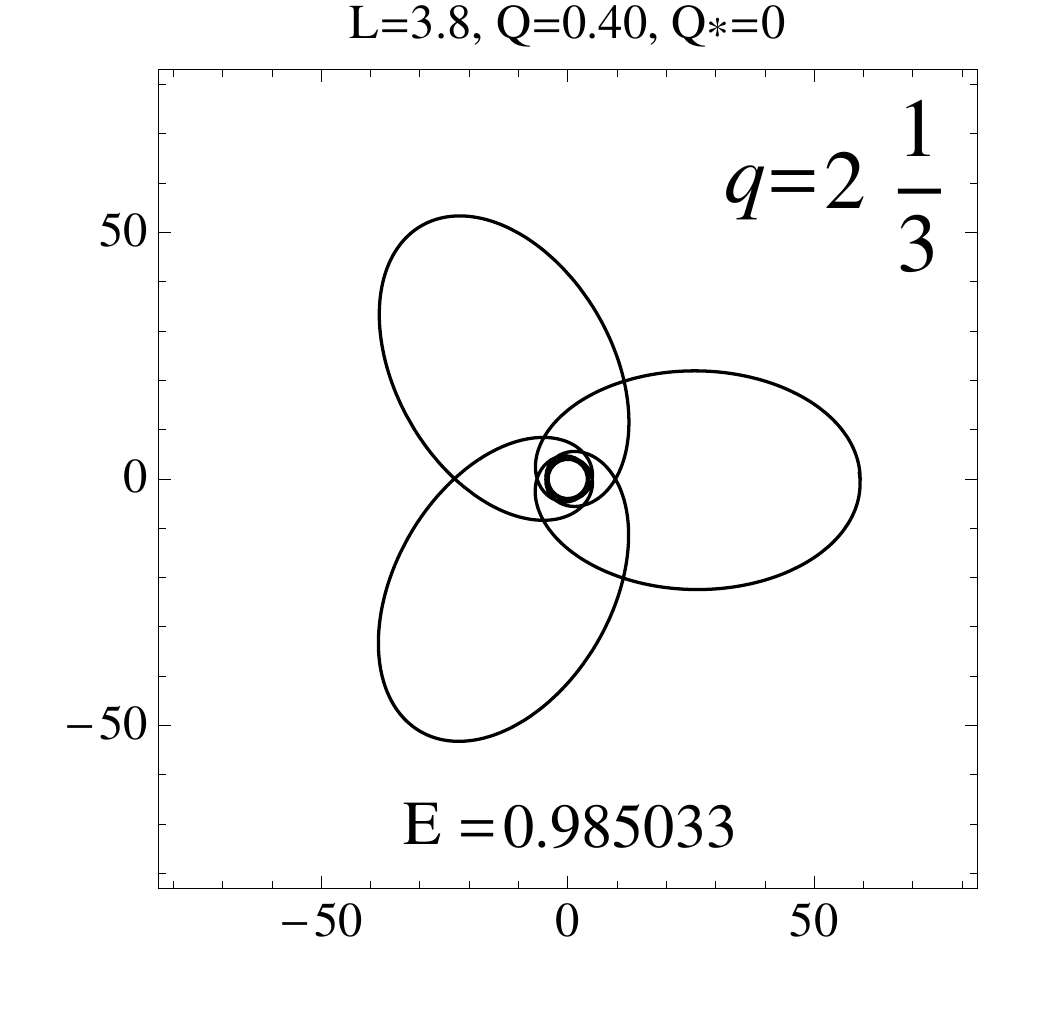} \hfill \\
	\includegraphics[width=0.32\textwidth]{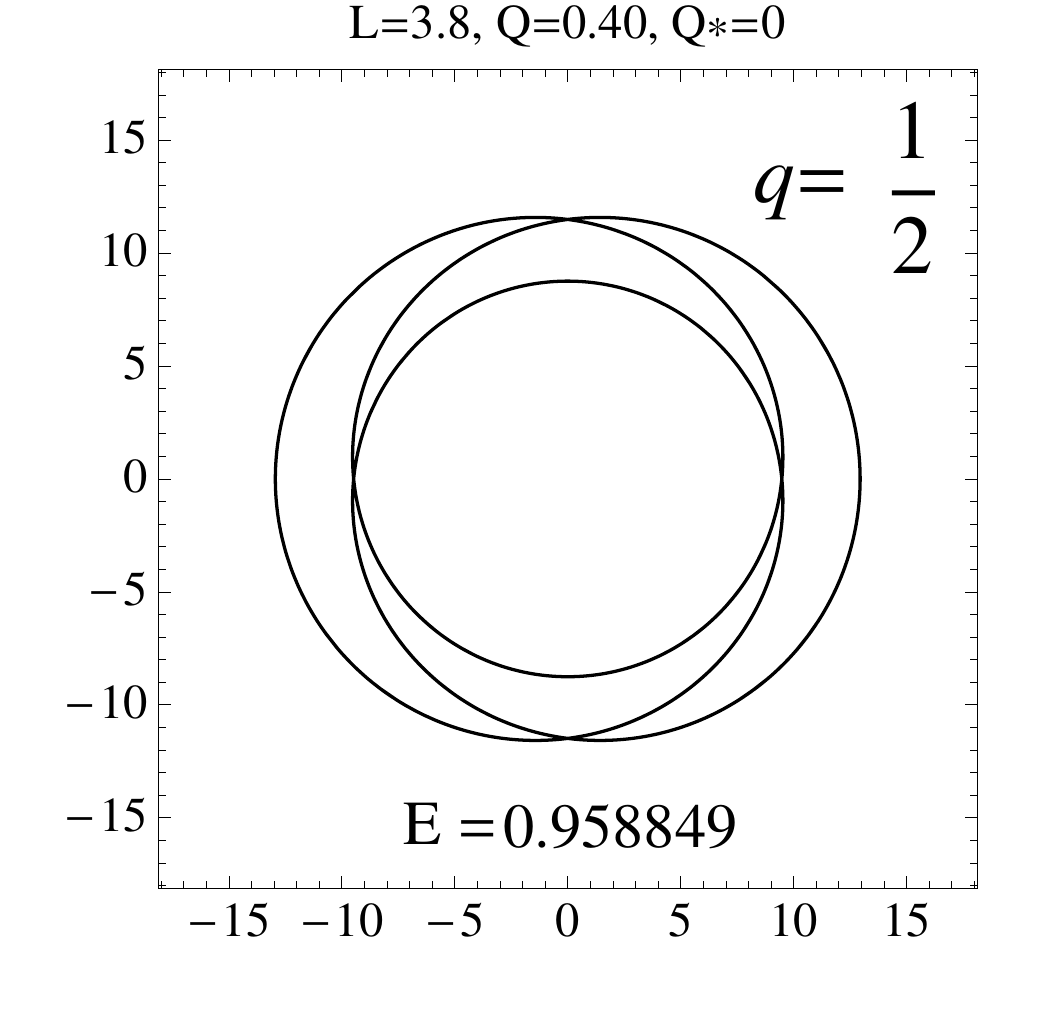}
	\hspace{-10pt}
	\includegraphics[width=0.32\textwidth]{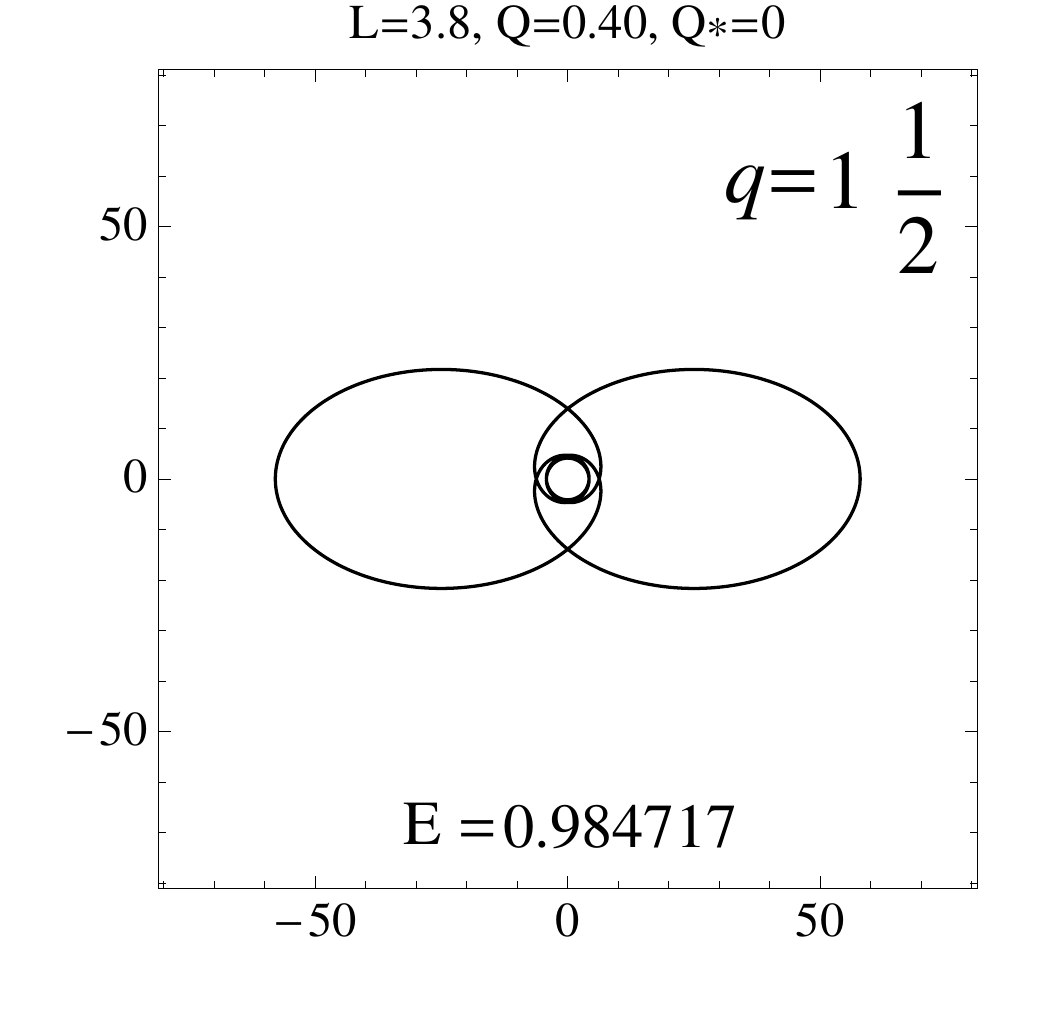}
	\hspace{-10pt}
	\includegraphics[width=0.32\textwidth]{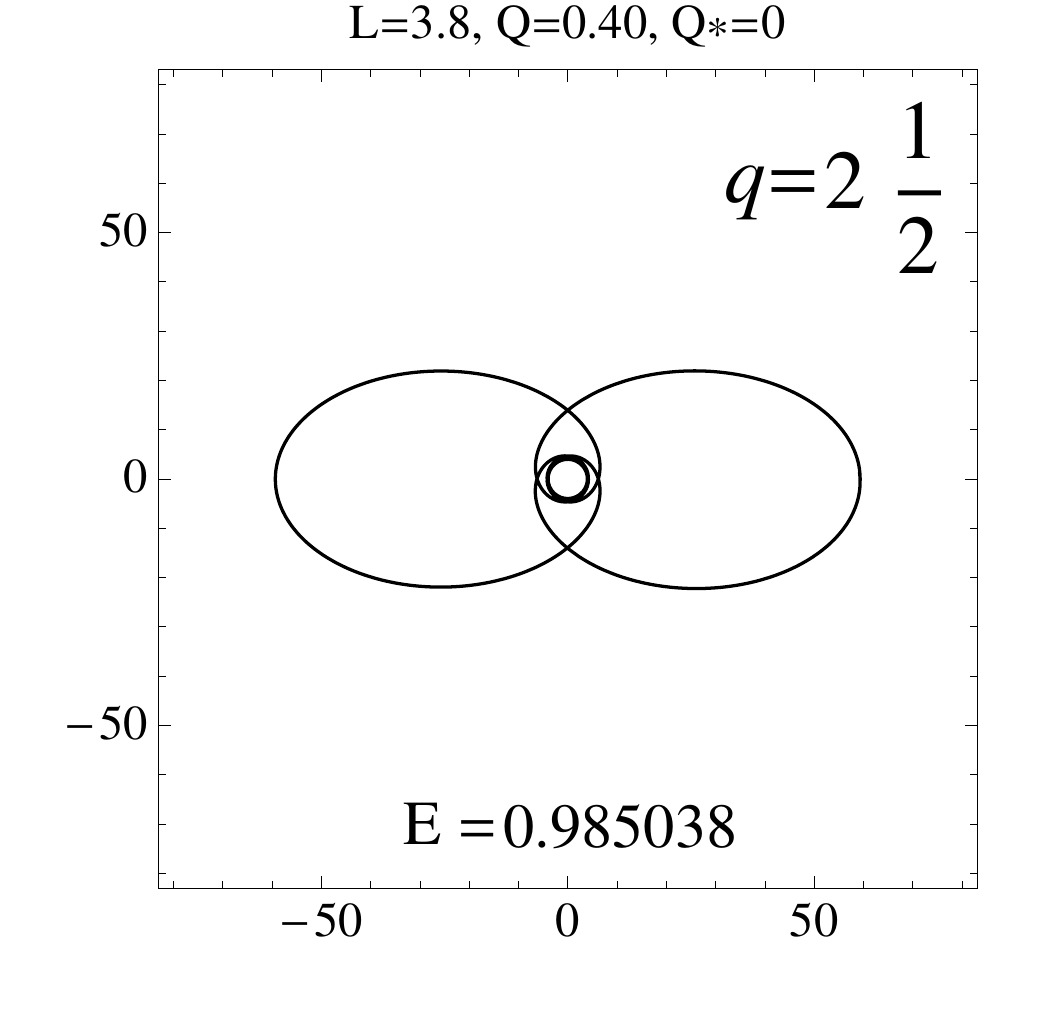} \hfill \\
	\includegraphics[width=0.32\textwidth]{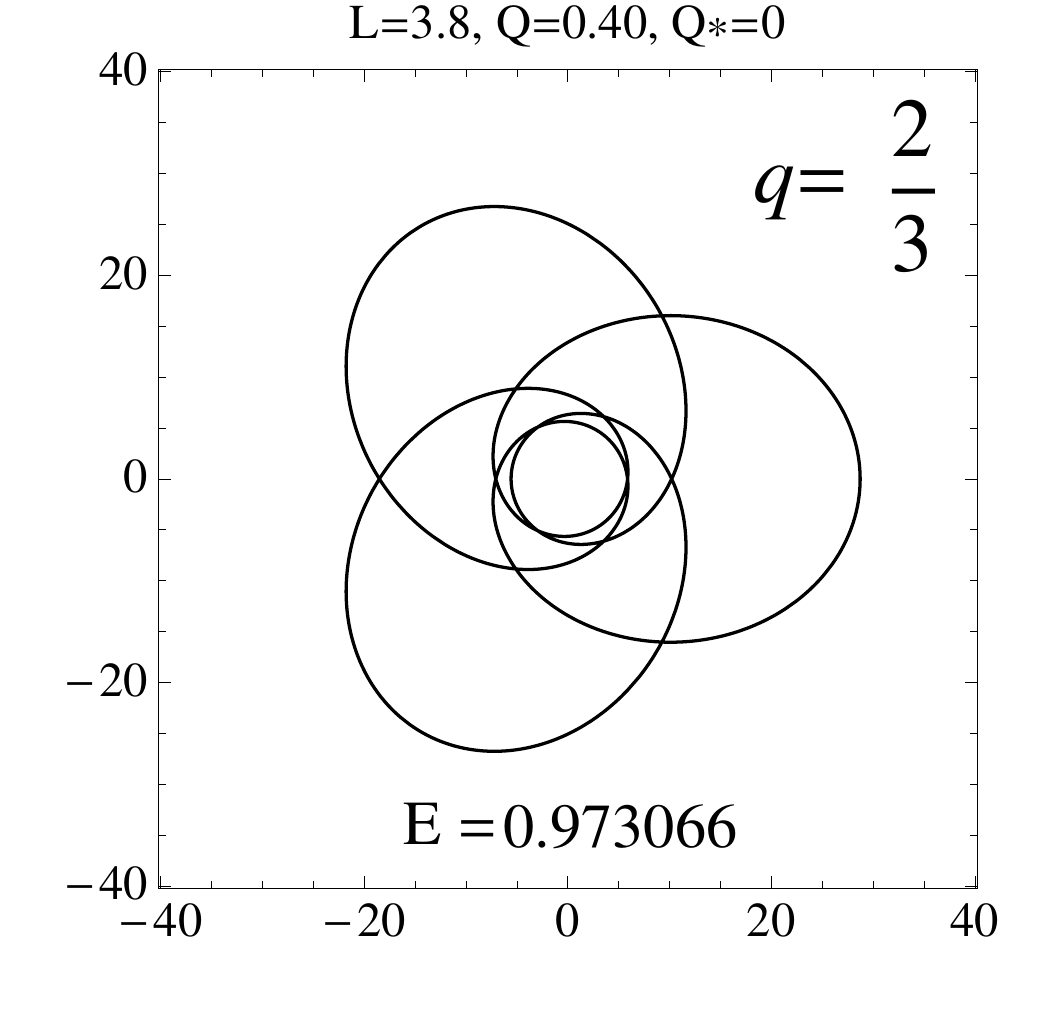}
	\hspace{-10pt}
	\includegraphics[width=0.32\textwidth]{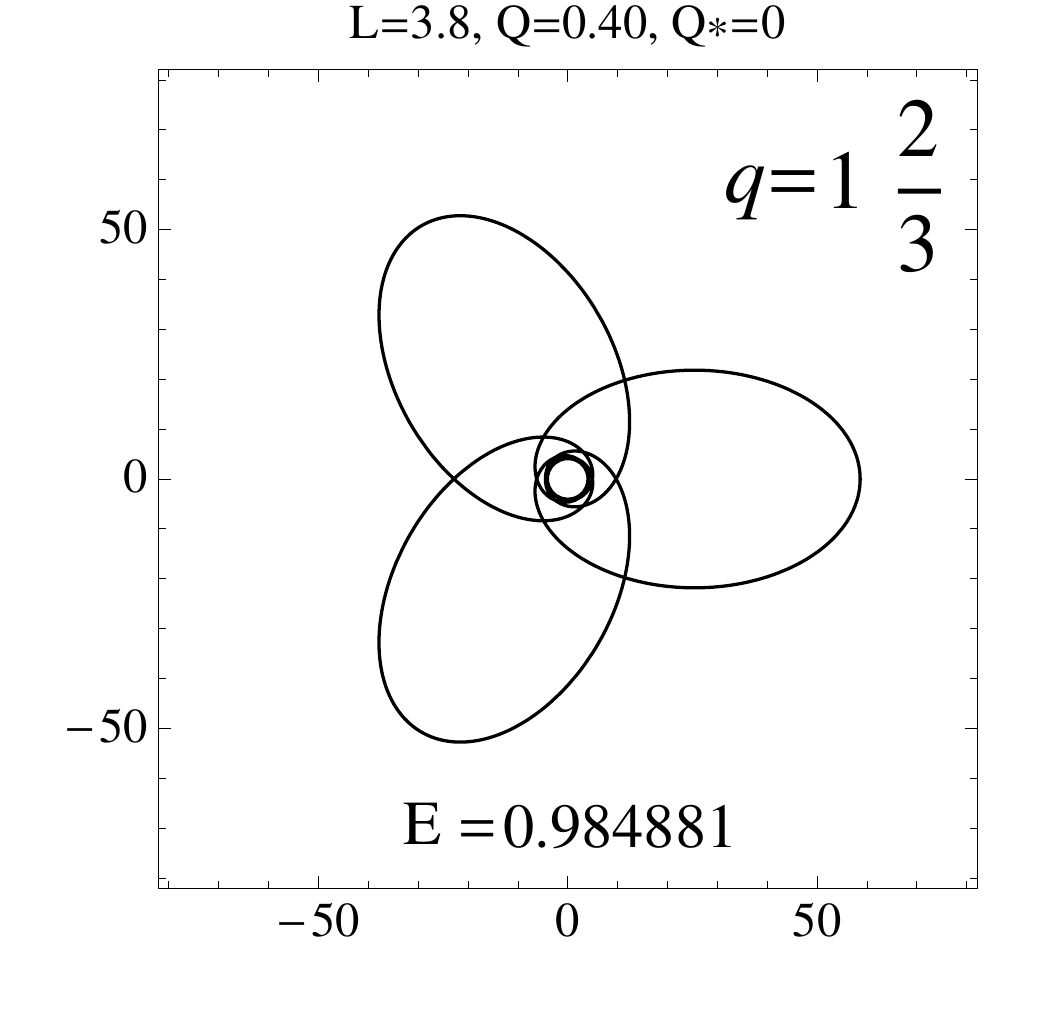}
	\hspace{-10pt}
	\includegraphics[width=0.32\textwidth]{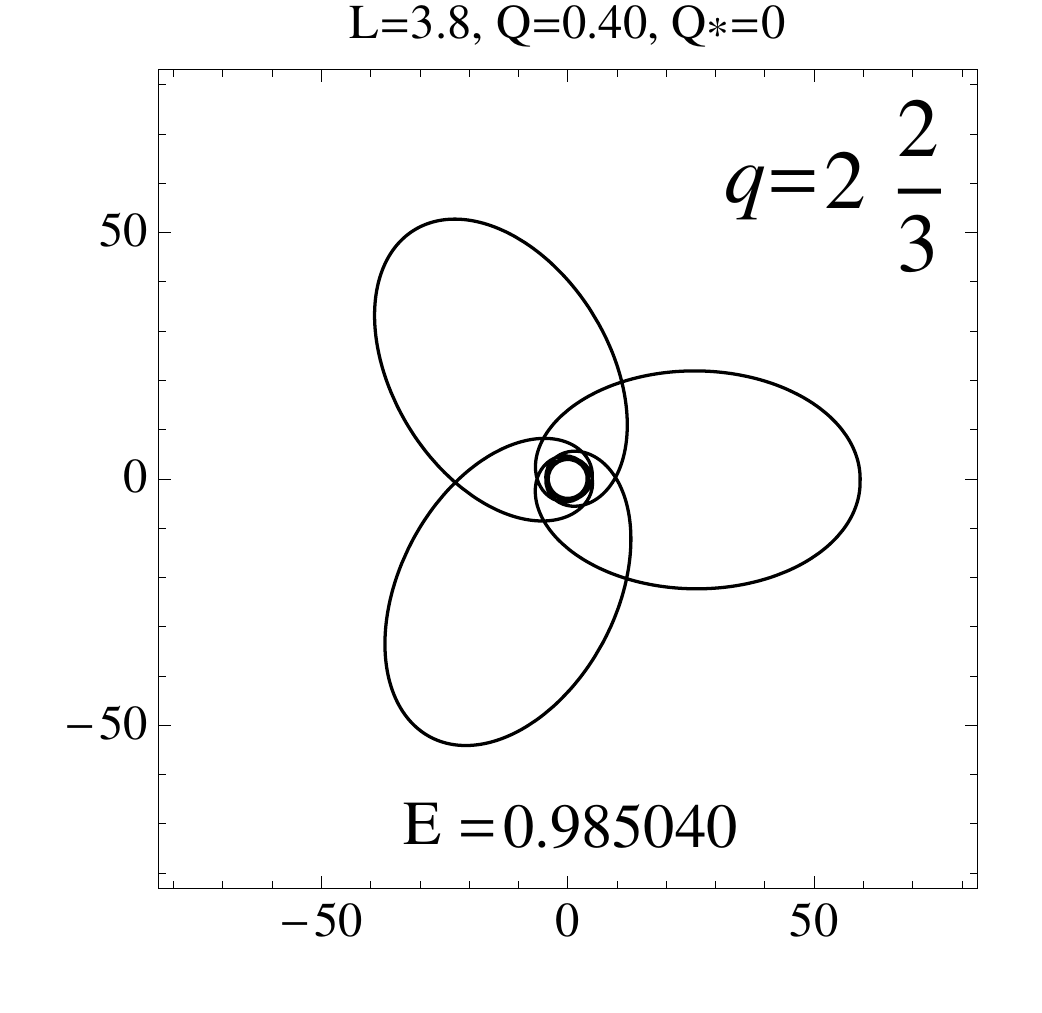} \hfill \\
	\vspace{-10pt}
	\caption{All RN $z = $ 1, 2, 3 orbits with $w = $ 0 for the first column, $w = $ 1 
    for the middle column, and $w = $ 2 for the last column.
    Orbits increase in energy from top to bottom and left to right.
    \label{fig:ptrn}}
    \end{minipage}
\end{figure*}

The taxonomy also gives us a way to visually inspect orbits in different spacetimes
to understand whether we can distinguish between them based solely on the 
dynamics of their periodic orbits.  For example,
an important question is whether it is possible to distinguish RN orbits
from Schwarzschild orbits in this way.  Figures \ref{fig:pt} and \ref{fig:ptschwarzschild}
depict periodic tables for the RN and Schwarzschild geomtries, respectively. By inspection,
it is evident that high $w$ orbits occur at higher energies in RN spacetime than
their Schwarzschild counterparts.  As a result, the apastra for RN orbits with 
$w > 0$ are consistently larger than for Schwarzschild orbits.  
Comparing orbits in Figure \ref{fig:ptschwarzschild} to those in Figure \ref{fig:ptrn} instead
allows us to compare orbits with equal $q$ for the same $L$; note however that here, the RN
geometry is not extremal, as $Q = 0.4$. Also note that while the $q = 1/2$ orbit does not
exist in the $L = 0.4$ Schwarzschild geometry, it can be found in the 
$L = 3.8$, $Q = 0.4$ spacetime.  We still find that high $w$ orbits occur at higher energies in
RN spacetime and that, again, the apastra for these orbits are consistently larger than those of 
their Schwarzschild counterparts.


\begin{figure*}
    \begin{minipage}{\textwidth}
	\centering
	\includegraphics[width=0.23\textwidth]{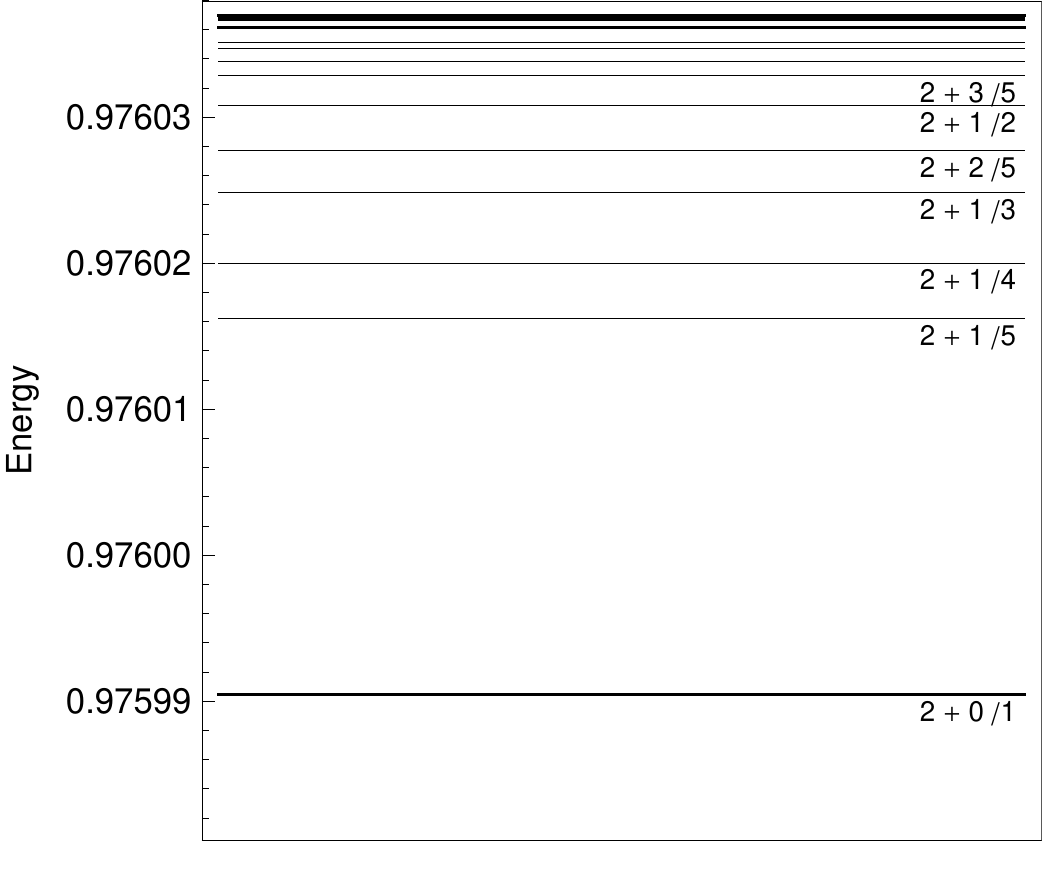}
    \hspace{2pt}
	\includegraphics[width=0.23\textwidth]{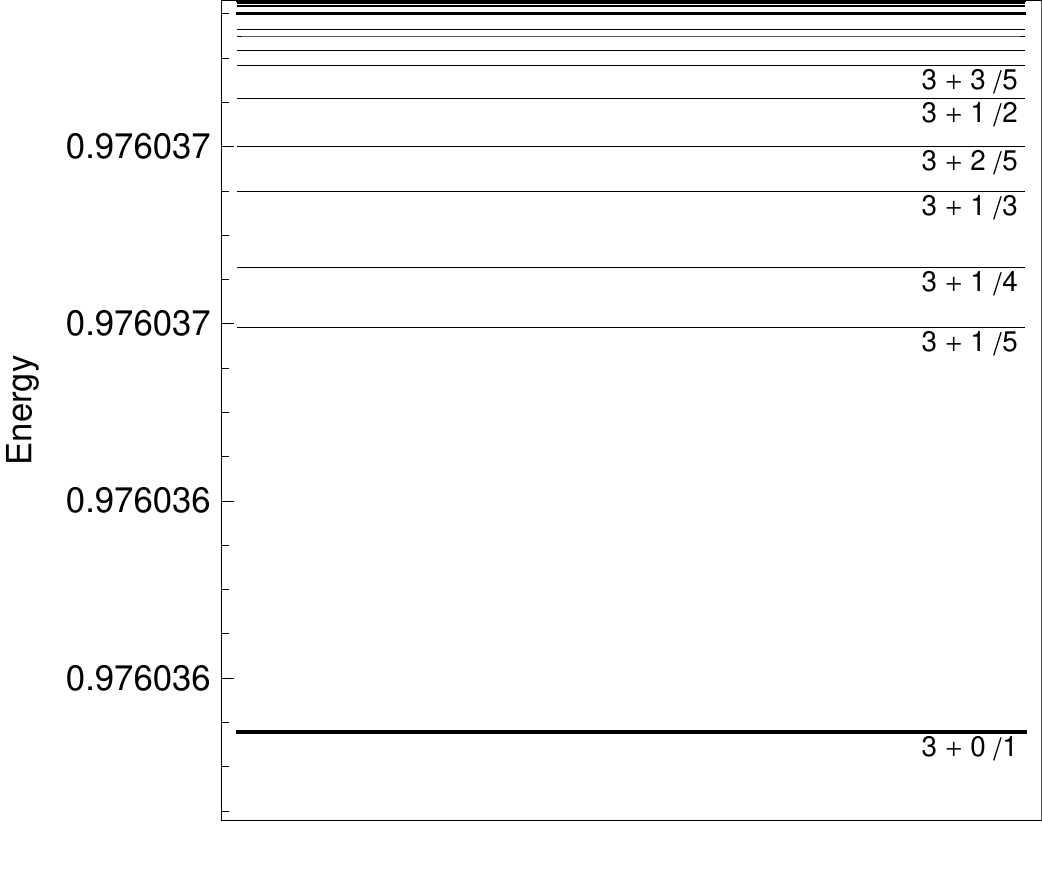}
    \hspace{2pt}
    \includegraphics[width=0.23\textwidth]{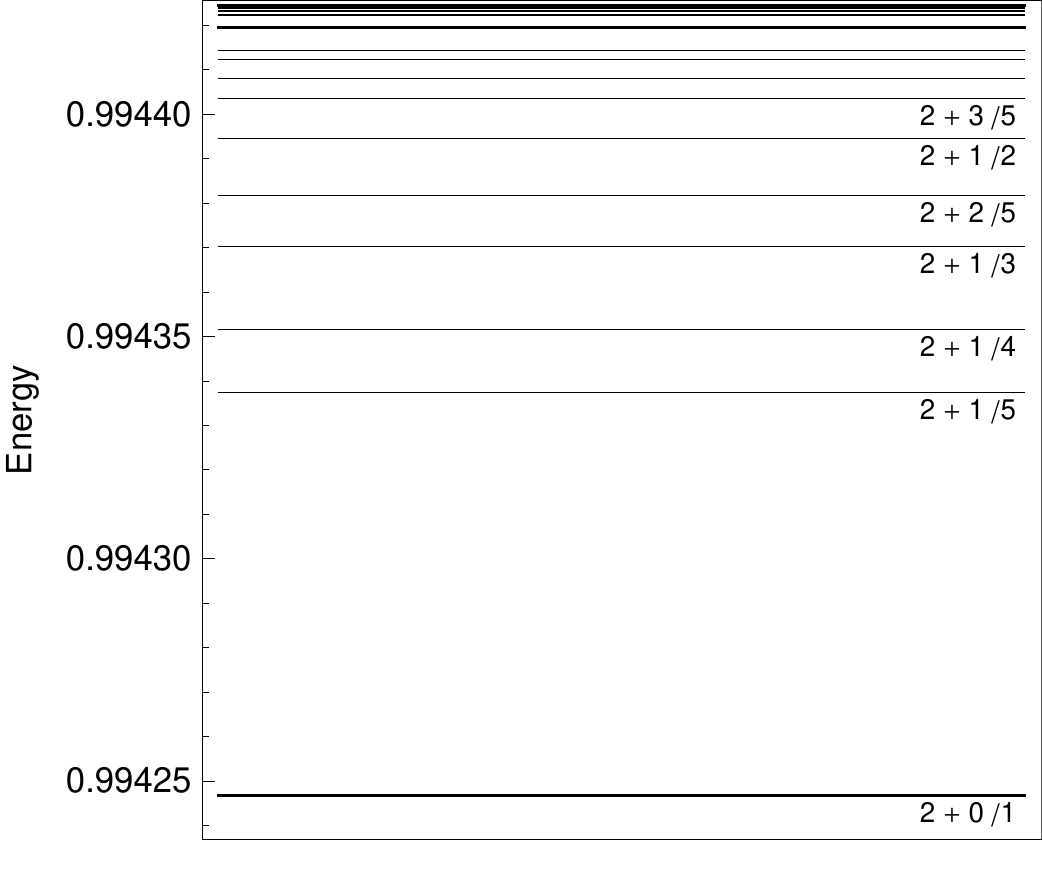}
    \hspace{2pt}
    \includegraphics[width=0.23\textwidth]{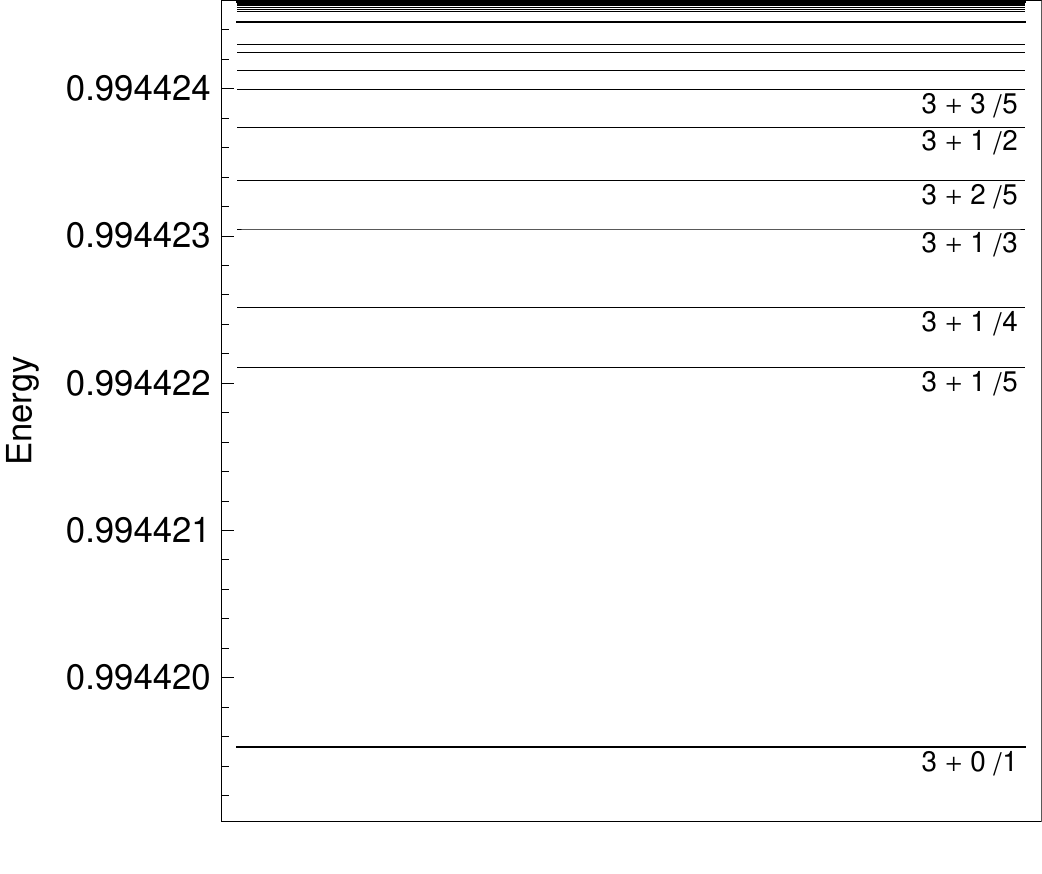} \hfill
	\caption{Energy levels for Schwarzschild orbits (first row) with $L = 3.8$ and RN Orbits (second row) with
$L = 3.3$, $Q = 1.0$, $Q_* = 0$.  The second and fourth figures show
detail of the first and third figures, respectively, 
indicating the rationals within the $w = 3$ band.  These diagrams show lines for all 
orbits with $w \in (2, 3, 4, 5, 6)$, $z \in (1, 2, 3, 4, 5)$ with $v < z$ and $v$ and $z$ coprime.\label{fig:energylevels}}
    \end{minipage}
\end{figure*}

\begin{figure}[htpb]
    \begin{minipage}{\columnwidth}
	\centering
	\includegraphics[width=\textwidth]{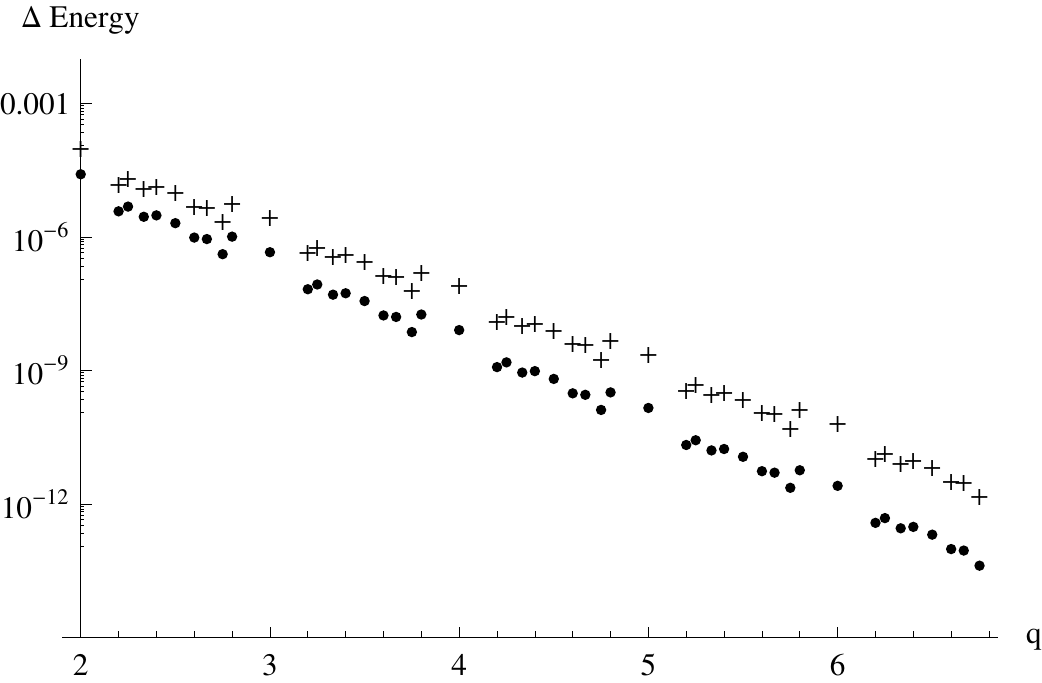} \hfill
	\caption{Differences in energies at which we find orbits of various $q$ in the RN
    (marked with ``$+$") and Schwarzschild (marked with ``$\cdot$") spacetimes.
    \label{fig:energydifferences}}
    \end{minipage}
\end{figure}

We may use energy level diagrams, such as the one in Figure \ref{fig:energylevels},
to understand the relationship between $q$ and $E$ in each spacetime (for a general discussion,
see \cite{levin_el}).  These diagrams
make it clear that RN orbits do appear at a higher set of energies.  Next we want to look 
at how the spacing between these energies varies.  Figure \ref{fig:energydifferences}
shows the differences between values of $E$ for successive orbits in the RN and Schwarzschild
cases. 

We find that not only are the RN energies collectively higher, they spacing between each pair
of energies is also consistently larger than in the Schwarzschild case.  Furthermore, the magnitude
of difference between spacings in each geometry increases.  The ratio of the energy difference in the
RN and Schwarzschild $q = 2$ and $q = 2\frac{1}{5}$ orbits is $3.52$, but this rises to $23.41$ 
when we compare the difference between the $q = 6$ and $q = 6\frac{1}{5}$ orbits in each geometry.
 
\section{Conclusions}
We have applied the taxonomy of \cite{levin} to the Reissner-Nordstr\"{o}m 
spacetime and demonstrated that within certain bounds for $Q$ and $L$, 
the orbital dynamics can be characterized using zoom-whirl behavior.  
These bounds were calculated for both the charged and uncharged particle cases.  
Furthermore it was demonstrated that the regions in which we find bounded 
orbits in RN spacetime are analogous to those in Schwarzschild spacetime, 
and that we do not find multiple stable circular orbits for any
choice of parameters within our predefined bounds.
Applying the taxonomy to RN spacetime for both charged and uncharged particles
enables us to differentiate between the set of periodic orbits in each geometry
based on their orbital dynamics.  We find not only that RN orbits occur at higher
energies than their Schwarzschild counterparts, but that this is a behavior that persists
for various $Q$.  Furthermore, we find that RN orbits of a given $q$ are more eccentric
than their Schwarzschild equivalents.

\section{Acknowledgments}
We are especially grateful to Alberto Nicolis for valuable discussions of this work.  
J. L. and V. M. acknowledge financial
support from NSF grant AST-0908365. This material is based 
in part upon work supported by a
scholarship from the Rabi Scholars program at Columbia University.

\appendix
\section{Hamiltonian formulation of RN geodesic motion}
For completeness we present a Hamiltonian formulation of RN
motion.  These were the equations integrated to generate the orbits in
the paper.
We begin with the Lagrangian density for a free particle in Reissner-Nordstr\"{o}m spacetime:
\begin{equation}
	\mathcal{L}=\frac{1}{2}g_{\alpha\beta}\dot{q}^\alpha\dot{q}^\beta.
\end{equation}
The $q^\alpha$ are dimensionless coordinates, as we have assumed that the orbiting particle is of unit mass.  Furthermore, for timelike trajectories, $\mathcal{L}=-1/2$.  Using
\begin{equation}
	p_\alpha\equiv\frac{\partial\mathcal{L}}{\partial\dot{q}^\alpha}
\end{equation}
we obtain the components of the momentum:
\begin{eqnarray}
	p_t 		& = & \frac{\partial\mathcal{L}}{\partial\dot{t}}=-\Delta\,\dot{t}\\
	p_r 		& = & \frac{\partial\mathcal{L}}{\partial\dot{r}}=\Delta^{-1}\,\dot{r}\\
	p_\theta 	& = & \frac{\partial\mathcal{L}}{\partial\dot{\theta}}=r^2\dot{\theta}\\
	p_\phi		& = & \frac{\partial\mathcal{L}}{\partial\dot{\phi}}=r^2\sin^2\theta\,\dot{\phi}.
\end{eqnarray}
The Hamiltonian is then defined
\begin{equation}
	\label{eqn:hamdef}
	\mathcal{H}=p_\mu\dot{q}^\mu - \mathcal{L}
\end{equation}
Like the $q^\alpha$, the particle's 4-momentum is also dimensionless and is here equivalent to the 4-velocity (since $\mu$, the particle mass, is set equal to 1).  So we may write Equation (\ref{eqn:hamdef}) as
\begin{equation}
	\mathcal{H}= \frac{1}{2}g^{\alpha\beta}p_\alpha p_\beta.
\end{equation}
If we compute $\mathcal{H}$ we find that
\begin{eqnarray}
	\mathcal{H} & = & \frac{1}{2}\Big(-\Delta^{-1}(-\Delta\dot{t})^2+\Delta^{-1}(\Delta\dot{r})^2 + \\ \nonumber
		        &   & r^{-2}(r^2\dot{\theta})^2+r^{-2}\sin^{-2}\theta(r^2\sin^2\theta\dot{\phi})^2\Big) \\ \nonumber
				& = & \frac{1}{2}\left(-\Delta\dot{t}^2+\Delta^{-1}\dot{r}^2+r^2\dot{\theta}^2+r^2\sin^2\theta\dot{\phi}^2\right) \\\nonumber
				& = & \mathcal{L}.
\end{eqnarray}
Because each quantity in the Hamiltonian is just half the contraction of the 4-momentum, it is identical to the Lagrangian, which is not surprising because our Lagrangian and Hamiltonian contain only kinetic terms \cite{levin}.

Next we wish to plug our Hamiltonian into Hamilton's equations,
\begin{equation}
	\dot{q}_i=\frac{\partial\mathcal{H}}{\partial p_i} \hspace{.5in} \dot{p}_i=\frac{\partial \mathcal{H}}{\partial q_i},
\end{equation}
which requires that we rewrite $\mathcal{H}$ in terms of the $p_i$.  Hamilton's equations give us explicitly
\begin{eqnarray}
	\frac{\partial \mathcal{H}}{\partial p_r} = \dot{r}		& = & \Delta\cdot p_r  \\ \nonumber
	\frac{\partial \mathcal{H}}{\partial p_t} = \dot{t}		& = & -\Delta^{-1} \cdot p_t \\ \nonumber
	\frac{\partial \mathcal{H}}{\partial p_r} = \dot{\theta}& = & r^{-2} \cdot p_\theta  \\ \nonumber
	\frac{\partial \mathcal{H}}{\partial p_r} = \dot{\phi}	& = & r^{-2}\sin^{-2}\theta \cdot p_\theta
\end{eqnarray}
for the $\dot{q}_i$ and, for the $\dot{p}_i$,
\begin{eqnarray}
	\dot{p}_t		=-\frac{\partial \mathcal{H}}{\partial t} 		& = & 0 \\ \nonumber
	\dot{p}_r		=-\frac{\partial \mathcal{H}}{\partial r} 		& = & -\dot{\theta}^2r-\sin^2\theta\dot{\phi}^2 r + \\ \nonumber
	& & \left(\frac{1}{r^2}-\frac{Q^2}{r^3}\right) \left(\frac{\dot{r}^2}{\Delta^2}+\dot{t}^2\right) \\ \nonumber
	\dot{p}_\theta	=-\frac{\partial \mathcal{H}}{\partial \theta} 	& = & -r^2\dot{\phi}^2 \sin\theta\cos\theta \\ \nonumber
	\dot{p}_\phi	=-\frac{\partial \mathcal{H}}{\partial \phi} 	& = & 0.
\end{eqnarray}

The equations above were used in the numerical results presented in the body of the paper.


\bibliographystyle{aip}
\bibliography{reissner}

\end{document}